\documentclass[fleqn,usenatbib]{mnras}
\usepackage{newtxtext,newtxmath}
\usepackage{lscape}
\usepackage{natbib}
\usepackage{url} 
\usepackage[T1]{fontenc}
\usepackage{ae,aecompl}
\usepackage{pdflscape}
\usepackage{footnote}
\usepackage{multicol}
\usepackage{supertabular}
\usepackage{longtable,lscape}
\usepackage{upquote}
\usepackage{units}
\usepackage{float}

\usepackage{graphicx}	
\usepackage{amsmath}	
\usepackage{amssymb}	
\usepackage{rotating}
\makeatletter

\usepackage[mathletters]{ucs}
\usepackage[utf8x]{inputenc}

\newcommand*{\rom}[1]{\expandafter\@slowromancap\romannumeral #1@}
\makeatother

\title[A new radio selected sample of Planetary Nebulae]{The Coordinated Radio and Infrared Survey for High-mass Star Formation. \rom{4}: A new radio selected sample of compact Galactic Planetary Nebulae}

\author[T. Irabor et al.]{T. Irabor,$^{1}$\thanks{E-mail: ee11ts@leeds.ac.uk (TI)}
M.G. Hoare,$^{1}$
R.D. Oudmaijer,$^{1}$
J.S. Urquhart,$^{2,3}$
S. Kurtz,$^{4}$
\newauthor S.L. Lumsden,$^{1}$
C.R. Purcell,$^{6}$
A.A. Zijlstra, $^{5}$
and  G. Umana,$^{7}$
\\
$^{1}$School of Physics \& Astronomy,University of Leeds, Leeds LS2 9JT, United Kingdom\\
$^{2}$Centre for Astrophysics and Planetary Science, University of Kent, Canterbury, CT2 7NH, United Kingdom\\
$^{3}$Max-Planck-Institut f$\ddot{u}$r Radioastronomie, Auf dem H$\ddot{u}$gel 69, D-53121 Bonn, Germany\\
$^{4}$Centro de Radioastronomia y Astrofisica, Universidad Nacional Autonoma de Mexico, Morelia, Michoacan, Mexico\\
$^{5}$School of Physics and Astronomy, University of Manchester, United Kingdom\\
$^{6}$Dept. of Physics and Astronomy, Macquarie University, NSW 2109, Sydney, Australia.\\
$^{7}$INAF-Osservatorio Astrofisico di Catania, Via S.Sofia 78, 95123, Catania, Italy
}

\date{Accepted XXX. Received YYY; in original form ZZZ}

\pubyear{2018}

\volume{{\rm in press}}
\begin{document}
\label{firstpage}
\pagerange{\pageref{firstpage}--\pageref{lastpage}}
\maketitle

\begin{abstract}
We present a new radio-selected sample of PNe from the CORNISH survey. This  is a radio continuum survey of the inner Galactic plane covering Galactic longitude, $10^\circ <l< 65^\circ$ and latitude, $|b| < 1^\circ$ with a resolution of 1.5$^{\arcsec}$ and sensitivity better than 0.4 mJy/beam. The radio regime, being unbiased by dust extinction, allows for a more complete sample selection, especially towards the Galactic mid-plane. Visual inspection of the CORNISH data, in combination with data from  multi-wavelength surveys of the Galactic plane, allowed the CORNISH team to identify 169 candidate PNe. Here, we explore the use of multi-wavelength diagnostic plots and analysis to verify and classify the candidate PNe. We present the multi-wavelength properties of this new PNe sample.  We find 90 new PNe, of which 12 are newly discovered and 78  are newly classified as PN. A further 47 previously suspected PNe are confirmed as such from the analysis presented here and 24 known PNe are detected. Eight sources are classified as possible PNe or other source types. Our sample includes a young sub-sample, with physical diameters < 0.12 pc, brightness temperatures (> 1000 K) and located closer than 7 kpc. Within this sample is a water-maser PN with a spectral index of $-0.55\pm 0.08$, which indicates non-thermal radio emission. Such a radio-selected sample, unaffected by extinction, will be particularly useful to compare with population synthesis models and should contribute to the understanding of the formation and evolution of PNe. 
\end{abstract}




\section{Introduction}
Planetary nebulae (PNe) are ionized envelopes of dust and gas, ejected during the AGB (Asymptotic Giant Branch) phase of intermediate mass stars ($0.8M_{\sun} \le M \le 8.0M_{\sun} $). These objects (PNe) are known to exhibit simple to complex morphologies \citep{bal1987,sahai2011}. More than $80 \%$ of observed PNe show non-spherical morphologies (\citealt{parker2006,sahai2007}. Also see \citealt{kwok2010}), which raises the question of how the deviation from spherical symmetry is achieved, especially with multi-polar, point-symmetric and irregular morphologies. Collimated fast winds or jets have been proposed as the primary shaping agents during the post AGB phase, where the jet characteristics determine the shape of the PNe when photoionization starts \citep{SahaiTrauger1998}. The genesis of the collimated jet activities and the driving mechanisms are still not well constrained. Other proposed shaping agents are rotation and/or magnetic fields in single stellar systems \citep{garcia1999,garcia2005}, binary companions and even triple stellar systems (see \citealt{soker2016,orsola2015,orsola2009} and references therein).

Understanding these objects and how they form and evolve is key to understanding the late phase evolution of intermediate-mass stars. Achieving this will require an uncontaminated and representative sample that covers the different stages of their evolutionary sequence, including very young PNe, where the physical processes associated with shaping and formation may still be active.  \cite{zi1991} predicted the Galactic disc PNe population to be $23000 \pm 6000$. From population synthesis, \cite{moe2006} estimated a total Galactic PNe population of $46000 \pm 13000$ PNe (with radii $<$ 0.9 pc), assuming single and binary stellar systems result in PNe. About $3540$ have been identified so far, including true, likely and possible PNe \citep{parker2017}.

Optical surveys are usually used to search for, and identify these objects from characteristic emission lines associated with their ionized regions. However, the optical regime is limited in various ways, especially in the Galactic plane. Optical surveys are strongly affected by dust extinction, which causes a bias in the sample selection of PNe from such surveys. An optical survey cannot always distinguish very young and compact PNe from other compact, ionized objects such as H \rom{2} regions \citep{parker2010} and symbiotic systems \citep{kwok2003}. To overcome dust extinction and incompleteness associated with optical surveys, it is necessary to observe PNe at longer wavelengths.

\par Samples of PNe themselves are not homogeneous because they possess different morphologies, central star masses, evolutionary stages, ionization characteristics and diffrent environments. This makes the identification of PNe more challenging. However, to eliminate possible sample contamination by other sources that share similar observational properties and apply a proper constraint on the source classification, there is a need for multi-wavelength techniques \citep{parker2010}. This is possible because PNe have associated ionized gas, dust emission and some even have molecular emission \citep{kwok2007,ramos2017}, making them strong emitters at infrared and radio wavelengths. Infrared surveys can be used to search for PNe that are not visible in the optical because of extinction and free-free emission from their associated ionized gas can be explored at radio wavelengths.

The CORNISH survey provides a high resolution and unbiased radio survey of the northern Galactic plane (\defcitealias{cornish2012}{Paper~I}\citealt{cornish2012}. Hereafter \citetalias{cornish2012}). It covers the northern survey region of the Spitzer GLIMPSE I survey \citep{churchwell2009, benjamin2003}. Hence, it provides a complementary radio data and, with its depth and resolution, it is well suited for the search and study of compact PNe. The CORNISH team have visually identified 169 candidate PNe (CORNISH-PNe) from this survey using their appearance at different wavelengths (a combination of the radio, infrared image data and millimetre dust continuum. See Section \ref{selection}).

The first large scale radio survey that was used to construct a sample of PNe unbiased by extinction is the NRAO VLA sky survey (NVSS) with a resolution of $45^{\arcsec}$ \citep{condon1998}. Within the coverage of this survey (full sky survey north of $ \delta -40$), 702 of the then known 885 PNe in the Strasbourg-ESO catalogue of Galactic planetary nebulae \citep{acker1992} were detected at the time \citep{condon1999, condon1998_2}. With a sensitivity of $ \sim 2.5\  mJy/beam$, this survey was also used to reject contaminants and detect free-free emission towards 454 candidate PNe in the IRAS point source catalogue, 332 of which were known PNe and 122 were identified as candidate PNe \citep{condon1999}.

In this paper, we examine and analyse the nature of the 169 CORNISH-PNe. We also explore their multi-wavelength properties to classify and identify possible contaminants within the sample. Section \ref{2} describes the data and PNe sample selection. The radio properties and use of different multi-wavelength diagnostic plots, including extinction and distances are presented in Section \ref{3}. Discussion of results is presented in Section \ref{4} and the CORNISH-PNe catalogue is described in Section \ref{5}.

\section{Candidate Radio PNe Sample} \label{2}
\subsection[]{The CORNISH Survey}
The CORNISH survey is a 5 $GHz$ radio continuum survey covering 110 $\ deg^2$ of the northern Galactic region, defined by $10^\circ < l < 65^\circ$ and $|b| < 1^\circ$. It probes free-free emission from the ionized regions of compact Galactic sources using the B and B\textquotesingle n\textquotesingle A configurations of the VLA (Very Large Array) \citep{cornish2012}. With its resolution of $1.5^{\arcsec}$ and sensitivity better than 0.4 mJy/beam, it forms the radio continuum part of a series of high-resolution, high-sensitivity multi-wavelength surveys of the northern Galactic plane. The CORNISH survey provides a combination of higher resolution and depth compared to previous radio surveys of the Galactic plane.

Objects identified in the CORNISH survey include H \rom{2}  regions, PNe, radio galaxies with lobes and radio stars. The resolution and sensitivity of this survey, together with other comparable surveys of the northern Galactic plane at different wavelengths, is able to  discriminate between compact ionized sources. In addition, it should aid identification of new PNe not detected by previous surveys. Details of the survey design, scientific motivation, data reduction and measurements of properties are provided in previous papers by \citetalias{cornish2012} and \defcitealias{cornissh2013}{Paper~II}\cite{cornissh2013} (hereafter \citetalias{cornissh2013}).

\subsection{PNe Candidate Selection}\label{selection}

PNe emit at radio and infrared wavelengths due to their ionized gas and warm dust. Despite the different morphologies, ionization characteristics and central star masses, PNe typically appear red and isolated in the infrared. However, they are sometimes confused with H \rom{2} regions and both can get mixed up in the same sample, depending on the distance and how compact and/or young the H \rom{2} regions are. This is because H \rom{2} regions also emit at radio and infrared wavelengths from associated ionized gas and dust.

In the infrared, these two classes of objects can be distinguished by looking at their local environment and morphology. H \rom{2} regions, unlike PNe, are associated with dense molecular material, which becomes very evident in the MIR as strong PAH (polycyclic aromatic hydrocarbon) emission. This also gives them strong millimetre (mm) dust continuum \citep{urq2013}, whereas PNe are usually very faint or not detected in dust continuum surveys. A comparison of the CORNISH and ATLASGAL survey \footnote{http://www3.mpifr-bonn.mpg.de/div/atlasgal/} revealed only a handful of PNe (see \citealt{urq2013}). H \rom{2} regions also show regions of strong dust extinction and often form in complexes. PNe, on the other hand, often do not show any form of association with dense molecular material or form in complexes. Furthermore, H \rom{2} regions often show a variety of irregular morphologies \citep{wood21989} as opposed to PNe whose morphologies usually show a good degree of symmetry \citep{sahai2011}. This difference in morphology is also clearly seen in the MIR. In the near-infrared, H \rom{2} regions are redder than PNe due to the amount of dust emission and sometimes they are not even detected, owing to very high extinction.

The SEDs (spectral energy distribution) of PNe usually peak between 20 $\umu$m and 100 $\umu m$. However, some PNe could peak above 100 $\umu m$ (see \citealt{urq2013}), due to the range of dust temperatures ($\mathrm{30\ K \ga T_d > 100\ K}$) in their circumstellar shell  \citep{vill2002,pottasch1984, kwok1986}.  H \rom{2} regions, on the other hand, peak at longer wavelengths of about $70\ \umu m$ and above because they possess a larger fraction of cooler dust \citep{anderson2012,wood11989}. This would result in an overlap in the SED of some PNe with H \rom{2} regions.  

\begin{figure*}

	\includegraphics[width=16cm]{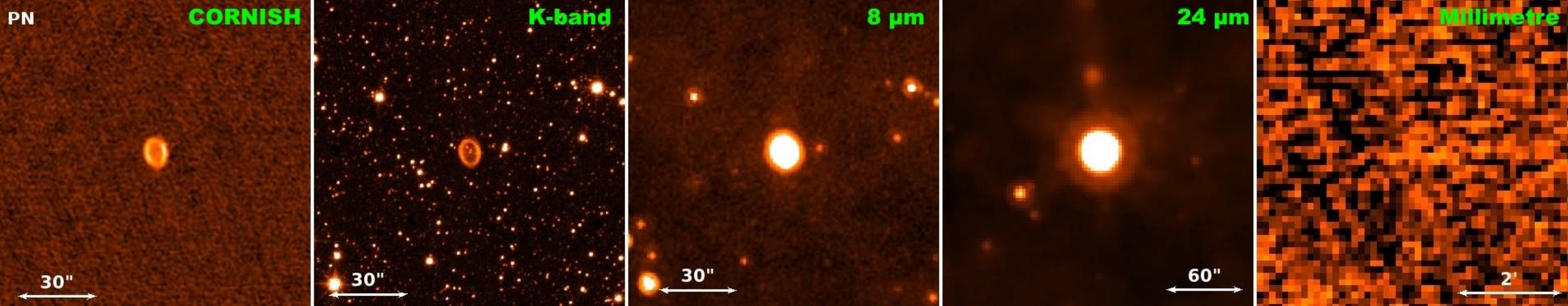}
	
	\includegraphics[width=16cm]{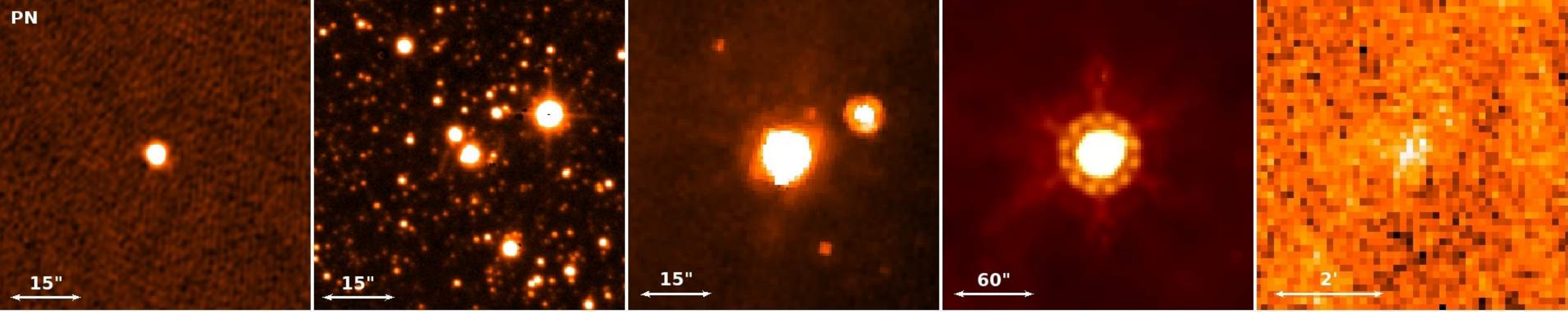}
	
	\includegraphics[width=16cm]{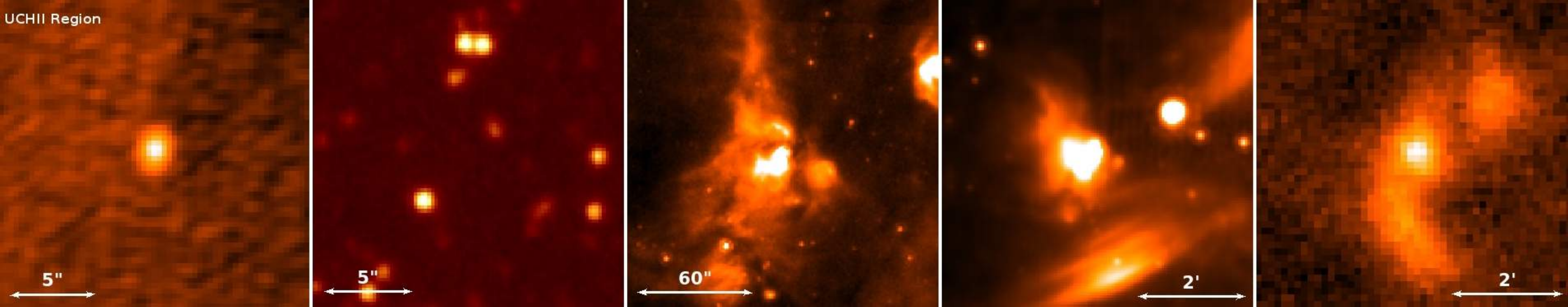}
	\includegraphics[width=16cm]{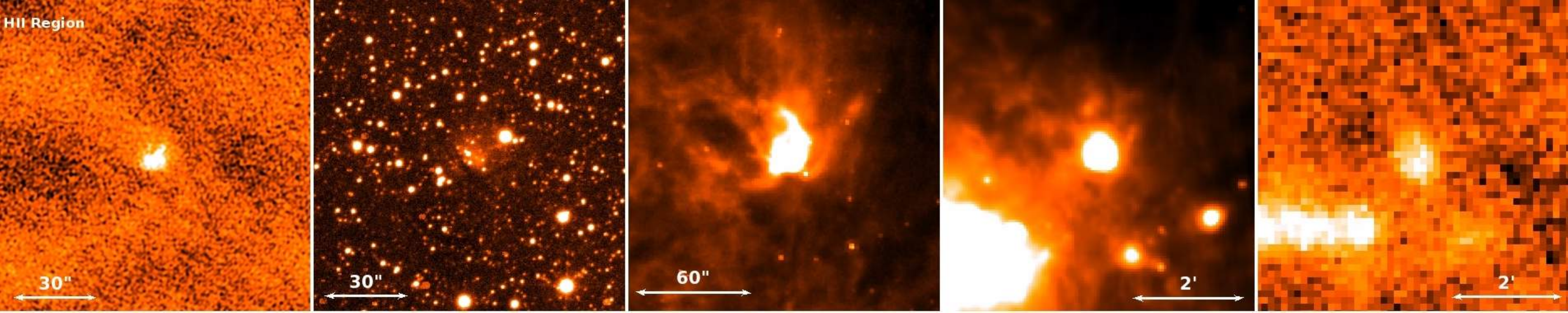}
	\includegraphics[width=16cm]{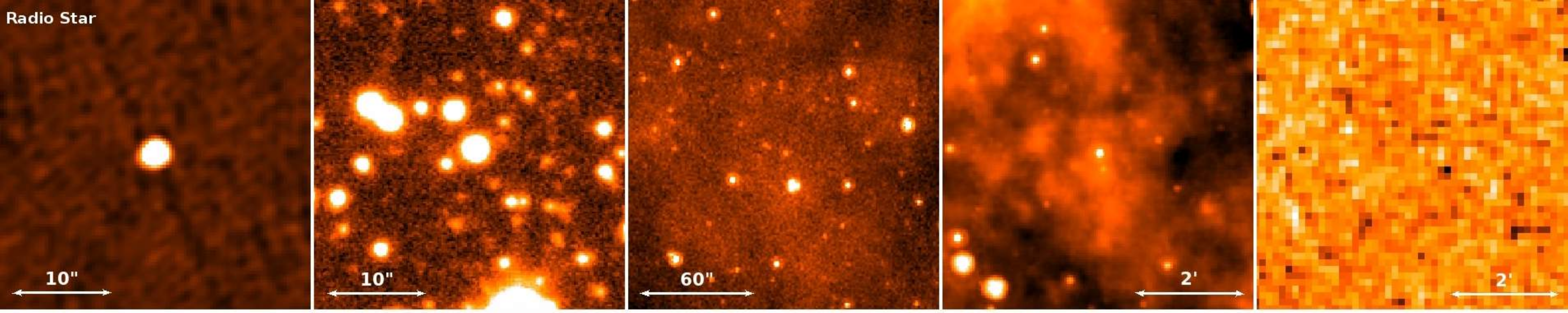}
	\includegraphics[width=16cm]{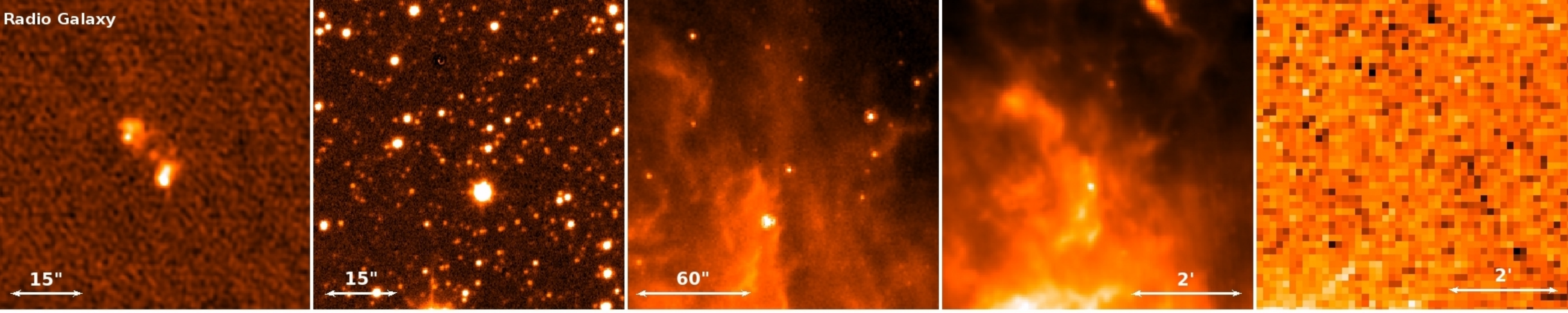}

    \caption{Multi-wavelength observational differences between PNe and other sources identified in the CORNISH survey. Surveys from left to right: CORNISH 5 GHz radio, UKIDSS-GPS (K band), GLIMPSE ($8\ \umu m$), MIPSGAL ($24\ \umu m$) and BGS 1.1 mm dust continuum. The sources from top to bottom are PNe: G051.5095$+$00.1686 and G035.4719$-$00.4365; UCH II region: G010.3204$-$00.2586; H II region: G053.1865$+$00.2085; Radio star: G045.3657$-$00.2193 and Radio galaxy: G054.1703$-$00.0092. These are the main multi-wavelength surveys used by the CORNISH team to select the CORNISH-PNe sample.}
    \label{sal2}
\end{figure*}

Using these observational differences across the different wavelengths over which PNe emit, the CORNISH team, from visual inspection of multi-wavelength image data, classified 169 of the detected sources in the $7 \sigma$ catalogue as candidate PNe (CORNISH-PNe). These criteria were also used by \cite{urq2009} to distinguish between PNe and H \rom{2} regions in the Red MSX Source survey \citep{lum2013}. The CORNISH-PNe make up $ \sim 6\%$ of the CORNISH catalogue. The multi-wavelength image data used include the Multi-array Galactic Plane Survey (MAGPIS) at 20 cm \citep{hel2006,white2005}, the Galactic Legacy Infrared Mid-Plane Survey Extraordinaire (GLIMPSE; \citealt{churchwell2009, benjamin2003}), MIPSGAL \citep{carey2009}, the United Kingdom Infrared Deep Sky Survey (UKIDSS-GPS;  \citealt{lucas2008}) and the Bolocam Galactic plane survey (BGPS; \citealt{rob2010}). The use of such multi-wavelength data provides a good constraint on the visual classification of compact objects (see Figure \ref{sal2}), especially towards the Galactic mid-plane, where source crowding is a problem. Observational differences at different wavelengths used by the CORNISH team are illustrated in Figure \ref{sal2}.

With the visual classification, it is expected that radio stars, radio galaxies and other extra-galactic sources have been eliminated. However, having done visual classification of these objects, there is a possibility that the sample is still not contaminant-free. Possible contaminants are dusty radio stars and H \rom{2} regions, especially when they are distant and still very young.  

\subsection{Multi-Wavelength Data} \label{otherdata}
The CORNISH survey provides information at a single frequency (5 GHz), which is not sufficient to determine the nature and discriminate between compact source types in the CORNISH catalogue. To search for CORNISH-PNe counterparts at other wavelengths, we have queried data from other surveys using the CORNISH coordinates (positional accuracy of 0.1$\arcsec$. See \citetalias{cornissh2013}). Data query and cross-matching is generally limited by coverage and resolution, and so the choice of surveys considered is based on comparable resolution and coverage. We briefly describe below the details of other surveys in the northern Galactic plane that we have used in the analysis presented here.

\subsubsection{MAGPIS (20 cm) }
The MAGPIS at 20 cm is a radio continuum survey of the northern Galactic plane, defined by $5^\circ < l < 48^\circ.5$ and $|b| < 0^\circ.8$. Observations were made using the VLA (Very Large Array) with an angular resolution of $\sim 5^{\arcsec}$ and rms noise level of $\sim 0.3 $ mJy. Details of this survey are given in \cite{hel2006}. We have queried the $> 5\sigma$ catalogue for CORNISH-PNe counterparts from the compact radio catalogues compiled by \cite{white2005}.

\subsubsection{GLIMPSE (Mid-Infrared)}
The GLIMPSE, as one of the Spitzer legacy science programs, is a survey of the inner Galaxy in the mid-infrared. It covers an area of 220 $deg^2$, defined by longitude $10^\circ < l <  65^\circ$, $-10^\circ > l > -65^\circ$ and latitude $|b| \leq 1^\circ$ with a spatial resolution of $ \sim 2^{\arcsec}$. This survey made use of the infrared array camera (IRAC) mounted on the Spitzer space telescope (SST), centred on $3.6\ \umu m$, $4.5\ \umu m$, $5.8\ \umu m$ and $8.0\ \umu m$. It achieved a $3\sigma$ point source sensitivity of 0.2, 0.2, 0.4 and 0.4 mJy for the four bands respectively \citep{churchwell2009, benjamin2003}.

\subsubsection{UKIDSS (Near-infrared)}
The GPS (Galactic plane survey), which is one of the five UKIDSS surveys, covers an area of 1868 $deg^2$ defined by Galactic latitude $|b| \le 5^\circ$ and longitude of $15^\circ < l < 107^\circ$, and $142^\circ < l < 230^\circ$ with a seeing of $\sim  1.0^{\arcsec}$. These regions were imaged in the J, H and K bands centred on $1.25\ \umu m$, $1.65\ \umu m$ and $2.20\ \umu m$, to a depth of 20, 19.1, 19.0 in magnitudes, respectively. Full details of the survey and data release are given in \cite{lucas2008}.

This survey has a positional accuracy of $\sim 0.1^{\arcsec}$ and we have queried the UKIDSS point source catalogue\footnote{http://wsa.roe.ac.uk:8080/wsa/crossID$\_$form.jsp} within a $2^{\arcsec}$ radius. We used the option of nearest objects only to ensure correct matches due to crowded field. 
\subsubsection{MIPSGAL and WISE (Mid/Far-infrared)}
The MIPSGAL is one of the Spitzer Galactic plane surveys covering 278 $deg^2$ of Galactic longitude $5^\circ < l < 63^\circ$, and $298^\circ < l < 355^\circ$ and latitude $|b| < 1^\circ$ regions. This survey was carried out using the MIPS (Multiband Infrared Photometer for Spitzer) on the Spitzer Space Telescope at $24\ \umu m$ and $70\ \umu m$, achieving a resolution of $6^{\arcsec}$ and $18^{\arcsec}$, respectively. From this survey, we used only the $24\ \umu m$ data within the $5^\circ < l < 63^\circ$ region with a point source sensitivity at $3\sigma$ of 2 mJy \citep{carey2009}. 

The Wide-field Infrared Survey Explorer (WISE) is an infrared all-sky survey in four bands, centred on 3.4, 4.6, 12, and $22\ \umu m$. This survey achieved a $5\sigma$ point source sensitivity better than 0.08, 0.11, 1.0, and 6.0 mJy with angular resolutions of $6^{\arcsec}.1$, $6^{\arcsec}.4$, $6^{\arcsec}.5$, and $12^{\arcsec}.0$ in the four bands, respectively  \citep{wright2010}. Among other objects, this survey has also uncovered PNe \citep{ress2010, kron2014}. From this survey, we have taken the 12 and $22\ \umu m$ data to complement the GLIMPSE and MIPSGAL data.


\subsubsection{Hi-Gal (Far-Infrared)}
The Hi-Gal (Herschel infrared Galactic Plane Survey) is a far-infrared survey of the Galactic plane, covering  $-70^\circ \le l \le +68^\circ;\  | b |\leq 1^\circ$ region in five wavebands. This survey used the PACS \citep{pog2008} and SPIRE \citep{griffin2009} photometric cameras of the Herschel space observatory centred on $70\ \umu m$, $160\ \umu m$, $250\ \umu m$, $350\ \umu m$ and $500\ \umu m$. It achieved a spatial resolution of 6.0$^{\arcsec}$, 12.0$^{\arcsec}$, 18.0$^{\arcsec}$, 24.0$^{\arcsec}$ and 35.0$^{\arcsec}$ in the five bands, respectively. Full details about observation, processing and data release are given in  \cite{molinari2010,molinari2016}.

\subsubsection{IPHAS (H$\alpha$)}

The INT (Isaac Newton Telescope) Photometric H$\alpha$ survey of the Galactic Plane (IPHAS) is a 1800 $deg^2$ CCD survey, covering the Galactic latitude $|b| < 5^\circ$ and longitude $30^\circ < l < 215^\circ$ region, in the broad-bands sloan r, i and narrow-band H$\alpha$ filters. The WFC (wide field camera) of the 2.5 m INT with a pixel scale of $0.33^{\arcsec}$  was used for this survey. It is the first fully photometric H$\alpha$ survey of the Galactic plane with mean $5\sigma$ depths of 21.2, 20.0 and 20.3 for the r, i and H$\alpha$ bands in the Vega magnitude system, respectively \citep{drew2005,barensten2014,gon2008}. This survey has a median seeing of $\sim 1.0^{\arcsec}$ and has detected both unresolved and resolved PNe, including PNe with lower surface brightness than previously known \citep{mampaso2006}. \cite{corradi2008} and \cite{viironen2009} have used the (r- H$\alpha$) vs (r - i) colour-colour plane in combination with NIR colours to identify and separate PNe from other source classes.

\subsection{Aperture photometry}\label{aperturephotometry}

To measure magnitudes and flux densities of extended sources (absent in published point source catalogues) from some of the surveys, we have performed aperture photometry. This was done using the corresponding CORNISH-PN aperture as a reference. Details on how the fluxes of extended sources in the CORNISH survey were measured are given in \citetalias{cornissh2013}, and we have used the same methods here. In measuring the integrated flux densities of the CORNISH sources, a 2D Gaussian was used for point sources and manually drawn polygons for extended sources.

The mid-infrared emission is expected from a region larger than the ionized gas region due to dust and molecular emissions. This informs the need for a larger aperture for the GLIMPSE data. We define a parameter that allows the size of the reference aperture to be adjusted accordingly by adding an optimal value. The optimal parameter value, which in turn defines an optimal aperture size, was determined from a curve of growth. The need for an optimal aperture was to avoid larger errors due to poor background subtraction (worst for point sources). We have used different apertures for the four bands due to the varying resolution and the varying strength of these emissions across the four bands. We increased the reference apertures  by $2.5^{\arcsec}$, $2.5^{\arcsec}$, $3.5^{\arcsec}$ and $4.5^{\arcsec}$ for the $3.6\ \umu m$, $4.5\ \umu m$, $5.8\ \umu m$ and $8.0\ \umu m$, respectively. For the near infrared and optical, we have increased the reference aperture size by $1^{\arcsec}$.

The background level for each source was measured by means of an annulus, which provides a good estimate of the background level around a source. Due to background variations across the different wavelengths, the median background level was used, since it is not sensitive to extreme values or bright sources in the background. An annulus width of $10^{\arcsec}$, with an offset of $5^{\arcsec}$ from the source aperture was used to estimate the median background level. This ensures that the area of the annulus is greater than the source aperture area, allowing a better estimation of the median background level \citep{reach2005}. Aperture photometry was performed using Equations 19 and 22 in \citetalias{cornissh2013}.

\begin{figure*}
	\includegraphics[width=\columnwidth, height=7cm]{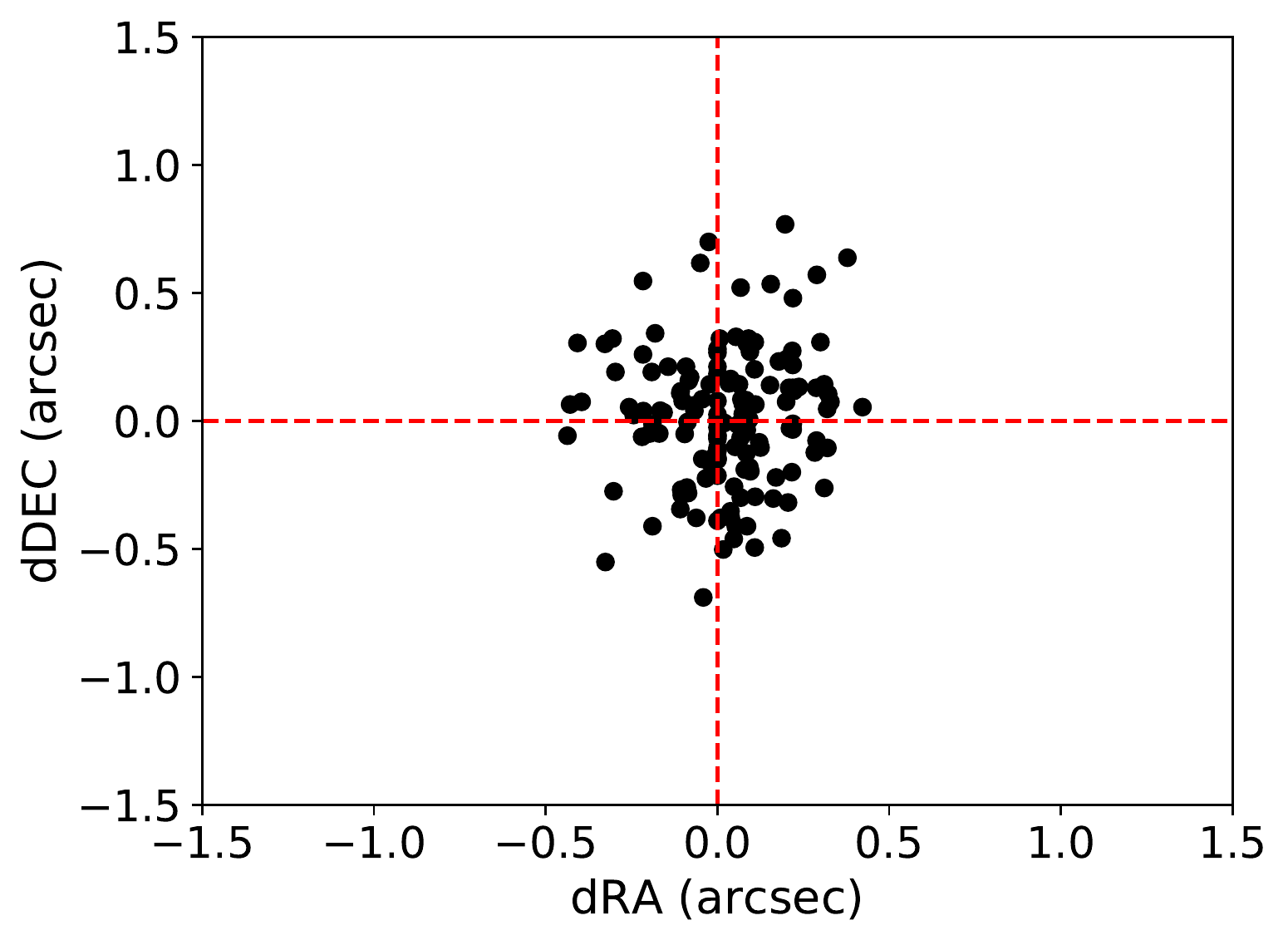}
	\includegraphics[width=\columnwidth, height=7cm ]{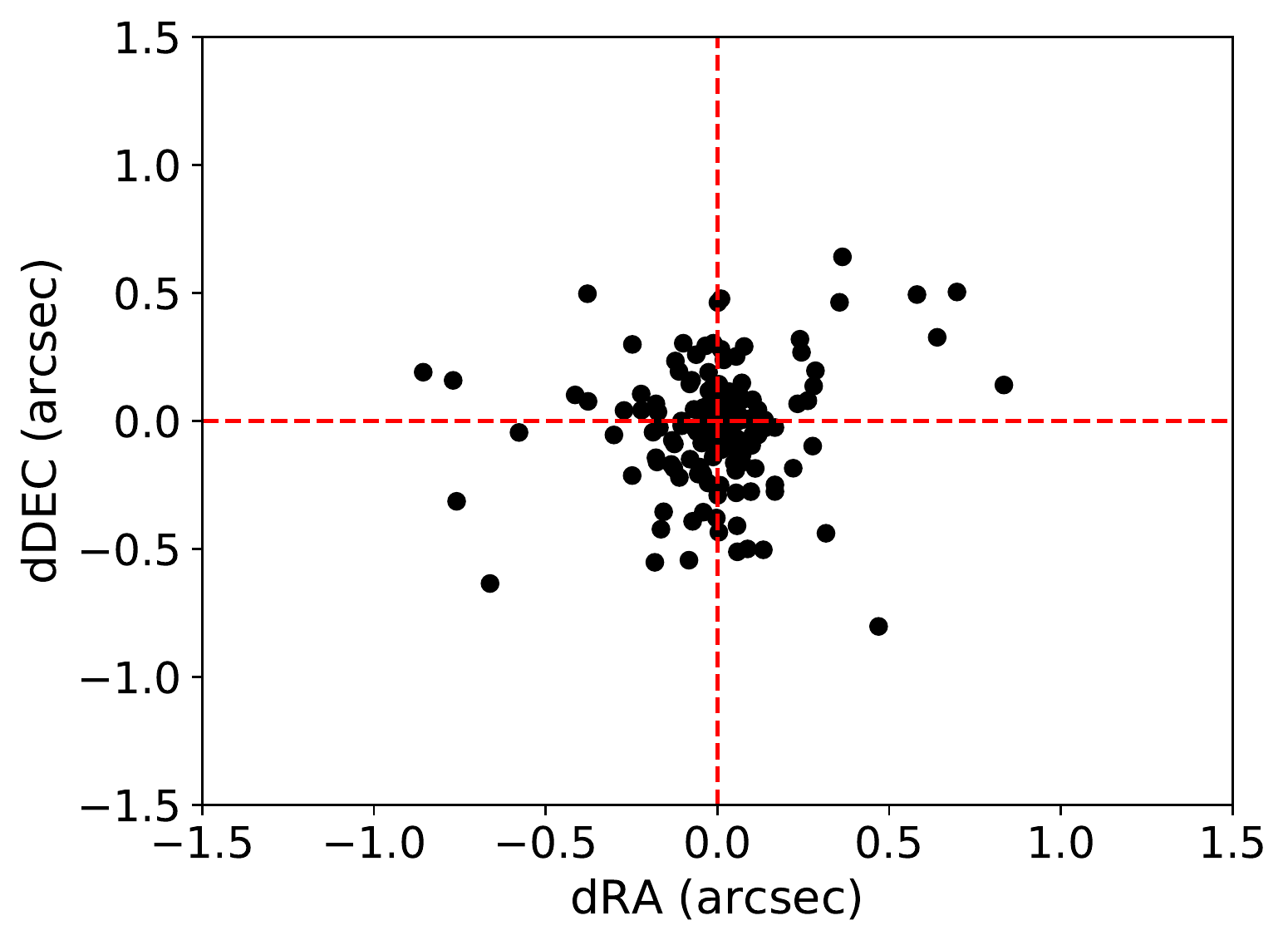}
    \caption{Offset positions in RA and DEC between the CORNISH-PNe and point source counterparts in the GLIMPSE (left) and UKIDSS (right).}
    \label{crossx}
\end{figure*}

For all data considered, the CORNISH-PNe positions were used to obtain uniform image cut-outs from the latest processed and final calibrated images, where stated or available.  Cross-matching of the CORNISH-PNe with individual survey catalogues (UKIDSS and GLIMPSE) was done to compare point source intergrated flux densities/magnitudes with results of our aperture photometry. In all cases, the results were similar and showed good agreement for point sources. The cross-matched positions revealed no systematic offsets (Figure \ref{crossx}). Aperture photometry was not done for the MAGPIS survey, where we have used the published catalogue values, due to its resolution of $\sim 6^{\arcsec}$ compared to the CORNISH resolution of $1.5^{\arcsec}$. This also applies to the Hi-Gal, MIPSGAL and WISE data.

\section{Results and Analysis}\label{3}
\subsection{Radio Properties}

The primary properties of a radio source are its flux density and angular size. Details of the angular sizes and flux densities measurements of sources in the CORNISH catalogue are given in \citetalias{cornissh2013} and we have briefly summarized it here. 

Source detection and photometry was performed using OBIT\footnote{http://www.cv.nrao.edu/∼bcotton/Obit.html} FndSou. FndSou works by finding and identifying neighbouring islands of emission above a set intensity cut-off and it attempts to fit one or more 2D Gaussians to these emissions. An elliptical aperture was used to measure the source properties, which extended to the 3$\sigma$ Gaussian major and minor axes. The rms noise and median background level of the sky were measured as described in Section \ref{aperturephotometry}. For extended sources, indicated by a cluster of Gaussian fits, the fitted parameters were replaced with single measurements under a manually drawn polygon around the emission. For such extended sources, the fitted Gaussian axes are replaced with intensity-weighted diameter given by Equation 16 in \citepalias{cornissh2013}.

\begin{figure}
	\includegraphics[width=\columnwidth, height=7cm]{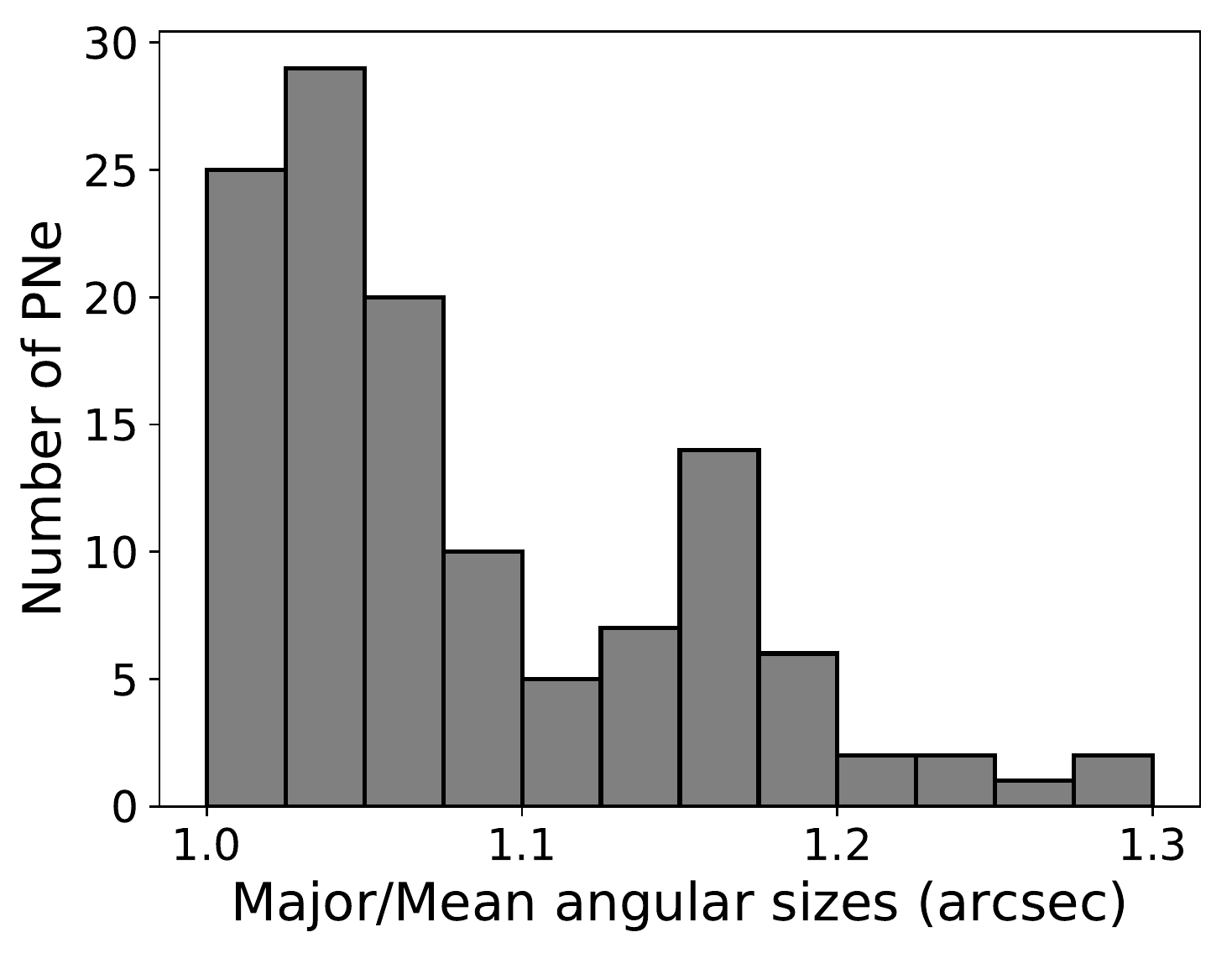}
    \caption{Distribution of the ratios of the major axes to the geometric mean sizes for the CORNISH-PNe.}
    \label{figure3}
\end{figure}

We have used the geometric mean of the fitted Gaussian axes ($\sqrt{\theta_{M} \theta_{m}}$, where $\theta_{M}$ and $\theta_{m}$ are the fitted major and minor axes, respectively) and intensity-weighted diameter for extended sources. Because the CORNISH-PNe are fairly round in the radio, there is not much difference between the geomtric mean sizes used and major axes as shown in Figure \ref{figure3}.  The intensity weighted sizes would make the extended sources appear smaller than they would in optical catalogues. The deconvolved sizes may also underestimate the sizes of extended sources (see \citealt{vanhoof2000}). To estimate the true angular sizes, we use a correction factor at 6 cm, according to \cite{vanhoof2000}. With this method, $\theta_{true}=\gamma \theta_{d}$ and $\gamma$ is estimated using Equation \ref{correction}, where $\beta =\theta_{d}/\theta_{beam}$ ($\theta_{d}$ is the deconvolved size).

\begin{equation}
\hspace{2cm}
\gamma (\beta)=\frac{0.3429}{1+0.7860\beta^2}+1.6067
\label{correction}
\end{equation}

Table \ref{correct_size_table} shows a comparison of the measured sizes (FWHM), deconvolved sizes, calculated true sizes and reported optical sizes for a few CORNISH-PNe with optical sizes from the MASH\footnote{Macquarie/AAO/Strasbourg H$\alpha$ Planetary Galactic Catalogue} \citep{parker2006,mis2008} and IPHAS \citep{sabin2014} catalogues. For three of the sources the optical sizes are 10 to 20 times larger. These sources are bipolar PNe (G027.6635-00.8267, G050.4802+00.7056 and G062.7551-00.7262) and the CORNISH survey has detected only the bright cores. On average, the corrected sizes show better agreement with the optical sizes. These corrected angular sizes will be used throughout this paper. Furthermore, we show in Figure \ref{figure3} that there is not much difference in using the major axis and the angular sizes we have adopted here.

\begin{table*}
\caption{The measured radio sizes ($\theta_{FWHM}$), deconvolved sizes ($\theta_{d}$), corrected sizes ($\theta_{true}$) compared to the optical angular sizes ($\theta_{Lit}$) from the MASH \citep{parker2006,mis2008} and IPHAS \citep{sabin2014} catalogues are presented. Where available, we show the major and minor optical sizes. Bipolar PNe are indicated with *}\label{correct_size_table}
\begin{tabular}{lllll}
\hline
\hline
  \multicolumn{1}{|l|}{CORNISH Name} &
  \multicolumn{1}{l|}{$\theta_{FWHM}$} &
  \multicolumn{1}{l|}{$\theta_{d}$} &
  \multicolumn{1}{l|}{$\theta_{true}$} &
  \multicolumn{1}{l|}{$\theta_{Lit}$} \\
 &(arcsec)&(arcsec)&(arcsec)&(arcsec)\\
\hline
G014.5851$+$00.4613 &$3.28\pm 0.04$& $2.91\pm 0.05$& $4.93\pm 0.08$& $6.0\times 5.0$\\
G018.2402$-$00.9152 &$8.08\pm 0.03$& $7.94\pm 0.03$& $12.88\pm 0.05$& $12.0\times 12.0$\\
G026.8327$-$00.1516 &$2.52\pm 0.07$& $2.03\pm 0.08$& $3.55\pm 0.15$& $7.0\times 5.0$\\
G027.6635$-$00.8267* &$1.67\pm 0.06$& $0.73\pm 0.14$& $1.38\pm 0.26$& $35.0\times 12.0$\\
G032.5485$-$00.4739 &$2.16\pm 0.04$& $1.55\pm 0.05$& $2.78\pm 0.09$& $9.0\times 9.0$\\
G035.5654$-$00.4922 &$6.24\pm 0.03$& $6.05\pm 0.03$& $9.88\pm 0.05$& $11.0\times 11.0$\\
G050.4802$+$00.7056* &$1.60\pm 0.06$& $0.56\pm 0.16$& $1.08\pm 0.31$& 19.0\\
G051.8341$+$00.2838 &$2.25\pm 0.03$& $1.68\pm 0.05$& $2.98\pm 0.08$& 6.0\\
G062.7551$-$00.7262* &$2.27\pm 0.07$& $1.71\pm 0.09$& $3.04\pm 0.16$& 27.0\\
\hline\end{tabular}
\end{table*}

\begin{figure*}
	\includegraphics[width=\columnwidth, height=7cm]{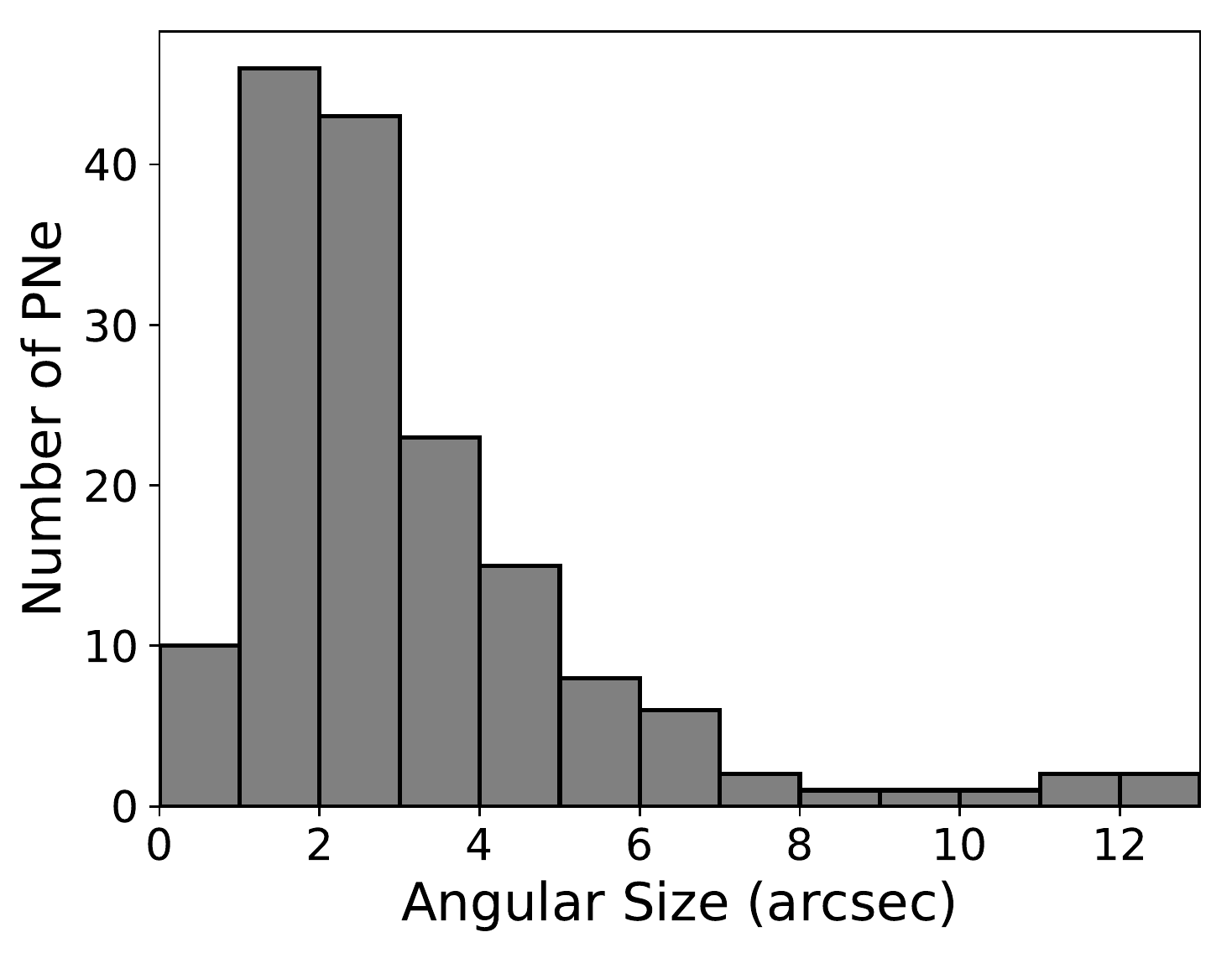}
	\includegraphics[width=\columnwidth, height=7cm ]{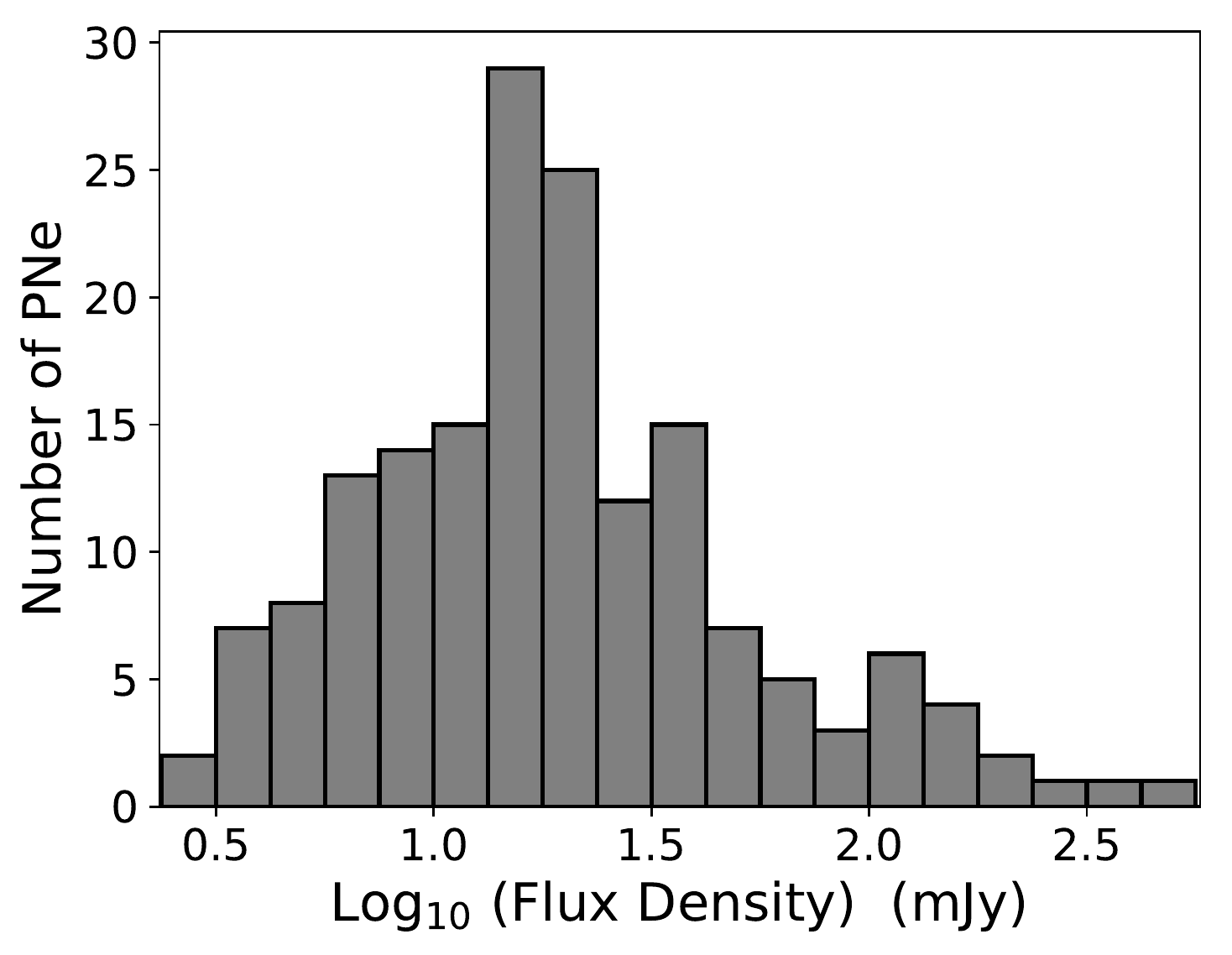}
    \caption{The angular size distribution of the CORNISH-PNe and integrated flux density (right) distribution of the CORNISH-PNe.}
    \label{fig:lb}
\end{figure*}

\begin{figure}
	\includegraphics[height=6.5cm, width=\columnwidth]{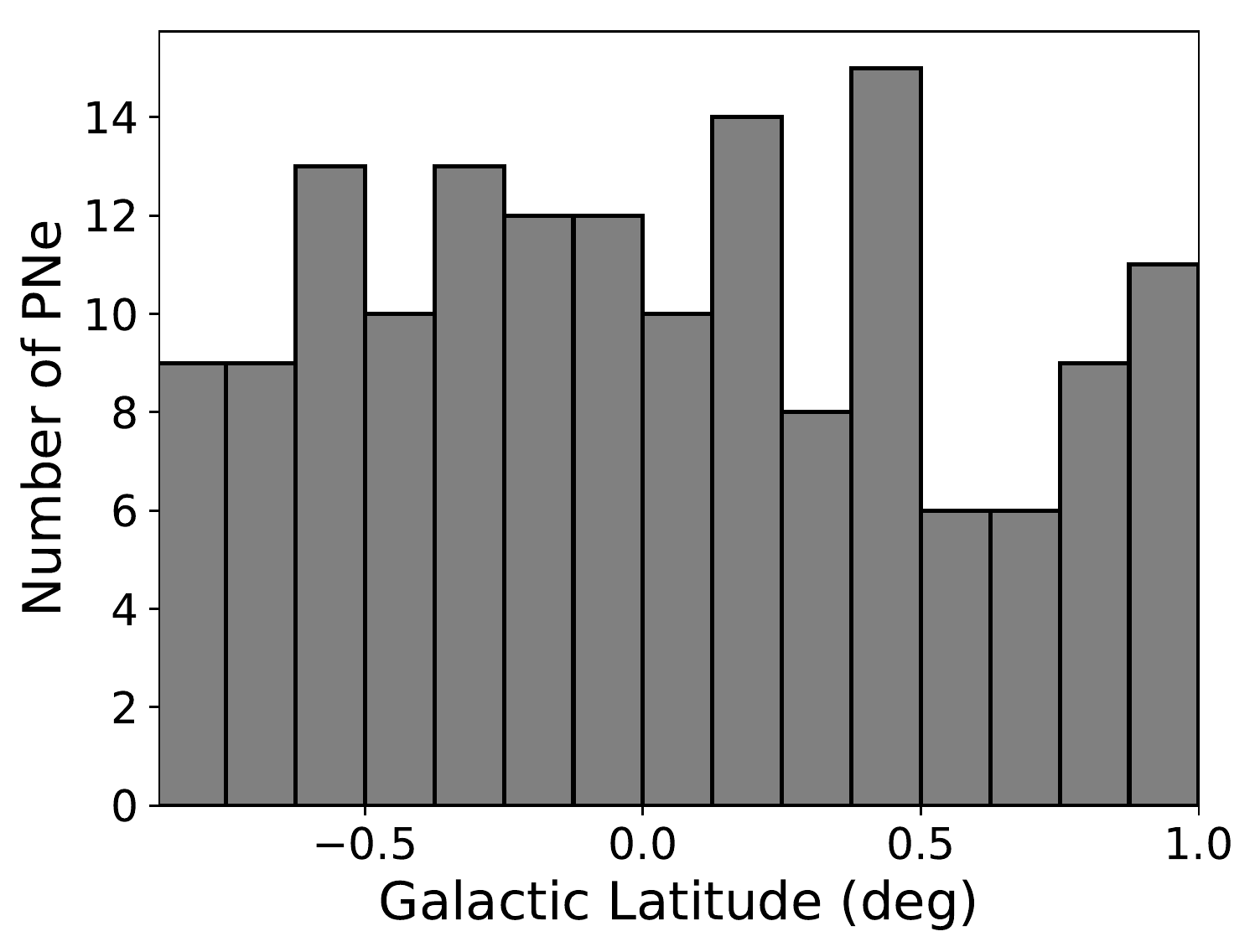}
	\includegraphics[height=6.5cm, width=\columnwidth]{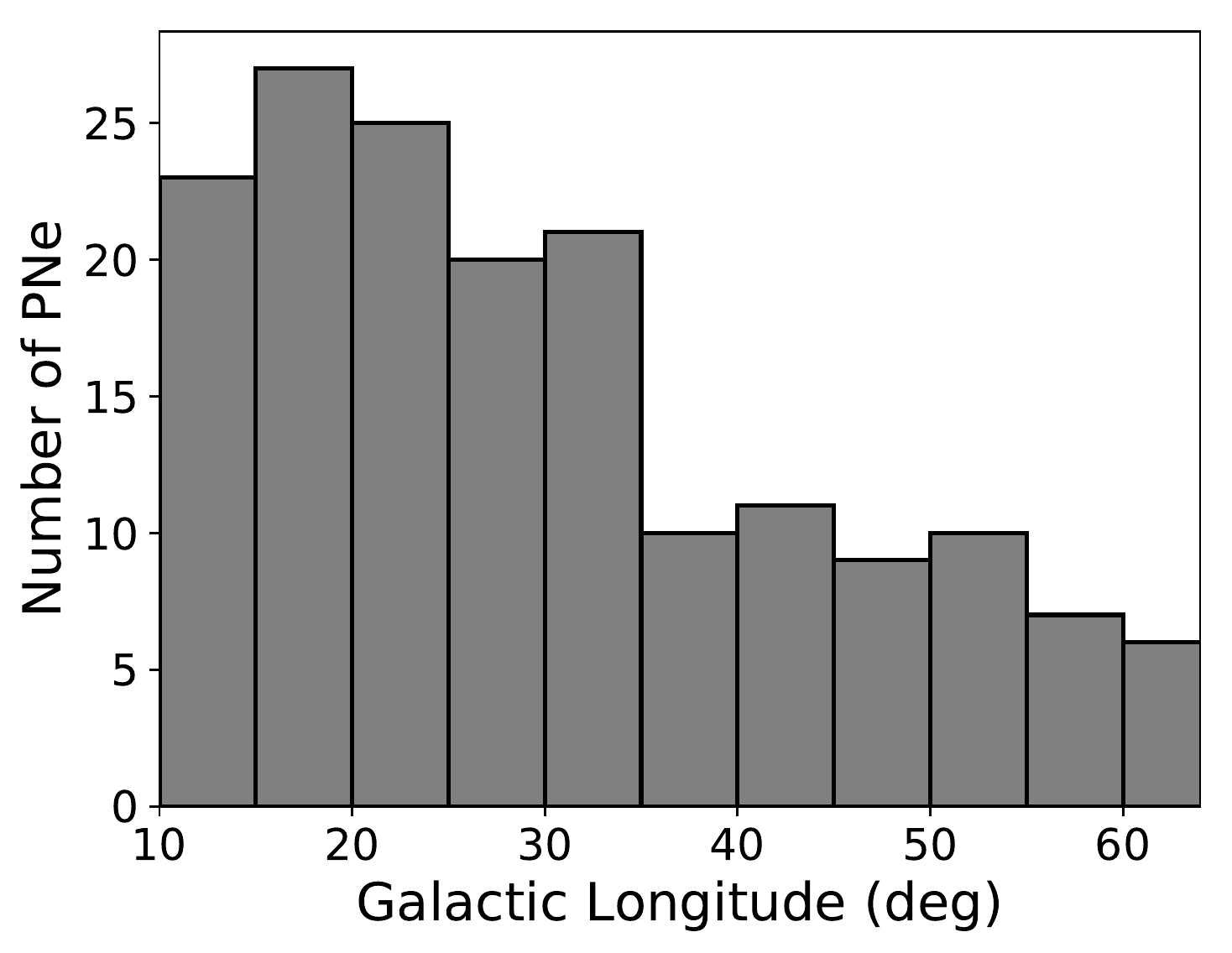}
    \caption{Galactic latitude (upper panel) and longitude (lower panel) distribution of the CORNISH-PNe.}
    \label{lat}
\end{figure}

All the CORNISH-PNe have peak fluxes $> 7 \sigma$, of which $91\%$ have flux densities $\leq$ 100 mJy (Figure \ref{fig:lb}, right panel), and $\sim 35\%$ are unresolved ($\theta < 1.8^{\arcsec}$). The distribution of the angular sizes with a median of  $2.5^{\arcsec}$ (see left panel in Figure \ref{fig:lb}) indicates a preferentially compact and/or distant sample. Figure \ref{lat} (upper panel) shows a Galactic latitude  distribution that is approximately flat (upper panel of Figure \ref{lat}), indicating a uniform detection of PNe within this region. At such low latitides we expect detection of PNe from more masive progenitors, hence, a more compact sample of PNe. The longitude distribution (Figure \ref{lat}, lower panel) shows an increase in the CORNISH-PNe population towards the Galactic bulge. This agrees with the observed distribution of PNe (see Galactic distribution of known PNe as compiled by \citealt{parker2017}\footnote{http://hashpn.space/}), as there are more PNe towards the Galactic centre.

The radio continuum emission from a PN can also be expressed in terms of its brightness temperature ($T_b$), a distance-independent property that can be used to infer the evolution of PNe. At 6 cm, most PNe are optically thin and so $T_b$ (proportional to the square of the electron density) decreases with continuous expansion as the nebulae evolve. Otherwise, PNe that are optically thick will show high $T_b$ ($ T_b=T_e(1-e^{-{\tau}_v})$, where $T_e$ is the electron temperature), as the optical depth approaches the optically thick limit ($\tau_v \propto n_{e}^2$). Therefore, more evolved and optically thin PNe should have lower brightness temperatures compared to younger and more compact PNe. Nonetheless, young PNe from low-mass progenitors could also show low $T_b$ because the envelope may have been mostly dispersed before the central star becomes hot enough to initiate ionization. 
\par

\begin{equation}
\hspace{3cm}
T_{b} = \frac{\lambda^2S_{\nu}}{2k\pi \theta^2}
\label{tbeqa}
\end{equation}

\begin{figure*}
	\includegraphics[width=\columnwidth, height=6.5cm]{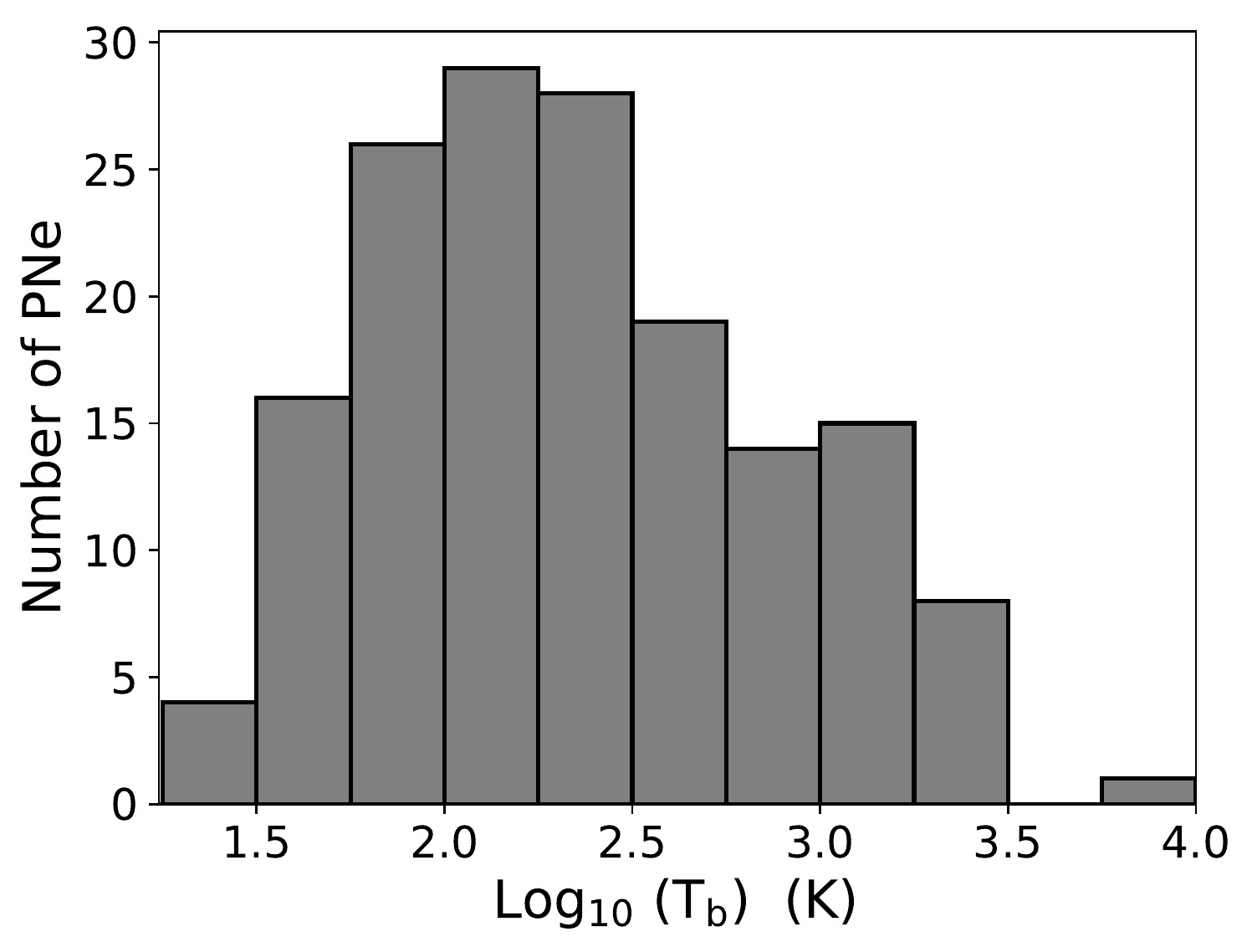}
	\includegraphics[width=\columnwidth, height=6.5cm ]{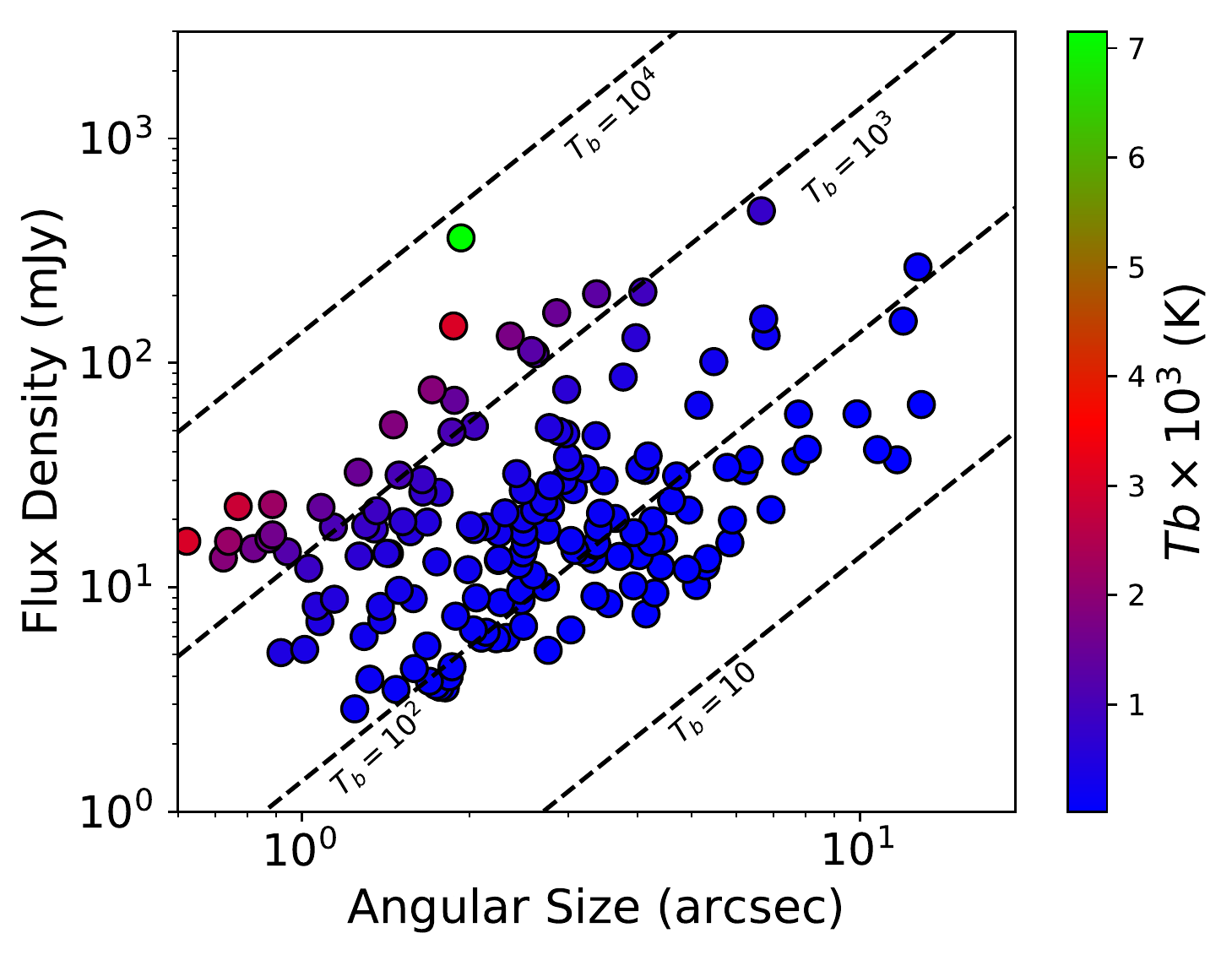}
    \caption{Left panel: Brightness temperature distribution. Right panel: Brightness temperature vs. and angular sizes, showing lines of constant temperature.}
    \label{angflux}
\end{figure*}

The average T$_{b}$ across each CORNISH-PNe at 5 GHz was estimated using Equation \ref{tbeqa}, where $\theta$ is the angular radius (arcsec) and $S_\nu$ is the integrated flux density (mJy). The CORNISH-PNe show T$_b$ between 20 K and 7000 K (Figure \ref{angflux}, left panel) with a  median of $\sim 200$ K. We also show the variation of the angular sizes with corresponding flux densities and lines of constant T$_b$ in Figure \ref{angflux} (right panel). It can be seen that there are a few PNe that can be considered younger. These younger CORNISH-PNe have small angular sizes and T$_b$ that lie within the region of $10^3 K < T_{b} < 10^4  K$ (see \citealt{kwok1985}).

\begin{table*}
\begin{center}
\caption{Radio properties of the CORNISH$-$PNe. For the angular sizes of sources whose deconvolved sizes could not be determined, we propagated an upper limit of 3$\sigma$  and they are preceded by `<'. Column 8 shows measured (aperture photometry) 1.4 GHz flux densities (MAGPIS) not present in the point source catalogue. Full table is available online as Table A1.}\label{Total}
\begin{tabular}{p{2.5cm}p{1.4cm}p{1.5cm}llllr}
\hline
\hline
CORNISH Name&RA&DEC &F$ _{5 GHz}$&F$ _{1.4 GHz}$& Angular Size &
$T_b$&Spectral\\
  &(J2000)&(J2000)&(mJy)&(mJy)&(arcsec)&(K)&Index\\
\hline
G009.9702$-$00.5292 &  18:09:40.54 &  $-$20:34:21.7 & 12.6 $\pm$ 1.6 &$11.55\pm 0.65$ &   2.45 $\pm$ 0.25 &     154.0 $\pm$ 29.0 &   0.07 $\pm$ 0.11\\
G010.0989$+$00.7393 &  18:05:13.13 &  $-$19:50:35.2 &  476.0 $\pm$ 43.6   &      &   6.66 $\pm$ 0.01 &     784.0 $\pm$ 72.0 &    0.20 $\pm$ 0.07 \\
G010.4168$+$00.9356 &  18:05:09.17 &  $-$19:28:11.3 &    14.4 $\pm$ 1.5 &     -&   0.94 $\pm$ 0.27 &   1182.0 $\pm$ 489.0 &-\\
G010.5960$-$00.8733 &  18:12:14.95 &  $-$20:11:24.7 &    34.5 $\pm$ 3.8&- &   3.02 $\pm$ 0.05 &     276.0 $\pm$ 31.0 &    0.01 $\pm$ 0.09 \\
G011.3266$-$00.3718 &  18:11:52.34 &  $-$19:18:31.0 &13.8 $\pm$ 2.4  &$11.63\pm 0.84$&   4.02 $\pm$ 0.11 &62.0 $\pm$ 11.0 &   0.14 $\pm$ 0.15\\
G011.4290$-$01.0091 &  18:14:27.14 &  $-$19:31:26.8 &14.6 $\pm$ 2.0 &-&   3.08 $\pm$ 0.31 &     112.0 $\pm$ 22.0 &   $-$0.40 $\pm$ 0.14 \\
G011.4581$+$01.0736 &  18:06:47.64 &  $-$18:29:37.7 & 7.4 $\pm$ 1.2&     -  &   1.89 $\pm$ 0.37 &152.0 $\pm$ 48.0&- \\
G011.7210$-$00.4916 &18:13:07.15 &  $-$19:01:12.4 & 5.9 $\pm$ 1.1 & 8.06 $\pm$ 0.48&2.10 $\pm$ 0.48 &98.0 $\pm$ 37.0 &  $-$0.24 $\pm$ 0.16\\
G011.7434$-$00.6502 &  18:13:45.22 &  $-$19:04:35.4 &  156.9 $\pm$ 15.0 & -&  6.72 $\pm$ 0.01 & 253.0 $\pm$ 24.0 &    0.00 $\pm$ 0.09 \\
G011.7900$-$00.1022 &18:11:48.89 &  $-$18:46:21.4 &41.2 $\pm$ 6.9 &49.10$ \pm$ 1.13&8.05 $\pm$ 0.08 &46.0 $\pm$ 8.0 &  $-$0.14 $\pm$ 0.13 \\
\hline		  			     	    	       		 	 	
\end{tabular}	  			     	    	       		 
\end{center}		     	    	       		 		
\end{table*}

\subsubsection{Spectral Index}
\begin{figure}
	\includegraphics[width=\columnwidth]{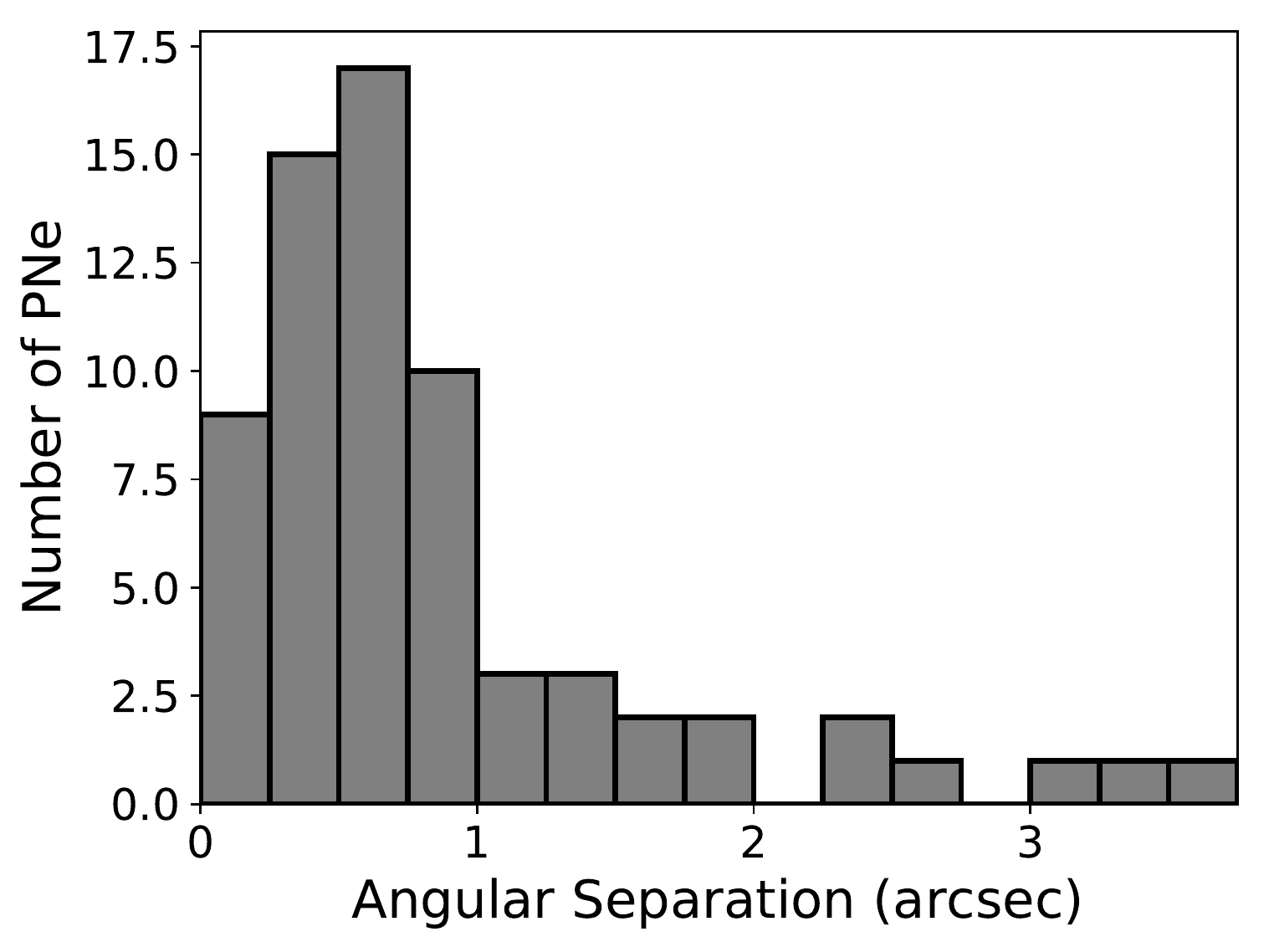}
    \caption{Distribution of the angular separation of the 67 CORNISH-PNe found in the MAGPIS survey within a cross-matching radius of $5^{\arcsec}$.}
    \label{fig:mag1}
\end{figure}

Having shown the possible existence of some young CORNISH-PNe in Figure \ref{angflux}, it is also important to determine the nature of their radio emission. We cross-matched the CORNISH survey at 6 cm with the MAGPIS radio survey at 20 cm within a radius of $5^{\arcsec}$. This resulted in a total of 67 matches out of the 169 CORNISH-PNe.  85$\%$ of the cross-matched sources have angular separations less than $1^{\arcsec}$ and a distribution that peaks about $0.5^{\arcsec}$ (see Figure \ref{fig:mag1}).

For the purpose of spectral indices estimation, there is a need to exclude very extended sources, whose emission could have been spatially filtered out due to the design of the CORNISH survey. Spatially filtered out emission will result in underestimated flux densities. To exclude such extended PNe, we compared the CORNISH-PNe integrated flux densities to their corresponding peak fluxes. Sources with angular sizes larger than 9$^{\arcsec}$, corresponding to $F_{int}/F_{peak} >10$ were considered too extended. Additionally, the flux densities and angular sizes of the CORNISH-PNe were compared with their counterparts in the 6 cm MAGPIS catalogue (at a lower resolution of $6^{\arcsec}$, \citealt{white2005}). This reduced our sample to 61, after excluding 6 sources considered to be very extended.

To get a more complete sample, we measured the flux densities of the CORNISH-PNe absent from the MAGPIS catalogue as described in Section \ref{aperturephotometry}, using the CORNISH-PNe positions on the MAGPIS 20 cm image data. This brought our sample to a total of 127 out of 169 PNe.  Measured 20 cm flux densities are given in Table \ref{Total}. The spectral indices and associated errors of the 127 CORNISH-PNe were estimated using Equations \ref{alpha} and \ref{alpha2}, where $S_1$ and $S_2$ are the flux densities in mJy, and $\nu_1$ and $\nu_2$ are the corresponding frequencies.

\begin{equation}
\hspace{3cm}
	\alpha =\frac{ln(S_1 / S_2)}{ln(\nu_1 / \nu_2)}
\label{alpha}
\end{equation}

\begin{equation}
\hspace{1cm}
	\delta\alpha =\frac{1}{ln(\nu_1/\nu_2)}\sqrt{\left( \frac{\delta S_1}{S_1}\right)^2 + \left( \frac{\delta S_2 }{S_2} \right)^2 }
\label{alpha2}
\end{equation}

\begin{figure}
\begin{center}
	\includegraphics[width=\columnwidth, height=6cm]{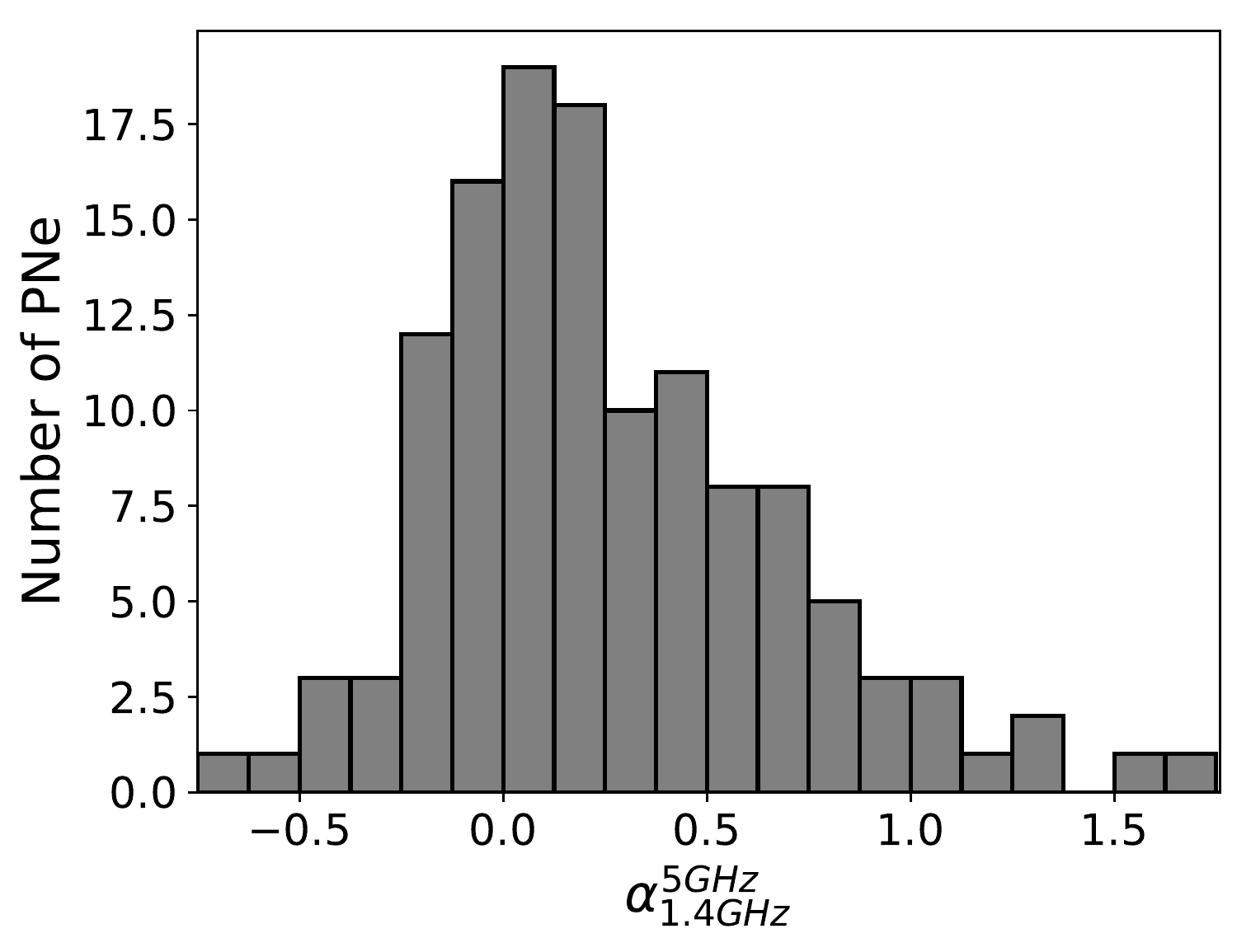}
	\includegraphics[width=\columnwidth]{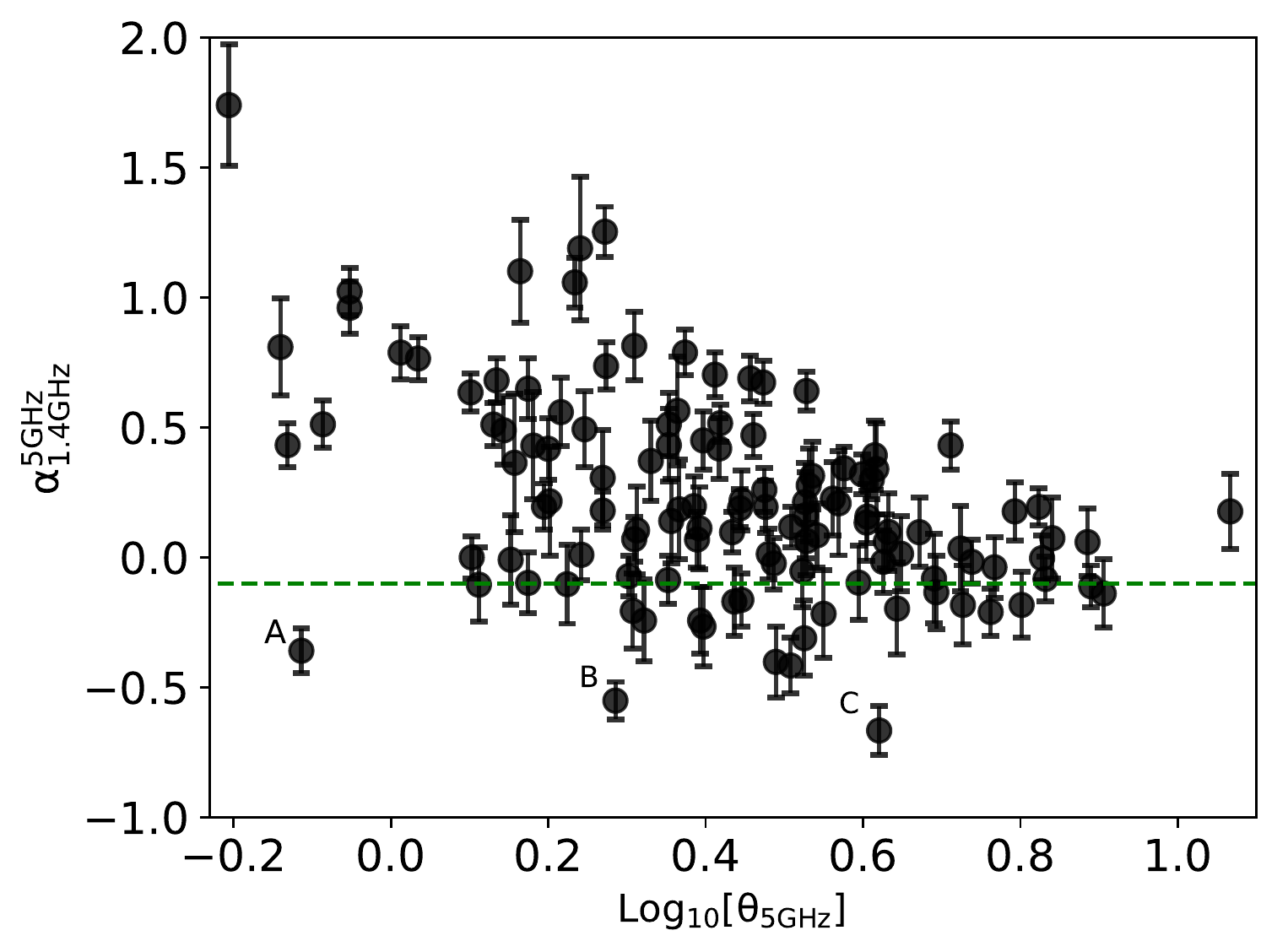}
	\includegraphics[width=\columnwidth, height=6cm]{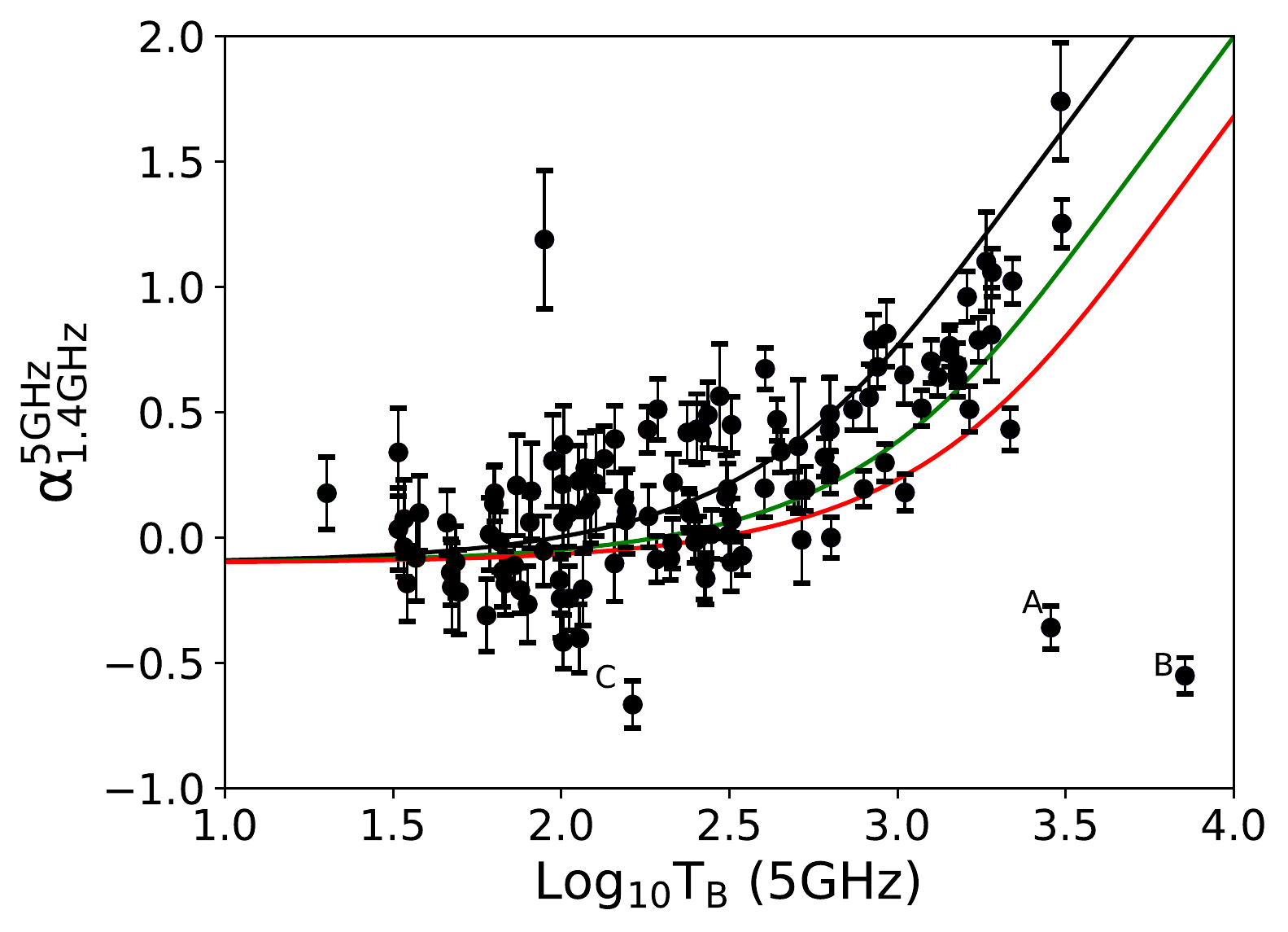}
    \caption{Top: Spectral index distribution of the CORNISH-PNe. Middle: Distribution of the spectral indices vs. the log of the angular sizes. The red filled circles are the PNe with non-thermal emission at a 3$\sigma$ significance level. Bottom: Shows our data overlaid on a model of spectral index (5 GHz to 1.4 GHz) against the brightness temperature at 5 GHz. The curves show predictions of the model for $T_e$ at $0.5\times 10^4$ K (black), $1.0\times 10^4$ K (green) and $1.5\times 10^4$ K (red). The observed radio emission of the outliers, marked A (G052.1498$-$00.3758), B (G030.2335$-$00.1385) and C (G019.2356$+$00.4951), cannot be explained by a simple model characterized by constant electron temperature and varying optical depth (see \protect\citealt{siodmiak2001}).}\label{fig:spec}
\end{center}
\end{figure}

Assuming the radio-continuum emission from a PN is free-free, the spectral index range is expected to be between $-$0.1 and 2, where it is optically thin and optically thick, respectively. Figure \ref{fig:spec} (top panel) shows the distribution of the CORNISH-PNe with a  peak at about 0.1, which agrees with optically thin free-free emission. In Figure \ref{fig:spec} (middle panel), $\sim 84\%$ of the 127 CORNISH-PNe have spectral indices within the theoretical range. However, only two PNe have spectral indices below $-$0.1 at a $3\sigma$ significance level, which is indicative of non-thermal emission (discussed in Section \ref{individual_sources}). These are marked as B (G030.2335$-$00.1385) and C (G019.2356$+$00.4951) in Figure \ref{fig:spec} (middle), where we plot spectral index against the log of the CORNISH angular sizes. The source marked A (G052.1498$-$00.3758) has a spectral index of -0.1 at a $3\sigma$ significance level and so its emission is treated as thermal.

If we assume the integrated flux density at 1.4 GHz and 5 GHz are from the same solid angle, we  can model the spectral index (1.4 GHz and 5 GHz) at constant electron temperatures of $T_e = 0.5 \times 10^4$ K,  $1.0 \times 10^4$ K and $1.5 \times 10^4$ K as in \cite{siodmiak2001} (see their Equations 2 and 5). Results are shown in Figure \ref{fig:spec} (lower panel), where we plot modeled spectral indexes against brightness temperatures at 5 GHz over our data. At 5 GHz, the CORNISH-PNe generally show a scatter around the theoretical, optically thin region of the model. Above $T_b=$ 1000 K, the CORNISH-PNe follow the trend of the model as spectral index increases with $T_b$ (excluding two outliers). There are a few PNe that approach the optically thick limit with increasing $T_b$.  This trend is also observed in Figure \ref{fig:spec} (middle panel) and Figure \ref{angflux} (right panel), where CORNISH-PNe with smaller angular sizes show higher $T_b$ and spectral indices. The observed radio emission of the outliers, marked A (G052.1498$-$00.3758) and B (G030.2335$-$00.1385), obviously cannot be explained by a simple model characterized by varying optical depth and constant electron temperature (see \citealt{siodmiak2001}). G019.2356$+$00.4951, G030.2335$-$00.1385 and G052.1498$-$00.3758 are further discussed in Sections \ref{individual_sources} and \ref{Combined}. Table \ref{Total} summarizes the radio properties of the CORNISH-PNe.

\subsection{Mid-infrared to radio continuum ratio}

It is expected that as the PN evolves with an increase in size, the ratio of the MIR (mid-infrared) to radio integrated flux density ($MIR/radio$) will decrease. This is a result of the increased ionized volume as compared to the amount of dust and excited PAH (polycyclic aromatic hydrocarbon) emission, present in the $8.0\ \umu m$ band. PAH emission in this band arises from the excited PAH molecules just outside the ionized regions of carbon-rich PNe \citep{cox2016}. As the ionization front spreads out, the PAH emission decreases \citep{cerr2009,guz2014,cohen2007a}, causing a decrease in the total flux in the $8\ \umu m$ band. Thus, compared to more-evolved PNe, younger PNe should have stronger PAH emission, resulting in higher values of the $MIR/radio$ ratio. This ratio can also be affected by the mass loss history in the AGB phase. However, irrespective of larger angular sizes as a result of evolution, bipolar PNe have high $MIR/radio$ values due to the larger amount of dust and the presence of a torus that could remain molecular for a longer time \citep{cox2016,guz2014}. 

\begin{figure}
	\includegraphics[width=\columnwidth, height=6.8cm]{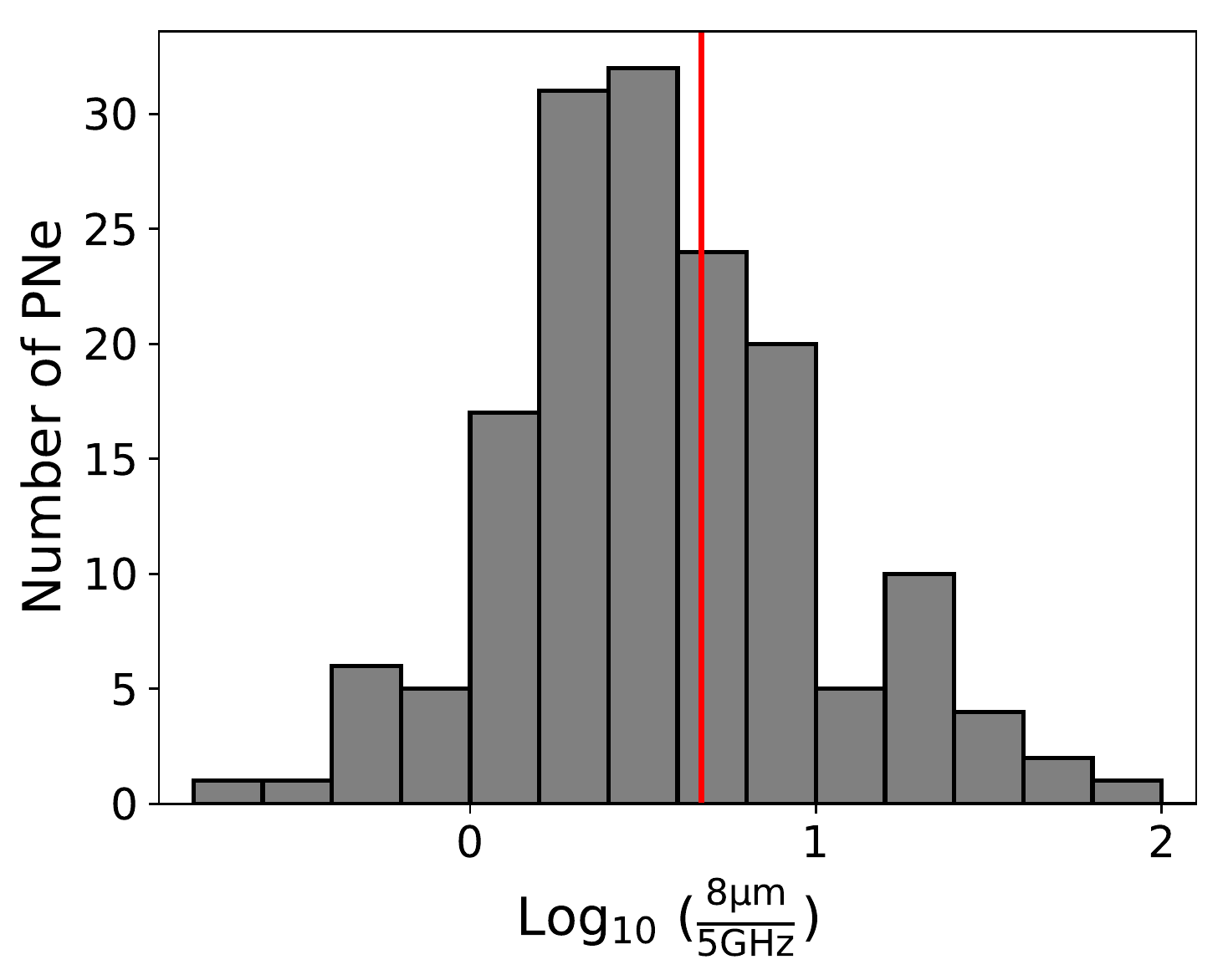}
    \caption{Distribution of the ${F_{8.0\umu m}/F_{5GHz}}$ showing the median (red dotted line) from \protect\citealt{cohenparker}.}
    \label{mirrad}
\end{figure}

\cite{cohenparker,cohen2007} used this ratio ($MIR/radio$) to discriminate between PNe and H\rom{2} regions. H\rom{2} regions are brighter in the MIR compared to PNe due to the larger amount of molecular gas associated with them. \cite{cohenparker} reported a median of $4.7 \pm 1.1$ for the MASH PNe sample using $F_{8.0\umu m}/F_{0.843 GHz}$. For the CORNISH-PNe, we have used the flux densities from the IRAC ${8\ \umu m}$ band and the CORNISH survey (${F_{8.0\umu m}/F_{5GHz}}$). The distribution is shown in Figure \ref{mirrad} with a median of $3.3 \pm 1.0$. If we convert the 5 GHz flux densities to 0.843 GHz, assuming the CORNISH-PNe are all optically thin, we have a median of $3.3 \pm 0.70$ (stated error is the standard error on the median). These estimates show a reasonable agreement within $1\sigma $ of the \cite{cohenparker} value. Using the same method, \cite{filli2009} found a median value of $9.0\pm 2.0$ (${F_{8.0\umu m}/F_{1.4GHz}}$) for 14 Magellanic cloud PNe, which is consistent with our value and \cite{cohenparker} within 2$\sigma$. For the CORNISH-PNe, there is no obvious trend or variation of the ${F_{8.0\umu m}/F_{5GHz}}$ with angular sizes.

\subsection{Multi-wavelength Colours}\label{colour}
The strength of multi-wavelength colour-colour diagrams to distinguish between PNe and their contaminants, in the absence of optical data, has been explored by \cite{cohenparker} and \cite{parker_2_2012}. Such plots allow separation of sources according to their colours. We have placed the CORNISH-PNe on different colour-colour plots and, where necessary, we have compared them with other Galactic sources that could contaminate our sample.

\subsubsection{Near-Infrared (NIR) Colours}\label{NIR_color}

The measured emission of PNe in the J, H and K bands is a combination of emission from the ionized nebulae, including free-bound and free-free emission, hot dust emission, stellar continuum from the central star and emission lines. The ionized gas and hot dust should dominate the emission, but in some cases, the stellar continuum could be mostly responsible \citep{white1985,garcia1997,phil2005,phil2009}.

\begin{figure}
	\includegraphics[width=\columnwidth, height=6.6cm]{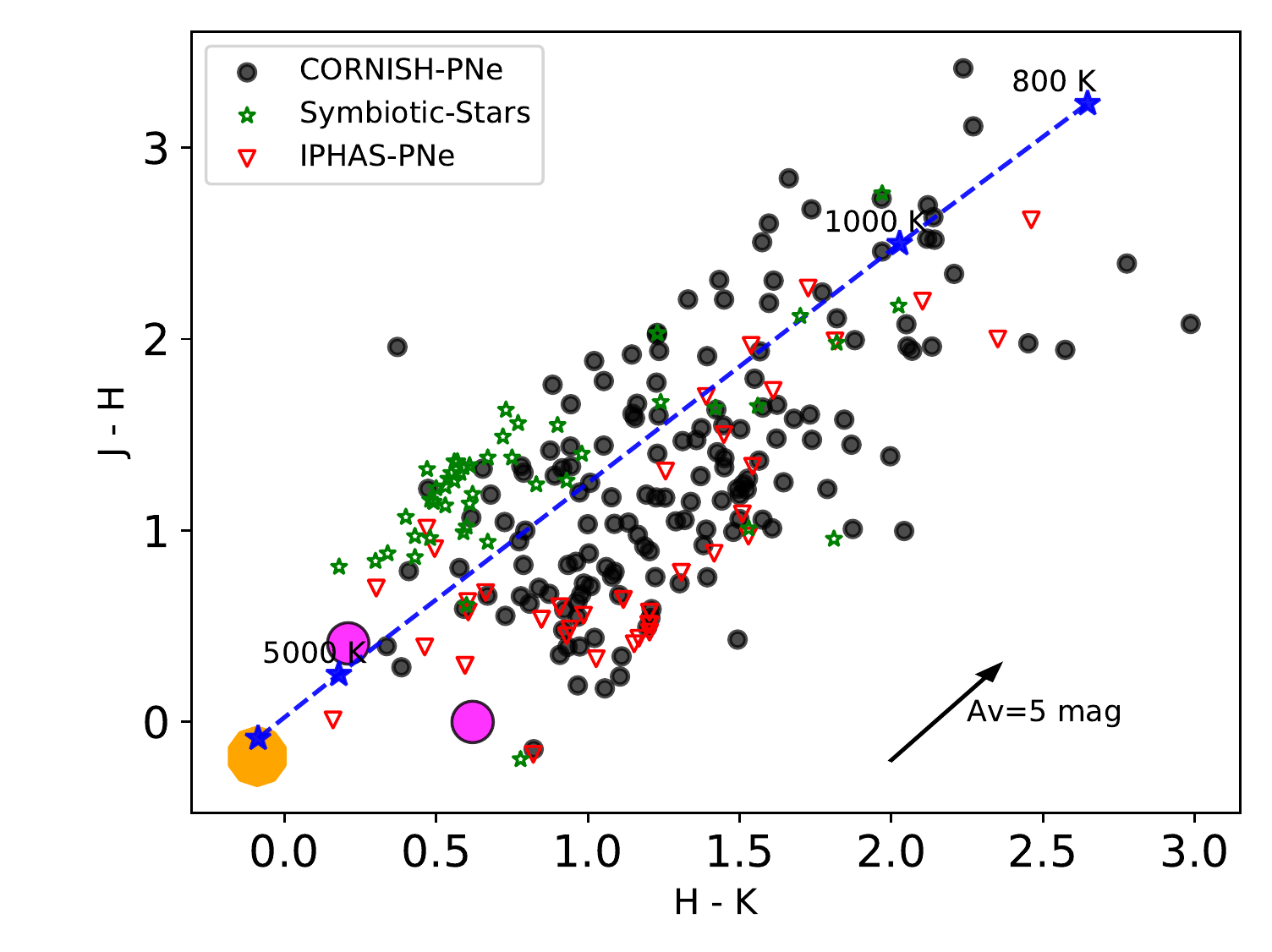}
    \caption{$J-H$ vs. $H-K$ colour-colour plot. We have taken PNe sample from the IPHAS catalogue \protect\citep{sabin2014,v2009,vii2009}; Symbiotic stars samples were taken from \protect\citealt{corradi2008,corradi2010,rog2014}  and new symbiotic stars from \protect\citealt{misz2013}. Modeled intrinsic colours of PNe (NGC 6720 and NGC 7027) shown in magenta (circles) are from \protect\citealt{Weid2013}. We also show the intrinsic colour of an O9 star (orange) from \protect\citealt{ducati2001} and blackbodies at different temperatures.}
    \label{fig:nir}
\end{figure}

Figure \ref{fig:nir} shows the distribution of the CORNISH-PNe on the $J-H$ vs. $H-K$ colour-colour plane (measured magnitudes are presented in Table \ref{NIR_table}). We also show the position of symbiotic stars, the intrinsic colours of an O9 star from \cite{ducati2001} and the modeled  intrinsic colours of NGC 7027 (0.21, 0.41) and NGC 6720 (0.62, 0.0) from \cite{Weid2013}. Some of the CORNISH-PNe show colours that are typical of symbiotic stars. This will make it difficult to rely only on the NIR colour-colour plane in differentiating between PNe and symbiotic stars. The CORNISH-PNe show a broad distribution that can be mostly explained by the range of PNe intrinsic colours (see \citealt{phil2009, phil2005}) and differing amounts of extinction.

Categorically, younger PNe with hot dust have redder colours, resulting in higher $H-K$ values. A contribution to the K-band could also arise from ${H_2}$ emission believed to be excited mainly by shocks and, in some cases, ultra-violet fluorescence \citep{mar2015}. The presence of this emission in the K-band could also lead to an increase in the $H - K$ colour.  The colour distribution of the CORNISH-PNe is similar to the IPHAS-PNe, but with more CORNISH-PNe displaying a higher reddening. Compared to the MASH sample analysed by \cite{phil2009} and PNe from the VVV (VISTA Variables in the Via Lactea) survey by \cite{Weid2013}, the CORNISH-PNe show a higher reddening. The median colours of the CORNISH-PNe are $1.23\pm 0.05$  and $1.25\pm 0.07$ for $H-K$ and $J-H$, respectively. 

\subsubsection{Optical Colours}
CORNISH-PNe counterparts were searched for within the region of the IPHAS survey that overlaps with the CORNISH survey, using the CORNISH positions. Owing to the Galactic position of these PNe, confusion with other sources in the background or foreground is possible. For this reason, the ${H\alpha}$ images were visually checked, aided with false 3-colour images. None of the images showed complexes or diffuse emission associated with H \rom{2} regions. A total of 22 out of the 76 CORNISH-PNe within the IPHAS survey region were found to have genuine counterparts. These include extended PNe not present in the point source catalogue. We measured the ${H\alpha}$, r and i magnitudes as discussed in Section \ref{aperturephotometry} and measurements are presented in Table \ref{optical_table}.

\begin{figure}
	\includegraphics[width=\columnwidth]{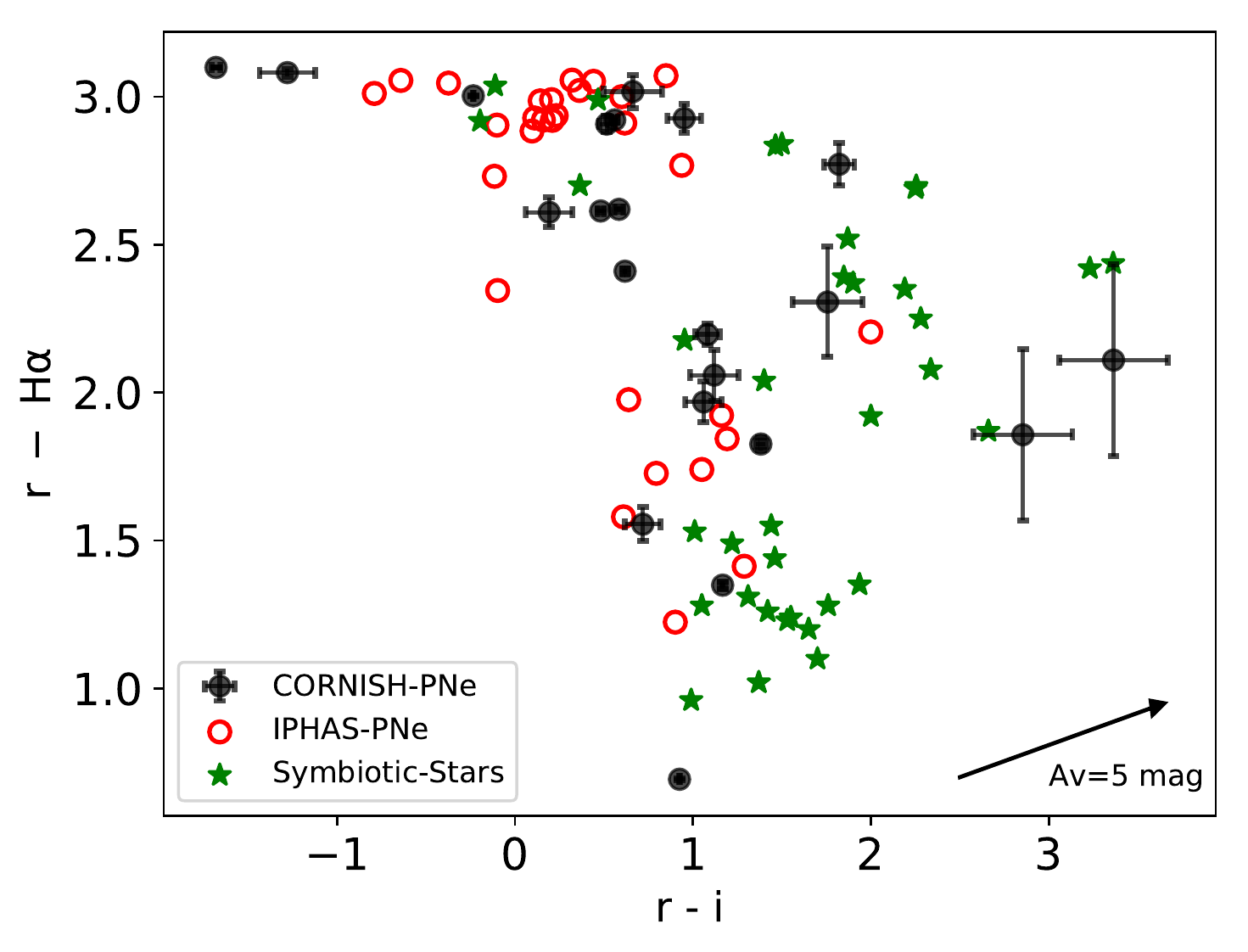}
    \caption{r-H$\alpha$ vs. r-i  colour-colour plane.  We show PNe sample from the IPHAS catalogue \protect\citep{sabin2014,v2009,vii2009}; Symbiotic stars samples from \protect\citealt{corradi2008,corradi2010} and \protect\citealt{rog2014}, and new symbiotic stars from \protect\citealt{misz2013}. The reddening vector is from the relationship in \protect\citealt{card1989}.}
    \label{fig:opt}
\end{figure}

PNe are known to be good emitters of ${H\alpha}$,  so they can be easily identified on the [r - H$\alpha$] vs. [r - i] colour-colour plane, where normal PNe have large r-H$\alpha$ colours. The positions of PNe in the optical colour-colour plane are well discussed in \cite{corradi2008} and \cite{viironen2009} using PNe from the IPHAS survey. PNe with H$\alpha$ excesses should be located within the upper region of the plot, while reddened PNe should be located towards the right, where higher $[r\ -\ i]$ values occur. We also show the positions of some Galactic PNe from the IPHAS survey and symbiotic stars in Figure \ref{fig:opt}. This is to show the distribution of the CORNISH-PNe within other PNe samples and possible contaminants like the symbiotic systems that are visible in the optical. 

Out of the 22 CORNISH-PNe with IPHAS counterparts, 21 seem to have normal PNe colours (Figure \ref{fig:opt}). An exception to this is the colour of G035.4719$-$00.4365 ($[r\ -\ i] \sim 0.92$, $[r\ -\ H\alpha] \sim 0.69$), where foreground stars in its photometry could be responsible. The general distribution of the CORNISH-PNe does not look different from the IPHAS-PNe.  Again, the CORNISH-PNe show colours similar to symbiotic stars and it will be difficult to separate PNe from such contaminants based on this optical colour-colour plots.

\begin{table*}
\begin{center}
\caption{\label{optical_table} ${H\alpha}$, r and i magnitudes and the ${H\alpha}$ line flux (${erg/cm^2/s}$) for the 22 CORNISH-PNe present in the IPHAS survey. These were measured using aperture photometry as discussed in Section \ref{aperturephotometry}. *Indicates PNe that are bipolar in the optical.}
\begin{tabular}{|l|l|l|l|l|l}
\hline
\hline
CORNISH Name   & ${H\alpha}$& r&i&LogF($H\alpha$) \\
&(mag)&(mag)&(mag)&(${erg/cm^2/s}$)\\
\hline

 G029.5780$-$00.2686 &  16.43 $\pm$ 0.01 &  19.44 $\pm$ 0.05 &  18.78 $\pm$ 0.15& $-13.38\pm 0.04$\\
 G032.5485$-$00.4739 &   13.76 $\pm$ 0.01 &  16.68 $\pm$ 0.01 &  16.12 $\pm$ 0.02& $-12.32\pm 0.01$\\
 G032.6136$+$00.7971 &  19.25 $\pm$ 0.11 &  21.11 $\pm$ 0.27 &  18.26 $\pm$ 0.08 &$-14.51\pm 0.40$\\
 G033.4543$-$00.6149 &   16.30 $\pm$ 0.01 &  18.13 $\pm$ 0.02 &  16.75 $\pm$ 0.02 &$-13.33\pm 0.05$\\
 G035.4719$-$00.4365 &   14.38 $\pm$ 0.01 &   15.08 $\pm$ 0.01 &   14.15 $\pm$ 0.01 &$-12.57\pm 0.01$\\
 G041.3540$+$00.5390 &  18.08 $\pm$ 0.12 &   20.19 $\pm$ 0.30 &  16.82 $\pm$ 0.04 &$-14.05\pm 0.50$\\
 G044.6375$+$00.4827 &  16.32 $\pm$ 0.01 &  19.09 $\pm$ 0.07 &  17.27 $\pm$ 0.05 &$-13.34\pm 0.05$\\
 G048.5619$+$00.9029 &  16.87 $\pm$ 0.03 &  18.42 $\pm$ 0.05 &  17.71 $\pm$ 0.09 &$-13.56\pm 0.10$\\
 G048.7319$+$00.9305 &  15.63 $\pm$ 0.01 &  18.24 $\pm$ 0.05 &  18.05 $\pm$ 0.12 &$-13.06\pm 0.04$\\
 G050.0405$+$01.0961 &  16.42 $\pm$ 0.02 &  18.39 $\pm$ 0.06 &  17.33 $\pm$ 0.08 &$-13.38\pm 0.09$\\
 G050.4802$+$00.7056 &  16.57 $\pm$ 0.01 &  18.77 $\pm$ 0.03 &  17.69 $\pm$ 0.06 &$-13.44\pm 0.05^*$\\
 G050.5556$+$00.0448 &  17.29 $\pm$ 0.03 &  19.35 $\pm$ 0.08 &  18.23 $\pm$ 0.11 &$-13.73\pm 0.10$\\
 G051.8341$+$00.2838 &  16.17 $\pm$ 0.01 &   19.10 $\pm$ 0.05 &  18.15 $\pm$ 0.08 &$-13.28\pm 0.04$\\
 G055.5070$-$00.5579 &   10.12 $\pm$ 0.01 &   13.12 $\pm$ 0.01 &   13.35 $\pm$ 0.01 &$-10.86\pm 0.01$\\
 G056.4016$-$00.9033 &   14.34 $\pm$ 0.01 &  16.75 $\pm$ 0.01 &  16.13 $\pm$ 0.01 &$-12.55\pm 0.01$\\
 G058.6410$+$00.9196 &  18.02 $\pm$ 0.06 &  20.33 $\pm$ 0.18 &  18.57 $\pm$ 0.09 &$-14.02\pm 0.20$\\
 G059.3987$-$00.7880 &  15.83 $\pm$ 0.01 &  17.18 $\pm$ 0.01 &  16.02 $\pm$ 0.01 &$-13.15\pm 0.02$\\
 G059.8236$-$00.5361 &    14.90 $\pm$ 0.01 &  17.81 $\pm$ 0.03 &  17.29 $\pm$ 0.03 &$-12.77\pm 0.01$\\
 G060.9866$-$00.5698 &   14.76 $\pm$ 0.01 &  17.37 $\pm$ 0.01 &  16.79 $\pm$ 0.02 &$-12.71\pm 0.01$\\
 G062.4936$-$00.2699 &   11.68 $\pm$ 0.01 &   14.78 $\pm$ 0.01 &  16.46 $\pm$ 0.02 &$-11.48\pm 0.01$\\
 G062.7551$-$00.7262 &    14.10 $\pm$ 0.01 &  17.18 $\pm$ 0.01 &  18.46 $\pm$ 0.16 &$-12.45\pm 0.01^*$\\
 G063.8893$+$00.1229 &   13.79 $\pm$ 0.01 &   16.40 $\pm$ 0.01 &  15.92 $\pm$ 0.02 &$-12.33\pm 0.01^*$\\
 \hline
\end{tabular}
\end{center}
\end{table*}

\subsubsection{Mid-Infrared (MIR) colour}

The emission of a PN in the MIR is a summation of dust continuum, free-free, free-bound, atomic line, molecular line and PAH emissions. All these contribute to the colour of a PN and its position on the MIR colour-colour plot, depending on which is dominant. The work of \cite{cohenparker}, using the GLIMPSE data,  showed that PNe can be well separated from H\rom{2} regions, using their MIR colours.  \cite{cohenparker} further showed that PNe have distinctive median colour indices compared to H\rom{2} regions.

\begin{figure*}
	\includegraphics[height=6.5cm,width=\columnwidth]{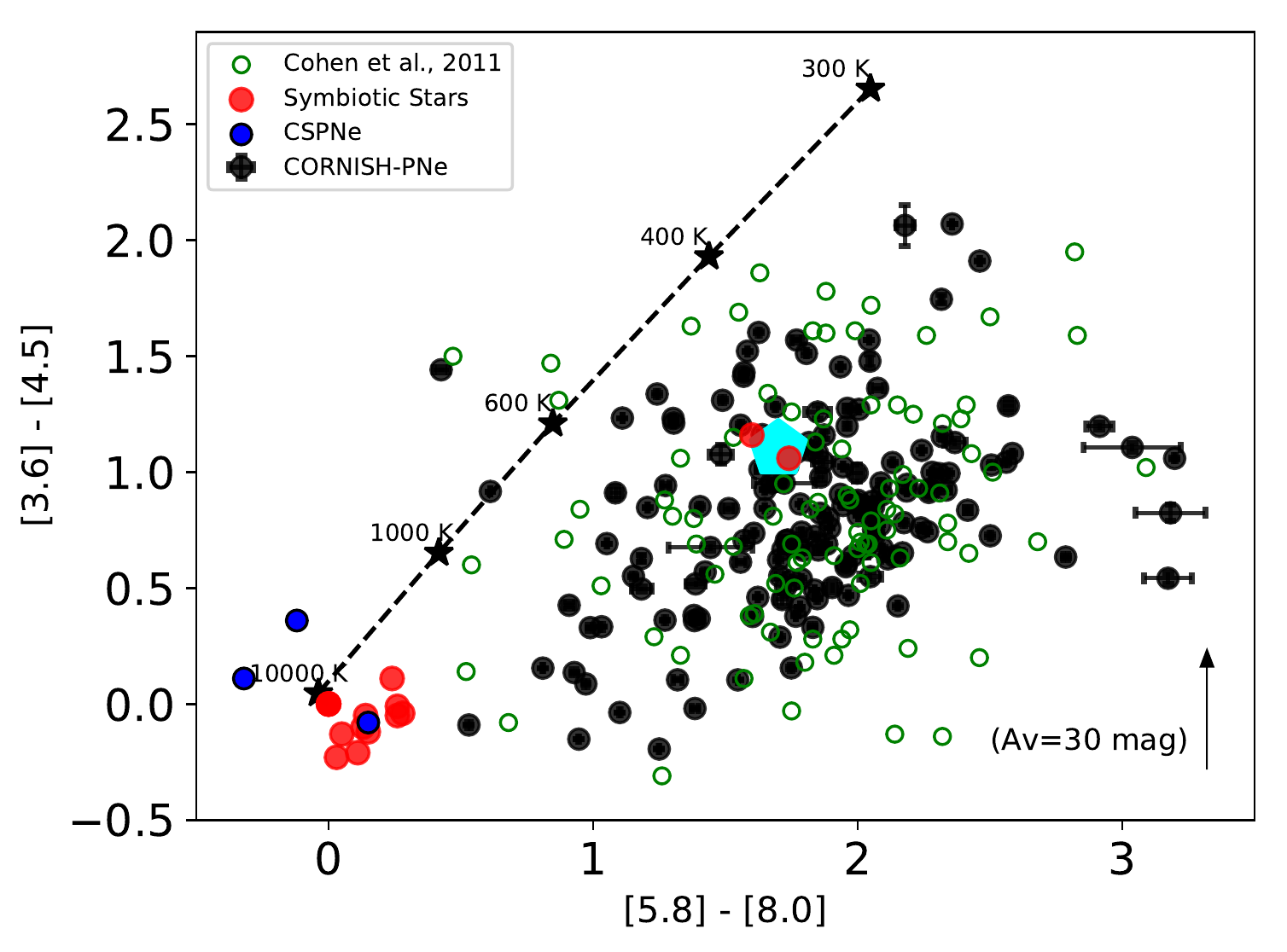}
	\includegraphics[height=6.5cm,width=\columnwidth]{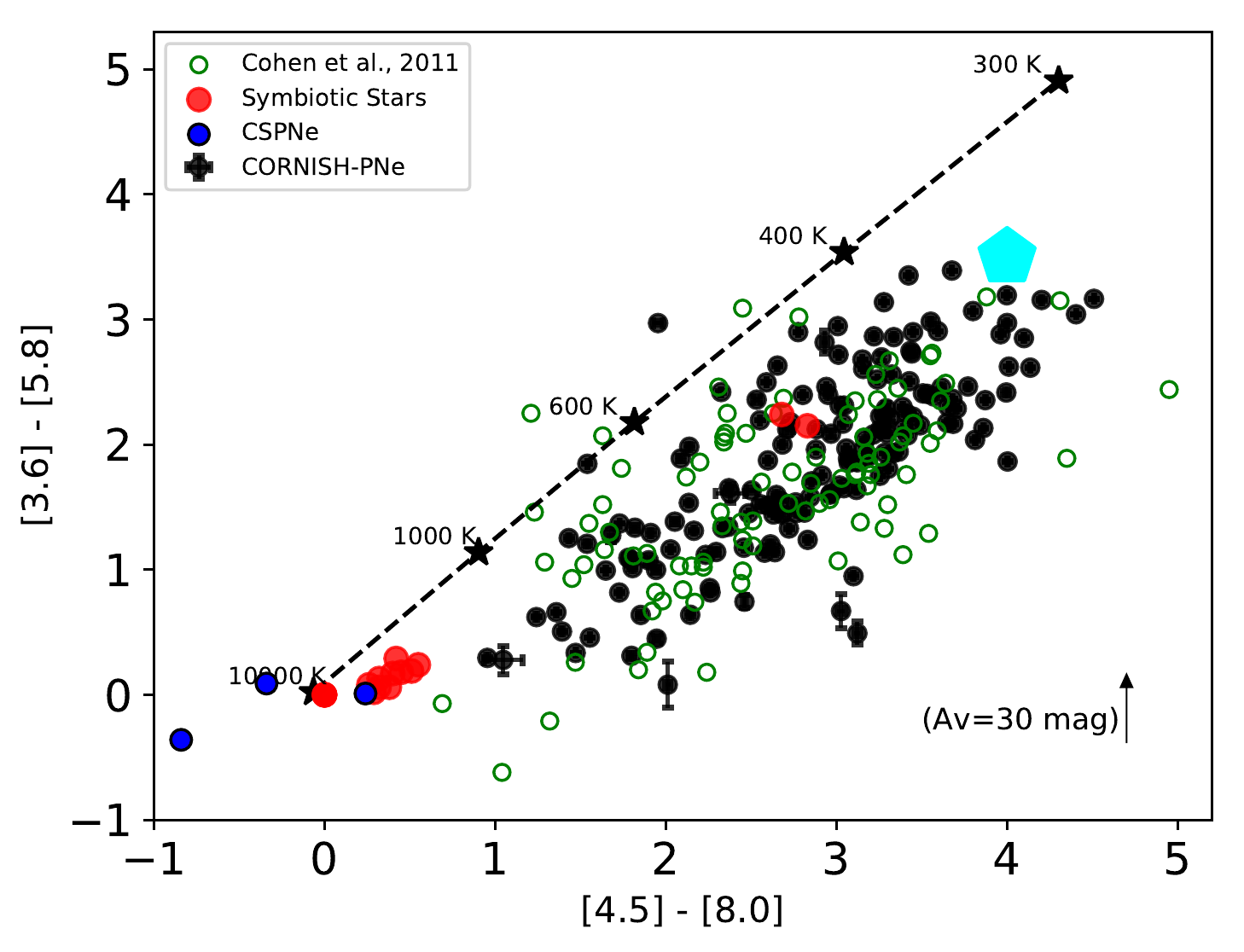}
	\caption{Left panel: $[3.6]-[4.5]$ vs.$[5.8]-[8.0]$. Right panel:$[3.6]-[5.8]$ vs.$[4.5]-[8.0]$. The green circles are PNe (GLIMPSE data) taken from \protect\citealt{cohenparker}, blue filled circles are CSPNe (central stars of PNe) taken from \protect\citealt{hora2004} and symbiotic stars from \protect\citealt{corradi2008} and \protect\citealt{misz2013} are represented as red filled circles. The cyan hexagon shows the mean colours of UCH IIs from the CORNISH survey \protect\citepalias{kalprep}. We also show blackbodies at different temperatures. The reddening vector is from the average extinction in \protect\citealt{inde2005}.}
	\label{fig:mid}
\end{figure*}

In Table \ref{colour_comp}, we compare median MIR colours of the CORNISH-PNe with the median colours of the MASH-PNe sample analysed by \cite{cohenparker}. The CORNISH-PNe median colours show good agreement within $2\sigma$. Furthermore, we show the colour distribution of the CORNISH-PNe on the [3.6]-[4.5] vs. [5.8]-[8.0] and [3.6]-[5.8] vs [4.5]-[8.0] colour-colour planes in Figure \ref{fig:mid}. Figure \ref{fig:mid} also shows the positions of PNe sample taken from \cite{cohenparker}, positions of the central stars of PNe \citep{hora2004}, symbiotic stars \citep{misz2013,corradi2008}, the mean colour of UCH II regions sample from the CORNISH survey \defcitealias{kalprep}{Paper~III}(\citealt{kalprep}, hereafter \citetalias{kalprep}) and the position of blackbodies at different temperatures.

\begin{table}
\begin{center}
\caption{Comparison of the median colours of 160 CORNISH-PNe with PNe samples from \citealt{cohenparker}.}\label{colour_comp}

\begin{tabular}{|c|c|c|c|c|}
\hline
\hline
 Colour   & \cite{cohenparker} & CORNISH-PNe \\
 Index    &107 &160\\
\hline
$[3.6] - [4.5]$ & $0.81\pm 0.08$ &$0.84 \pm 0.08$  \\ 
$[3.6] - [5.8]$&  $1.73\pm 0.10$ &$ 1.94 \pm 0.19 $  \\
$[3.6] - [8.0]$&  $ 3.70\pm 0.11$ &$ 3.83 \pm 0.37$\\
$[4.5] - [5.8]$& $ 0.86\pm 0.10$  &$ 1.08 \pm 0.11$\\
$ [4.5] - [8.0]$&  $ 2.56\pm 0.11$ &$ 3.01 \pm 0.30$ \\
$[5.8] - [8.0]$ &  $1.86\pm 0.07$&$1.84 \pm 0.18$\\
\hline		  			     	    	       		 	 	
\end{tabular}	  			     	    	       		 
\end{center}		     	    	       		 		
\end{table}

The CORNISH-PNe show the same broad distribution shown by the \cite{cohenparker} PNe sample and they are well separated from the S-type symbiotic stars and central stars of PNe, which was not seen in Figure \ref{fig:nir} and \ref{fig:opt} (NIR and optical colour-colour plane). On the  $[3.6]-[4.5]$ vs.$[5.8]-[8.0]$ plane, the colours of the CORNISH-PNe, dusty (D-type) stars and UCH II regions seem inseparable. This looks different on the $[3.6]-[5.8]$ vs.$[4.5]-[8.0]$ plane, where the CORNISH-PNe seem separated from the mean colour of the CORNISH UCH II regions. This could be as a result of the PAH emission strength in the 8 ${\umu m}$ band, compared to bands 3 ${\umu m}$ and 4 ${\umu m}$.

Generally, this broad distribution could be a result of different dominant emission mechanisms across the bands \citep{phil_b2009}. A combination of the evolutionary state and PAH strength across the different bands for carbon rich PNe could also contribute to the wide spread of the CORNISH-PNe colours. Most of the CORNISH-PNe show the red colours typical of PNe, including the ones dominated by dust continuum, ${H_2}$ and PAH emission. However, there are a few to the left in Figure \ref{fig:mid} (both colour-colour planes), where stellar continuum could be contributing to their emission. The colours of such PNe could also be a result of dominant ionized gas emission with little or no dust emission. The measured mid-infrared flux densities using aperture photometry are given in Table \ref{NIR_table}.

\begin{table*}
\caption{Measured NIR magnitudes and MIR flux densities for the CORNISH$-$PNe. These are measured using aperture photometry as described in Section \ref{aperturephotometry}. In the NIR, some of the CORNISH-PNe were not detected in one or more bands. For such sources,the corresponding columns are left blank. For non-detections in one or more of the bands and where the measured flux density is less than 3$\sigma$ in the MIR, we have defined an upperlimit of 3$\sigma$ limit preceeded by $` < '$. Full table is available online as Table A2.}\label{NIR_table}

\begin{tabular}{lrrrrrrr}
\hline
\multicolumn{1}{l}{CORNISH Name} &
\multicolumn{1}{r}{J} &
\multicolumn{1}{r}{H} &
\multicolumn{1}{r|}{K}&
\multicolumn{1}{|r}{$F_{3.6}$} &
\multicolumn{1}{r}{$F_{4.5}$} &
\multicolumn{1}{r}{$F_{5.8}$} &
\multicolumn{1}{r}{$F_{8.0}$} \\
  &(mag)&(mag)&(mag)  &(mJy)&(mJy)&(mJy)&(mJy)\\
\hline
G009.9702$-$00.5292 &  16.35 $\pm$ 0.03 &  14.69 $\pm$ 0.02 &  13.07 $\pm$ 0.02 &     8.6 $\pm$ 0.18 &   10.15 $\pm$ 0.14 &    15.32 $\pm$ 0.22 &    41.53 $\pm$ 0.47 \\
G010.0989+00.7393 &  10.66 $\pm$ 0.02 &  10.27 $\pm$ 0.03 &   9.34 $\pm$ 0.03  &  228.81 $\pm$ 0.18 &  621.73 $\pm$ 0.16 &  1105.25 $\pm$ 0.43 &  4046.75 $\pm$ 0.82 \\
G010.4168$+$00.9356 &  14.18 $\pm$ 0.02 &  13.51 $\pm$ 0.03 &  12.64 $\pm$ 0.03 &    8.06 $\pm$ 0.09 &    9.25 $\pm$ 0.07 &      7.9 $\pm$ 0.16 &    57.35 $\pm$ 0.16 \\
G010.5960$-$00.8733 &  16.16 $\pm$ 0.03 &  14.05 $\pm$ 0.01 &  12.23 $\pm$ 0.04  &   48.09 $\pm$ 0.09 &   46.99 $\pm$ 0.07 &    98.96 $\pm$ 0.21 &    245.86 $\pm$ 0.30 \\
G011.3266$-$00.3718 &  16.58 $\pm$ 0.03 &  14.27 $\pm$ 0.01 &  12.84 $\pm$ 0.01 &   12.44 $\pm$ 0.12 &   13.22 $\pm$ 0.09 &    20.91 $\pm$ 0.23 &    33.71 $\pm$ 0.34 \\
G011.4290$-$01.0091 &  11.32 $\pm$ 0.01 &  11.46 $\pm$ 0.02 &  10.64 $\pm$ 0.01  &   27.49 $\pm$ 0.11 &   23.82 $\pm$ 0.08 &    14.76 $\pm$ 0.24 &    20.47 $\pm$ 0.32 \\
G011.4581$+$01.0736 &  15.56 $\pm$ 0.02 &  14.12 $\pm$ 0.01 &  13.18 $\pm$ 0.01 &   $<$0.15 &    9.46 $\pm$ 0.08 &    <0.48 &    47.58 $\pm$ 0.21 \\
G011.7210$-$00.4916 &   - &    - &   -  &    $<$0.3&    0.91 $\pm$ 0.09 &    <0.63 &    <1.08 \\
G011.7434$-$00.6502 &  10.96 $\pm$ 0.01 &  10.78 $\pm$ 0.01 &   9.73 $\pm$ 0.01  &  157.13 $\pm$ 0.18 &  175.46 $\pm$ 0.15 &   536.58 $\pm$ 0.34 &  1814.38 $\pm$ 0.58 \\
G011.7900$-$00.1022 &  11.86 $\pm$ 0.01 &  11.27 $\pm$ 0.01 &  10.68 $\pm$ 0.01 &   25.03 $\pm$ 0.31 &    28.5 $\pm$ 0.23 &    13.65 $\pm$ 0.64 &    53.31 $\pm$ 1.68 \\
\hline		  			     	    	       		 	 	
\end{tabular}	  			     	    	       		 	   		
\end{table*}

\subsubsection{Far-Infrared (FIR) Colours}
The dust content in the circumstellar envelopes of PNe can be better viewed in the FIR, especially when combined  with MIR data. This dust, lost during the AGB phase, has temperatures in the range $\sim 10 \leq T_d  \leq 100$ K \citep{van2017,van2015, vill2002}.

The heating mechanism of dust grains in PNe is believed to be by direct stellar emission from the central stars, which can extend beyond the ionized region, and by Lyman $\alpha$ radiation within the ionized region \citep{pottasch1984,pottasch1986, kwok1986,van2017,hoare1990}. Following this, compact and younger PNe, where dust grain heating is mainly by direct stellar emission from their relatively cooler central stars, could show warmer dust temperatures in the range of ${\sim 110\ K < T_d <  200\ K}$  \citep{pottasch1984,pottasch1986, kwok1986,hoare1990,van1997}, and more evolved PNe, where the major source of heating is from Lyman $\alpha$ radiation, would have relatively cooler dust temperatures in the range of ${\sim 20\ K\lesssim T_d \lesssim 120\ K}$  \citep{van2015,phillips2011}.

The term `evolved'  as used here is relative and according to the analysis of \cite{phillips2011}, PNe with physical diameters $\lesssim 0.08$ pc show dust temperatures as high as 180 K, whereas PNe with larger physical diameters (more evolved) show a range of dust temperatures (${\sim 60\ K< T_d < 120\ K}$).

How bright PNe are in the FIR will be determined by the density and mean temperature of the dust in their circumstellar envelopes. Young PNe from massive stars will be brighter in the FIR. This is because of their association with cooler dust (outside the ionized region), ejected during the AGB phase \citep{kwok1982, cox2011}. This is also the case for bipolar PNe whose progenitors are believed to have experienced strong mass loss during the AGB phase. They are also believed to have a torus that remains neutral for a longer time after ionization \citep{guz2014}. However, H II regions are brighter in the FIR and at longer wavelengths because of the larger amount of dust they possess and lower mean dust temperatures, which can be as low as $\sim 25$ K \citep{anderson2012}.

The use of the Hi-Gal survey together with the MIPSGAL 
${24\ \umu m}$, and WISE ${12\ \umu m}$ and ${22\ \umu m}$ should help us identify H II regions present in the CORNISH-PNe, if there are any. The CORNISH-PNe were cross-matched with the Hi-Gal point source catalogue (individual bands) using a generous radius of $20^{\arcsec}$  and their images were checked to eliminate mis-matches and identifications due to background emission. The CORNISH-PNe were also cross-matched with the WISE ${12\ \umu m}$ and ${22\ \umu m}$ \footnote{http://irsa.ipac.caltech.edu/cgi-bin/Gator/nph-scan?submit=Select\&projshort=WISE}, and MIPSGAL 24 ${\umu m}$  point source catalogue  \citep{rob2015} using the same generous radius of $20^{\arcsec}$.

In Table \ref{anderson_colour}, we show the mean colours of CORNISH-PNe compared to the mean colours of PNe sample analysed by \cite{anderson2012}. The mean colours of the CORNISH-PNe show a good agreement with the \cite{anderson2012} sample within $1\sigma$.

\begin{figure*}
	\includegraphics[height=7cm,width=\columnwidth]{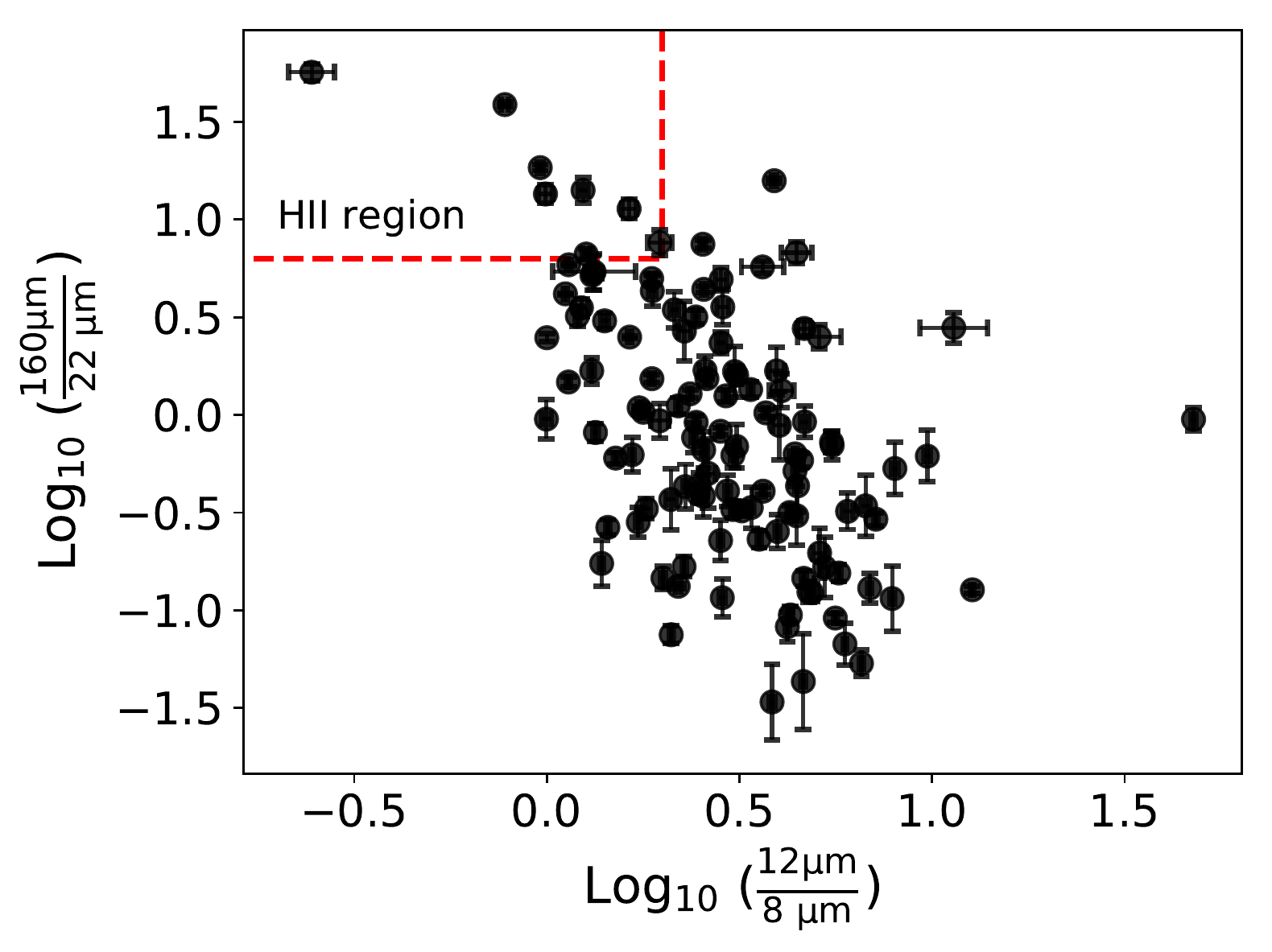}
	\includegraphics[height=7cm,width=\columnwidth]{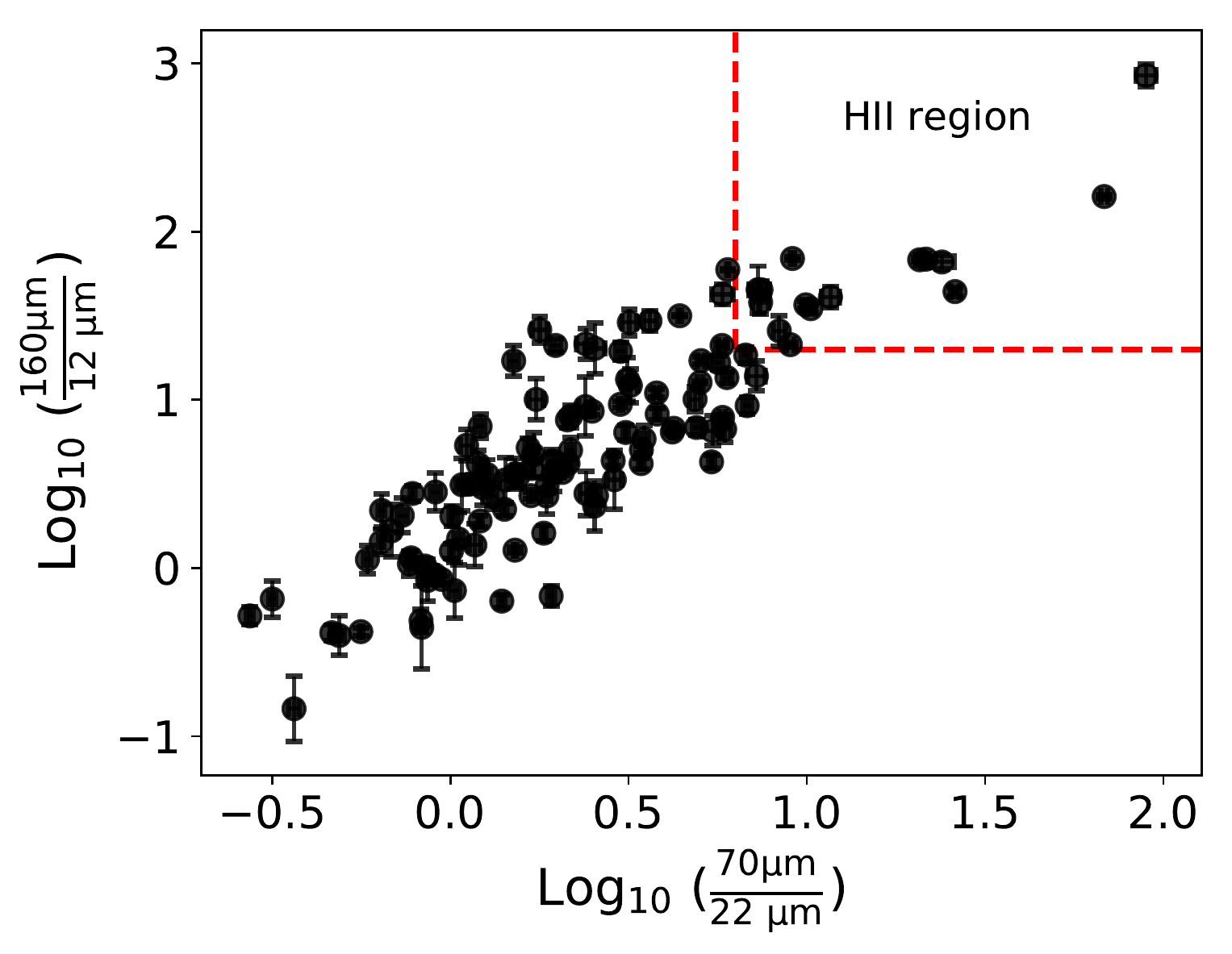}
    \caption{FIR colour-colour plots. Left: $Log_{10}[F_{160}/F_{24}]$ vs. $Log_{10}[F_{12}/F_{8}]$. Right:  $Log_{10}[F_{160}/F_{12}]$ vs. $Log_{10}[F_{70}/F_{22}]$. On both plots we show the cut-off colours (red dotted lines) from \protect\cite{anderson2012} for H II regions (see Table \protect\ref{anderson_colour}) on the CORNISH-PNe.}\label{higal2}
\end{figure*}

Due to the different mean dust temperatures of the PNe and H II regions, they should occupy distinctly different positions on the FIR colour-colour plots. Based on the analysis of \cite{anderson2012} on discriminating between H II regions and PNe (see Table 3 in \citealt{anderson2012} ), we chose the $Log_{10}[F_{160}/F_{24}]$ vs. $Log_{10}[F_{12}/F_{8}]$ and $Log_{10}[F_{160}/F_{12}]$ vs. $Log_{10}[F_{70}/F_{22}]$  colour-colour planes. The distributions of the CORNISH-PNe on these colour-colour planes are shown in Figure \ref{higal2}. A few of the CORNISH-PNe FIR colours extends into the H II region area. This is likely due to sample selection bias as the \cite{anderson2012} sample is made up of optically detected PNe. Hence, the CORNISH-PNe within the H II region area could be compact PNe with cooler dust outside their ionized regions. At a $3\sigma$ significance level, three of the CORNISH-PNe (see Section \ref{Combined}) are within the H II regions area on both colour-colour planes. 

\begin{table}
\begin{center}
\caption{\label{anderson_colour} Mean far-infrared colours of the CORNISH-PNe are shown in column 3 and the mean colours in column 2 are from \protect\cite{anderson2012} PNe samples. The error is the standard deviation.}
\begin{tabular}{|l|c|c|c|}
\hline
\hline
 Colour   &  \cite{anderson2012}  & CORNISH-PNe \\
\hline
$Log_{10}[12/8]$ &  $0.57\pm 0.29$& $0.45\pm 0.30 $ \\  

$Log_{10}[70/22]$ &$ 0.39\pm 0.51$& $0.34\pm 0.46 $ \\

$Log_{10}[160/12]$ & $0.80\pm 0.49$& $0.77\pm 0.67 $ \\ 
$Log_{10}[160/24]$ & $0.22\pm 0.55$& $0.02\pm 0.62 $ \\ 

\hline		  			     	    	       		 	 	
\end{tabular}	  			     	    	       				 	 
\end{center}			     	    	       		 		
\end{table}

\subsection{Extinction}\label{ext_sec}

Extinctions were estimated using the H$\alpha$ line emission, where available, and near-infrared (NIR) magnitudes. The different methods used are discussed below.

\subsubsection{$H\alpha/5 GHz$ ratio}
\begin{figure}
	\includegraphics[height=7cm,width=\columnwidth]{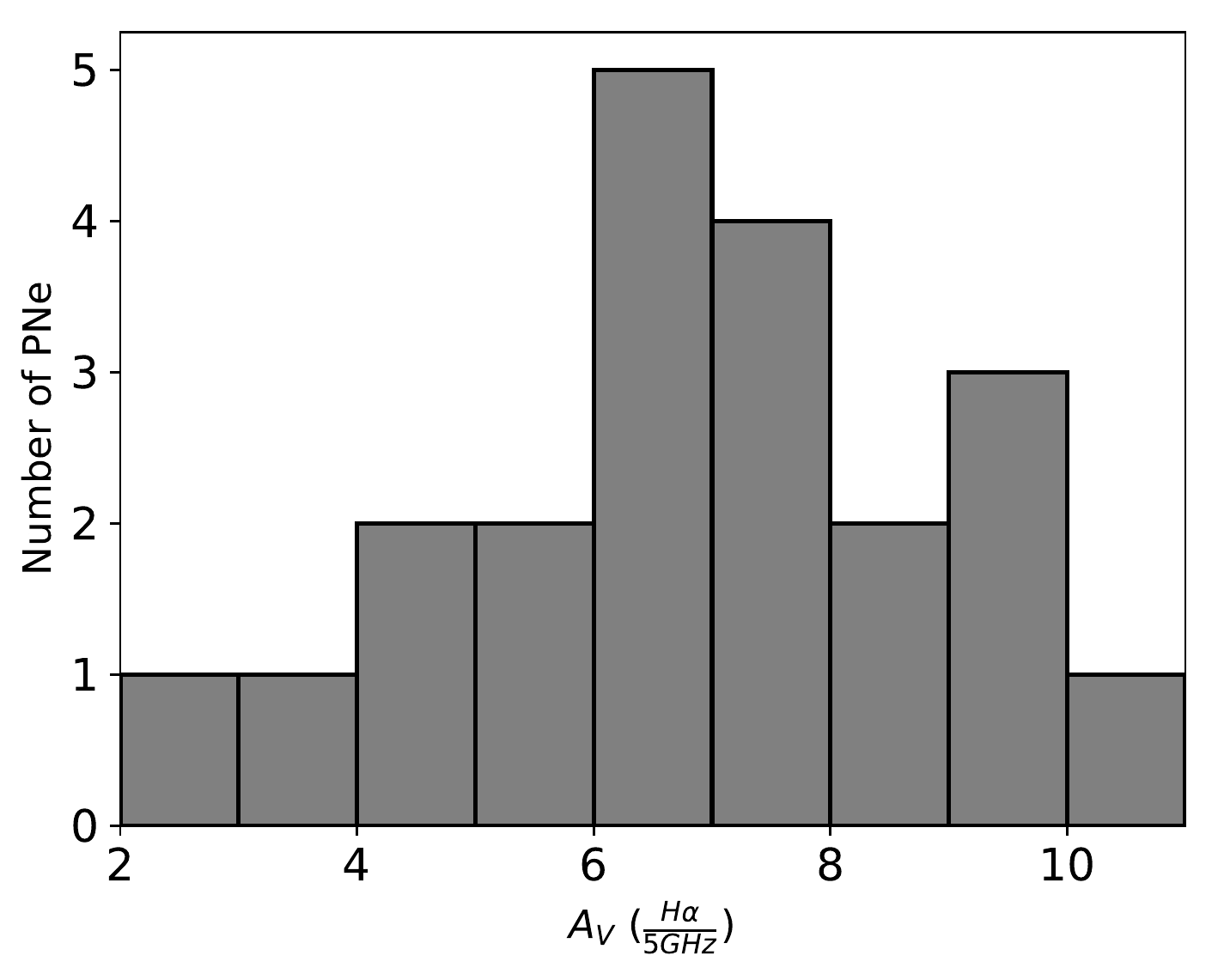}
	\includegraphics[height=7cm,width=\columnwidth]{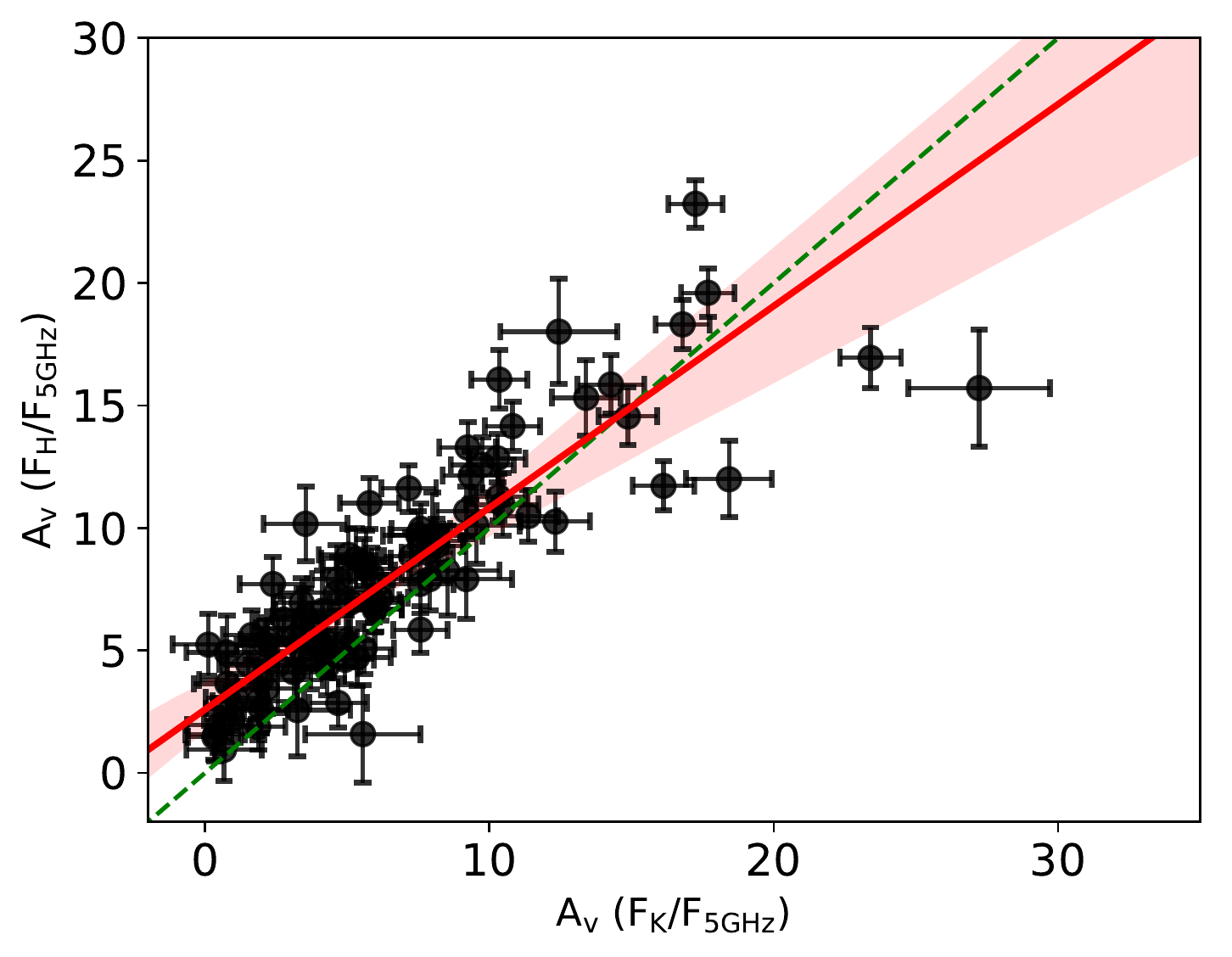}
    \caption{Upper panel: $A_V$ from $H{\alpha}/F_{5GHz}$ method for the 22 CORNISH-PNe with IPHAS counterparts. Lower panel: $A_V$ from the $F_H/F_{5GHz}$ method against $A_V$ from the $F_K/F_{5GHz}$ method. The regression line is shown in red with a $95\%$ confidence band. Less weight is placed on larger errors. The green line is the one-to-one fit.}
    \label{fig:H_alpha}
\end{figure}

Both the H$\alpha$ and 5 GHz emissions are assumed to come from the same effective volume of ionized gas, but unlike the hydrogen line emission, the free-free emission is not affected by interstellar extinction. Following this, the extinction was estimated by comparing the radio-continuum emission at 5 GHz to the $H\alpha$ line flux. In estimating the $H\alpha$ line fluxes (Column 5 in Table \ref{optical_table}), the continuum contribution was accounted for based on the Sloan filter profile\footnote{http://svo2.cab.inta-csic.es/svo/theory/main/}, while ignoring the [NII] contribution. Equation \ref{h} (see \citealt{ruf2004}) was used to estimate the extinction in magnitudes $(c_r)$, which was converted to visual extinction ($A_V$) using $A_V=1.2c_r$ (calculated from the \citealt{card1989} extinction curve with $R=3.1$). The distribution of $A_V$ from this method (for 22 PNe) is shown in Figure \ref{fig:H_alpha}.

\begin{equation}
c_{r}\ (mag)=-2.5log_{10}\left[ \frac{F(H_\alpha)(erg/cm^2/s)}{9.20\ \times\ 10^{-13}F_{5GHz}(mJy)}\right]
\label{h}
\end{equation}

\subsubsection{$K/5GHz$ and $H/5GHz$ ratio}\label{flux_ratio}
Extinction can also be determined from the ratio of the NIR to the radio integrated flux density. For this method, the intrinsic flux ratio determined by \cite{willner1972} (${F_K}/F_{5GHz}=0.3$ and ${F_H}/F_{5GHz}=0.26$) was used to determine $A_K$ and $A_H$ in Equation \ref{willner}. The relationship between the NIR extinction ($A_K$ and $A_H$) and $A_V$ was further derived using the relationships: $A_J/A_V=0.283$, $A_H/A_V=0.184$ and $A_K/A_V=0.113$ (calculated from the \citealt{card1989} extinction curve for $R=3.1$). This method assumes that emission is due to ionized gas (free-free emission, free-bound and hydrogen line emissions). In instances where hot dust dominates the emission, which is the case for some very young/dense and compact PNe, $A_V$ will be under-estimated using this method. This will result in negative $A_V$ magnitudes for some PNe. Approximately $38\%$ of the CORNISH-PNe with reliable K band magnitudes have negative $A_V$ (as low as $\sim -33$ mag) and $\sim 23\%$ with reliable H band magnitudes have negative $A_V$ (as low as $\sim -18$ mag). These sources are not shown on the plot (Figure \ref{fig:H_alpha}), but are given in Table \ref{extincttab}. This could be the effect of hot dust emission, resulting in larger H and K bands flux densities, compared to the level of ionized gas. For such CORNISH-PNe, where the effect is more on the K-band, $H/5GHz$ will give a better estimation. A comparison between $K/5GHz$ and $H/5GHz$ is shown in Figure \ref{fig:H_alpha}. The difference between the regression line and equality line is $< 5$ mag in $A_V$.

\begin{equation}
A_K=-1.086ln\frac{F_K}{0.3F_{5GHz}}; \\ A_H=-1.086ln\frac{F_H}{0.26F_{5GHz}}
\label{willner}
\end{equation}

\subsubsection{ E[H-K] and E[J-H]}\label{avh}
Extinction determined from the colour excess method normalizes observed colours to intrinsic or expected colours. Using Equation \ref{willner} (see Section \ref{flux_ratio}) and the flux ratio determined by \cite{willner1972} and in \citetalias{kalprep} (${F_J}/F_{5GHz}=0.43$), we determined intrinsic colours of $[H-K]_0\sim 0.68$ and $[J-H]_0\sim -0.1$. These colours agree with the modeled intrinsic colours of NGC 6270 (0.62, 0.0) determined by \cite{Weid2013} and the average intrinsic colours of $[H-K]_0\sim 0.65$ and $[J-H]_0\sim -0.1$ from the Galactic PNe sample analysed by \cite{phil2005}.

\begin{figure}
	\includegraphics[height=7cm,width=\columnwidth]{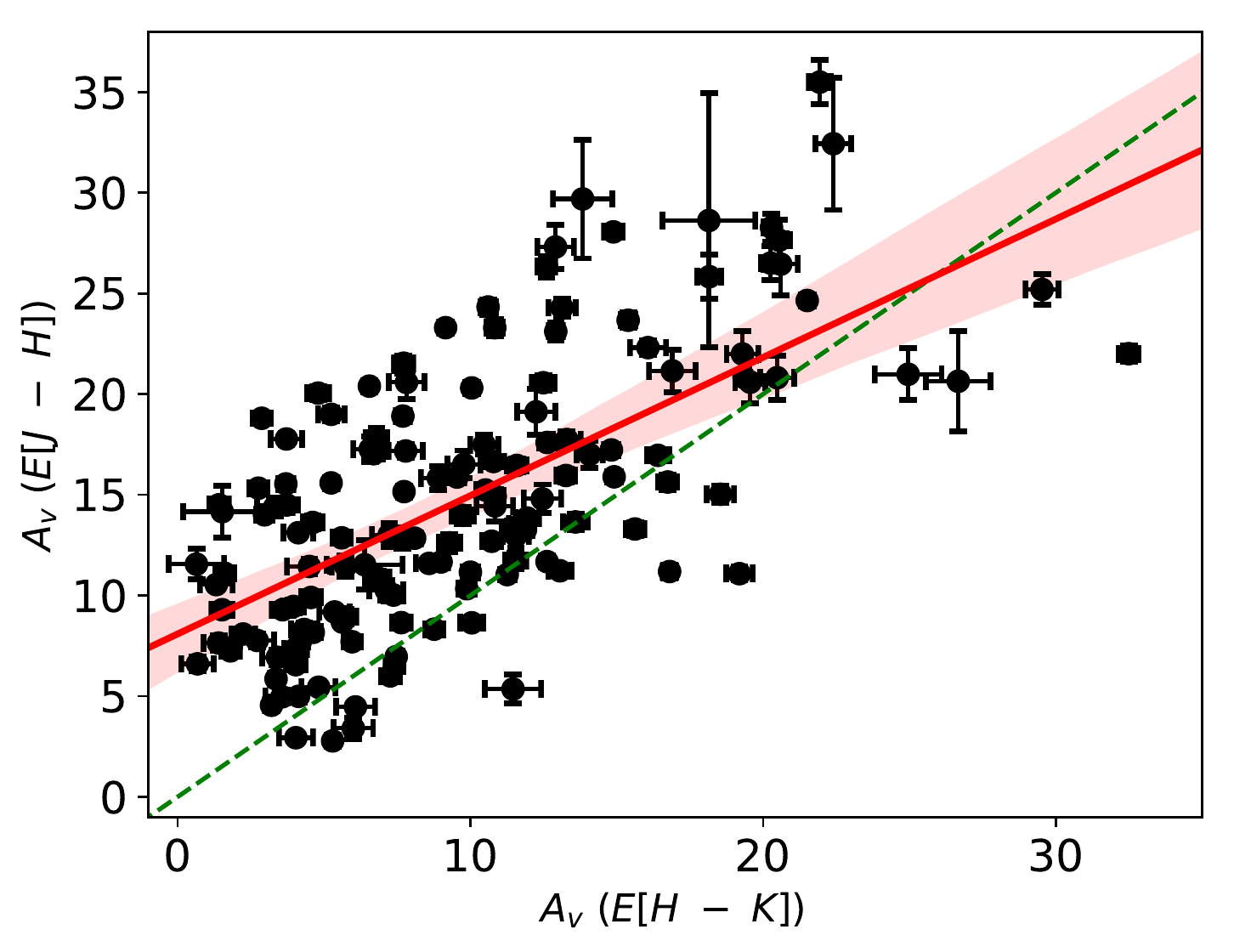}	
    \caption{$A_V$ from the colour excess methods, showing $A_V$ from E(J-H) against E(H-K). The regression line is shown in red with a $95\%$ confidence band. The green line is the one-to-one fit. Less weight is placed on larger errors.}
    \label{tit}
\end{figure}

The CORNISH-PNe NIR colours were normalized to the derived intrinsic colours of $[H-K]_0\sim 0.68$ and $[J-H]_0\sim -0.1$ and $A_V$ was determined using the relationships in Section \ref{flux_ratio}. In Figure \ref{tit}, we present comparison between of $A_v$ obtained using E[J-H] and E[H-K] colour excesses. The effect of hot dust is also seen, which is reflected in the difference between the regression line (red) and the equality line (green).

From the analysis presented in Section \ref{colour}, with the likelihood of the CORNISH-PNe being compact and young, we expect some CORNISH-PNe to be dominated by hot dust. In addition to hot dust, the large excesses in the $H-K$ and ${F_K}/F_{5GHz}$ could also result from $H_2$ emission within the K band. Hence, larger NIR excesses compared to the expected level for thermal ionized gas. For such sources, their intrinsic colours could be higher than 0.68 ($H-K$) and $-$0.1 ($J-H$). According to \cite{pena1987} the intrinsic $H-K$ colour for high density and younger PNe could be as high as 0.8. This is also reflected in the range of intrinsic modeled colours for NGC 7027 (0.21, 0.41) and NGC 6720 (0.62, 0.0) by \cite{Weid2013}, shown in Figure \ref{fig:nir}.

In table \ref{pots}, we compare $A_V$ from the different methods for a few sources. We also did a literature search for the $A_V$ for a few well known PNe within the CORNISH-PNe sample. The $F_K/F_{5GHz}$ method seem to underestimate $A_V$s more, compared to other methods. No one method can be said to be  accurate for all the CORNISH-PNe. However, there are a few CORNISH-PNe whose $A_V$ from the different methods seems to agree.

The varying extinctions from these methods can be attributed to the effect of different dominant emissions (see \citealt{phil2005}) as observed in Figure \ref{fig:nir}, resulting in the different $A_V$ using the different methods. $A_V$ within lower Galactic latitudes could be as high as $\sim 30$ mag (see the extinction map of \citealt{gonza2012,gon2008}), which agrees with the maximum estimated $A_V$ for the CORNISH-PNe using the different methods. $A_V$ from the different methods for the CORNISH-PNe are given in Table \ref{extincttab}.

\begin{table*}
\begin{center}
\caption{Comparison of the $A_V$ from the different methods for some extended sources.  Where available, we use literature $H\alpha$ or $H\beta$ fluxes to calculate $A_V$. Where c (extinction constant) is given in the literature, we converted to  $A_V$ using $c/E(B-V)=1.46$. * indicates that the $H\alpha$ is from a larger area compared to the radio emission.}\label{pots}

\begin{tabular}{|l|l|l|l|l|l|l|l|}
\hline
\hline
 Name   & Known&$A_V$ &$A_V$ & $A_V$& $A_V$&$A_V$&$A_V$\\
 &Name& ${H\alpha}/radio$ & $E[H-K]$ & $E[J-H]$ &$K/radio$&$H/radio$&Literature\\
\hline
G010.0989$+$00.7393 &NGC 6537&- & $3.56\pm 0.56$&$4.97\pm 0.34$&$1.96\pm 0.95$ &$2.62\pm 0.95$&$3.71^1$,  $3.78^2$, $4.3^3$\\

G035.4719$-$00.4365&-&$8.68\pm 0.29$&$5.63\pm 0.20$&$8.68\pm 0.14$&$4.32\pm 0.95$&$4.87\pm 0.95$&-\\

G050.5556$+$00.0448&-&$11.60\pm 0.41$&$11.26\pm 0.25$&$11.04\pm 0.20$&$7.25\pm 0.96$&$8.85\pm 0.97$&-\\
G051.8341$+$00.2838&- & $9.53\pm 0.31$&$8.76\pm 0.32$&$8.33\pm 0.15$&$5.92\pm 0.91$&$7.07\pm 0.90$&-\\
G055.5070$-$00.5579&M 1-71&$3.45\pm 0.11$&$4.04\pm 0.58$&$2.95\pm 0.14$&$0.96\pm 0.99$&$2.18\pm 0.93$&$4.46^4$\\
G059.8236$-$00.5361&-&$6.81\pm 0.22$&$7.26\pm 0.32$&$6.01\pm 0.27$&$3.94\pm 0.92$&$5.26\pm 0.91$&-\\
G062.4936$-$00.2699& M 2-48& $2.30\pm 0.08$*&$6.34\pm 0.35$&$6.55\pm 0.21$&$5.61\pm 1.03$&$5.08\pm 1.03$& $4.67^4$\\
G062.7551$-$00.7262&-&$4.95\pm 0.18$&$5.96\pm 0.29$&$7.70\pm 0.21$&$4.95\pm 1.03$&$5.38\pm 1.04$\\
\hline		  			     	    	       		 	 	
\end{tabular}
    \begin{minipage}{\textwidth}
   	\small
    \item $^1$\cite{mat2005}, $^2$\cite{pottasch2000}, $^3$\cite{kaler1983}, $^4$\cite{tylenda1992}. 
\end{minipage}	
\end{center}			     	    	       		 		
\end{table*}

\begin{table*}
\caption{\label{extincttab} Extinction table showing $AV$ from the different 
    methods in Section \ref{ext_sec} for all the CORNISH-PNe. Sources whose 
    $H\alpha$ emission area is larger than radio, we have indicated using *. 
    Full table available online as Table A3.}
\begin{tabular}{lcrrrr}

\hline
\hline
  \multicolumn{1}{l}{CORNISH Name} &
  \multicolumn{1}{c}{$H\alpha/F_{5GHz}$} &
  \multicolumn{1}{r}{$E(H-K)$} &
  \multicolumn{1}{r}{$E(J-H)$} &
  \multicolumn{1}{r}{$F_{H}/F_{5GHz}$} &
  \multicolumn{1}{r}{$F_{K}/F_{5GHz}$} \\
\hline

G009.9702$-$00.5292 & $-$ &$13.29\pm 0.47$ &$17.74\pm 0.4$ &$5.24\pm 1.26$ &$0.12\pm 1.26$\\
G010.0989$+$00.7393 &$-$ &$3.56\pm 0.57$ &$4.97\pm 0.34$ &$2.62\pm 0.96$ &$1.97\pm 0.96$\\
G010.4168$+$00.9356 &$-$ &$2.71\pm 0.58$ &$7.77\pm 0.35$ &$-0.4\pm 1.09$ &$-2.39\pm 1.09$\\
G010.5960$-$00.8733 &$-$ &$16.07\pm 0.62$ &$22.31\pm 0.37$ &$7.71\pm 1.12$ &$2.39\pm 1.18$\\
G011.3266$-$00.3718 &$-$ &$10.6\pm 0.26$ &$24.32\pm 0.38$ &$3.49\pm 1.75$ &$-1.03\pm 1.75$\\
G011.4290$-$01.0091 &$-$ &$1.99\pm 0.33$ &$-0.43\pm 0.24$ &$-11.49\pm 1.39$ &$-19.89\pm 1.37$\\
G011.4581$+$01.0736 &$-$ &$3.7\pm 0.25$ &$15.54\pm 0.2$ &$-0.99\pm 1.55$ &$-3.96\pm 1.55$\\
G011.7210$-$00.4916 &$-$ &- &-&$-$ &$-$\\
G011.7434$-$00.6502 &$-$ &$5.29\pm 0.2$ &$2.78\pm 0.14$ &$-1.15\pm 0.96$ &$-5.21\pm 0.96$\\
G011.7900$-$00.1022 &$-$ &$-1.24\pm 0.2$ &$6.98\pm 0.14$ &$-6.42\pm 1.67$ &$-9.66\pm 1.67$\\
\hline
\end{tabular}
\end{table*}

\subsection{Heliocentric Distances}\label{distance}
Heliocentric distances to the CORNISH-PNe were estimated using the statistical distance scale described in \cite{frew2016}. The integrated flux density of each CORNISH-PNe was  converted to $H\alpha$ surface brightness ($erg\ cm^{-2}\ s^{-1}$) and Equation 8 in \cite{frew2016} was used to estimate the physical radii and corresponding distances. It should be noted that Equation 8 provides a mean distance and we have not treated optically thin or thick PNe separately (See \citealt{frew2016} for review and details of this statistical distance calibration).

\begin{figure*}
	\includegraphics[width=\columnwidth]{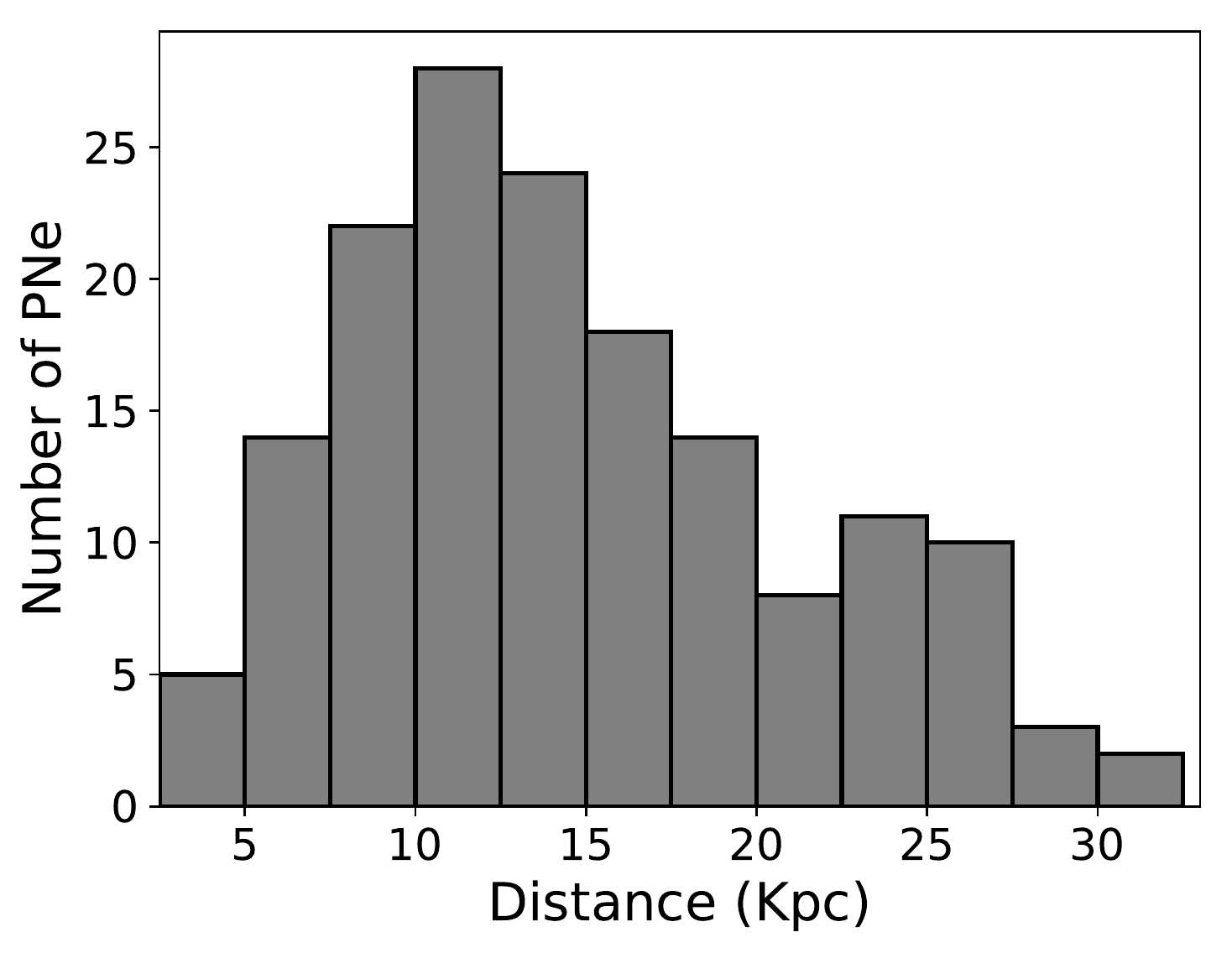}
	\includegraphics[width=\columnwidth]{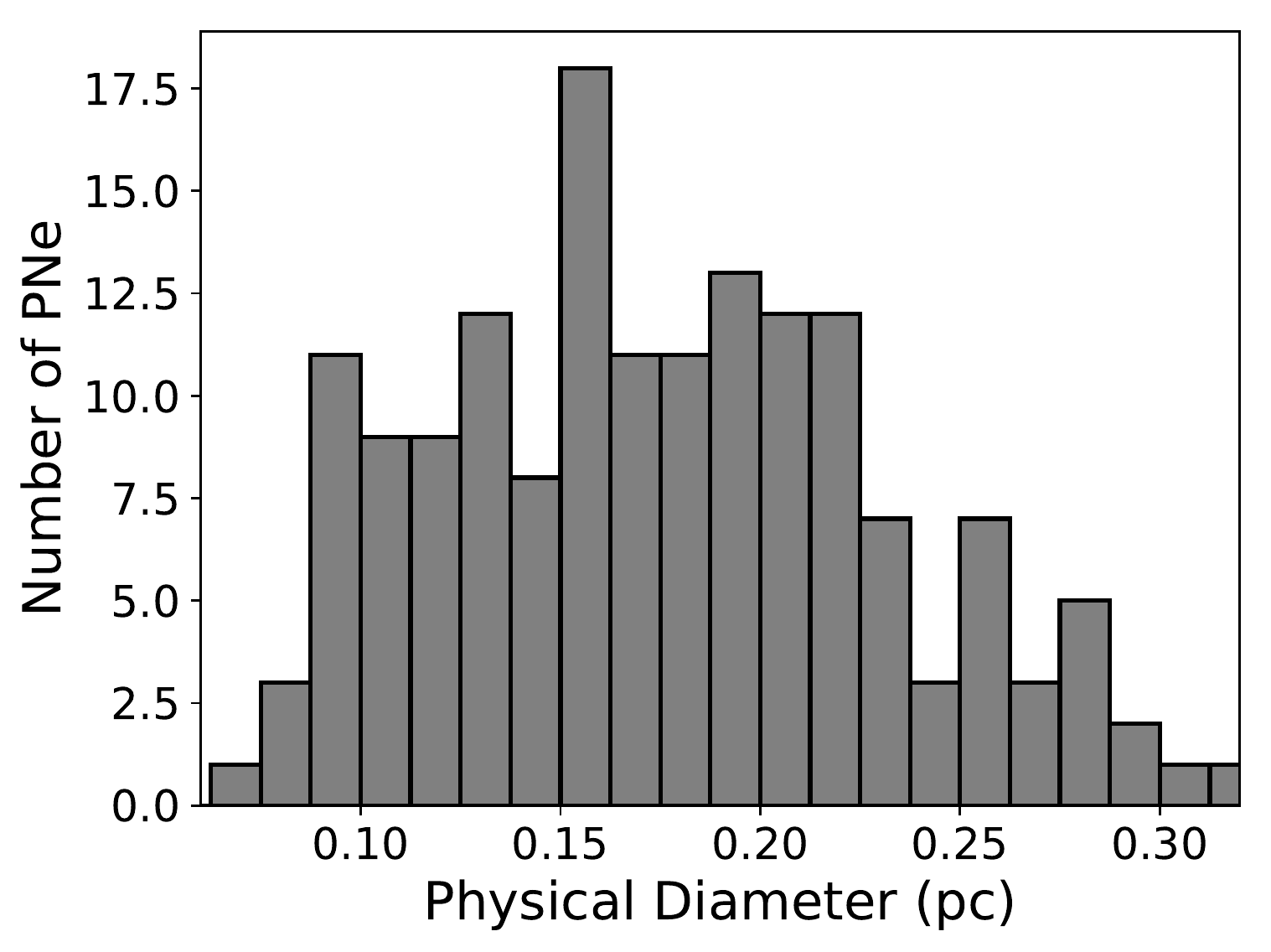}
    \caption{Distribution of the distances in kpc (left) and physical diameters in pc (right).}
    \label{dist}
\end{figure*}

The distributions of the CORNISH-PNe distances and physical sizes are presented in Figure \ref{dist}. The distance distribution shows a peak at $\sim 11$ kpc and gradually falls off to 32 kpc. The general physical diameter distribution shows a generally compact sample with a peak at $ 0.17\pm 0.06$ pc. The mean physical diameter and heliocentric distance are 0.14 pc and 14 $\pm$ 6 kpc, respectively. 

\begin{figure}
	\includegraphics[width=\columnwidth]{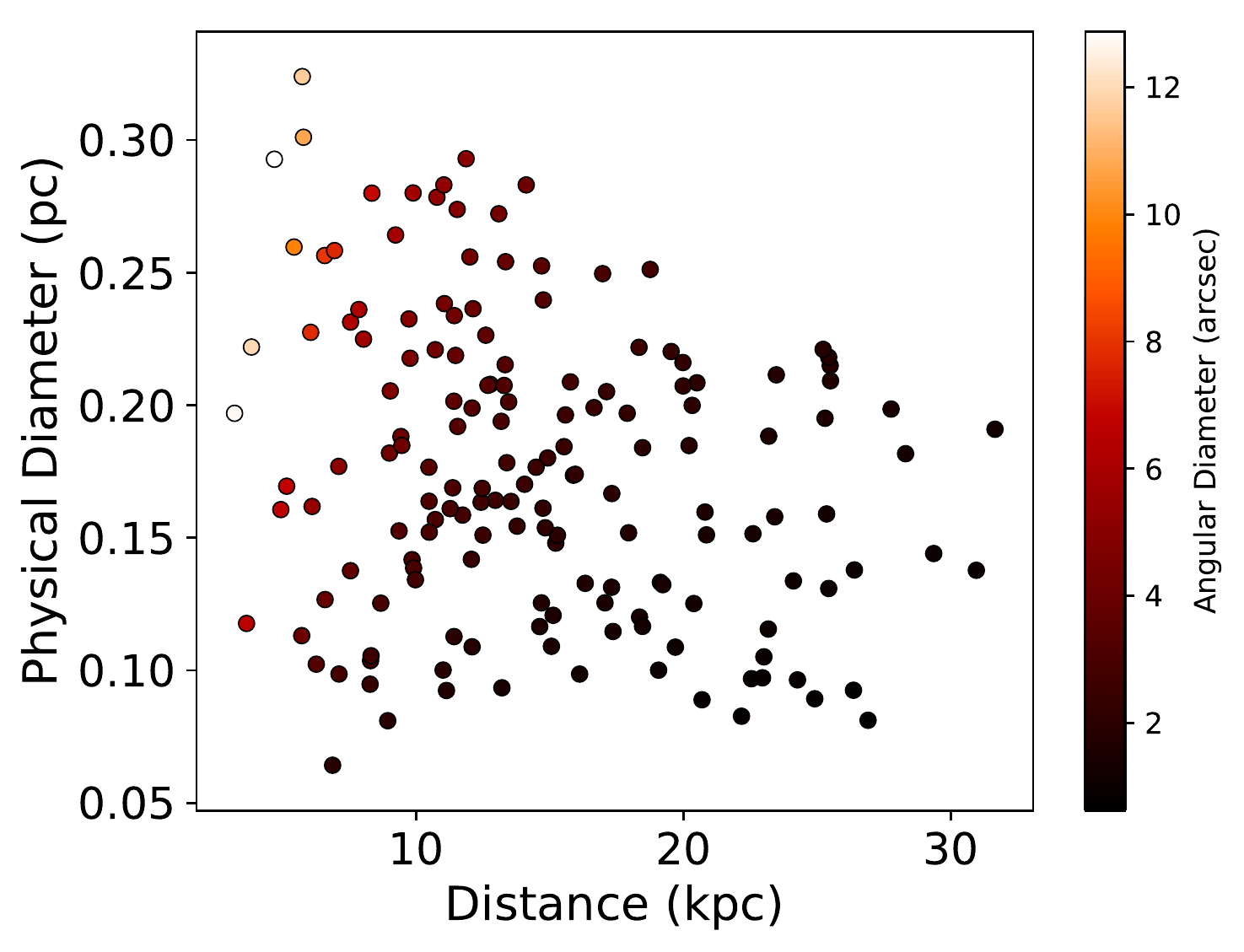}
    \caption{Distances against physical sizes for the CORNISH-PNe. The colour and size of each bubble represents the corresponding angular size.}
    \label{dist_size}
\end{figure}

In Figure  \ref{dist_size}, we show a scatter plot of the distances against the physical diametres and it can be seen that the CORNISH-PNe with smaller angular, having corresponding small physical sizes (< 0.2 pc) are spread over a range of distances and are also the CORNISH-PNe seen at larger distances. The distances to these angularly small CORNISH-PNe are possibly over-estimated, given that we have not taken into account the PNe that are optically thick. Estimated distances and physical diametres are presented in Table \ref{dist_table}.

\begin{table}
\centering
\caption{Distances and physical diameters estimated for the CORNISH-PNe.  For 
    the angular sizes whose deconvolved sizes could not be determined, we have 
    stated the upper(<) and lower limits (>) for the distances and physical 
    diameters respectively. Full table availbale online as Table A4.}\label{dist_table}
\begin{tabular}{lrr}
\hline
\hline
  \multicolumn{1}{l}{CORNISH Name} &
  \multicolumn{1}{c}{$Log_{10}R$ (pc)} &
  \multicolumn{1}{r}{Distance (kpc)}\\
\hline

G009.9702$-$00.5292 &$-0.73\pm 0.06$& $15.52\pm 2.29$\\
G010.0989$+$00.7393 &$-0.93\pm 0.04$& $3.65\pm 0.33$\\
G010.4168$+$00.9356 &$-0.98\pm 0.05$& $23.00\pm 2.65$\\
G010.5960$-$00.8733 &$-0.80\pm 0.05$& $10.71\pm 1.20$\\
G011.3266$-$00.3718 &$-0.63\pm 0.08$& $12.12\pm 2.15$\\
G011.4290$-$01.0091 &$-0.70\pm 0.07$& $13.45\pm 2.29$\\
G011.4581$+$01.0736 &$-0.73\pm 0.08$& $20.20\pm 3.79$\\
G011.7210$-$00.4916 &$-0.68\pm 0.10$& $20.50\pm 4.90$\\
G011.7434$-$00.6502 &$-0.79\pm 0.04$& $4.93\pm 0.47$\\
G011.7900$-$00.1022 &$-0.59\pm 0.07$& $6.57\pm 1.10$\\
\hline
\end{tabular}
\end{table}

\section{Discussion}\label{4}
\subsection{Reliability and Completeness}\label{completeness}
Understanding the evolution and formation of PNe requires the reasonable sampling of different stages of their evolution, including the very young PNe where the shaping mechanism may still be active. Surveys at optical wavelengths have greatly improved the number of known PNe, but these samples are biased by extinction. From the number of currently catalogued PNe ($3540$; \citealt{parker2017} ) compared to the predicted number from population synthesis ($46000 \pm 13000$; \citealt{moe2006}) in our Galaxy, it is clear that many Galactic PNe are yet to be detected, especially towards the Galactic mid-plane where extinction is severe. To improve on the number of known PNe, it is important that PNe are searched for at longer wavelengths. The CORNISH observations at 6 cm provides us with a sample of PNe that is not affected by extinction.

The reliability of this sample has been demonstrated by the multi-wavelength visual identification performed by the CORNISH team and the multi-wavelength properties presented in this work. To demonstrate the completeness and depth of the CORNISH survey, we have compared the CORNISH-PNe with samples of known PNe in the HASH database \citep{parker2017}. This catalogue includes PNe samples from the Strasbourg-ESO catalogue of Galactic PNe \citep{acker1992}, catalogue of Galactic PNe by \cite{koh2001}, Macquarie/AAO/Strasbourg Halpha planetary Nebula catalogues (MASH I, \citealt{parker2006} and MASH II, \citealt{mis2008}) and IPHAS catalogue \citep{sabin2014}.

\begin{figure}
	\includegraphics[width=\columnwidth, height=7cm]{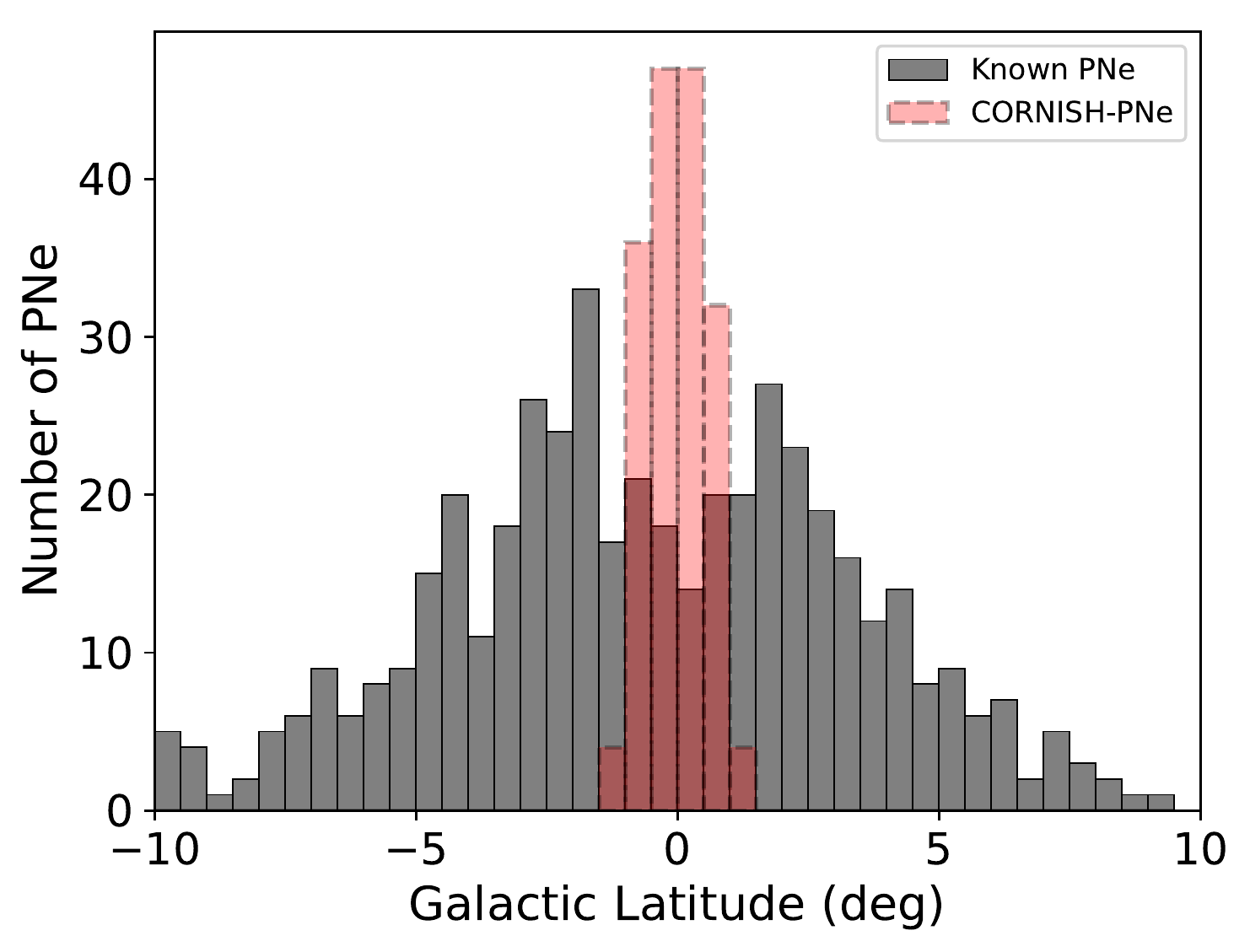}
    \caption{Galactic latitude distribution of the CORNISH-PNe (red) and known PNe (grey). Known PNe samples are from HASH database \protect\citep{parker2017} and includes PNe samples from the Strasbourg-ESO Catalogue of Galactic PNe \protect\citep{acker1992}, Catalogue of Galactic PNe by \protect\citealt{koh2001}, Macquarie/AAO/Strasbourg Halpha Planetary Nebula Catalogues (MASH I, \protect\citealt{parker2006} and MASH II, \protect\citealt{mis2008}) and IPHAS catalogue \protect\citep{sabin2014}.}
    \label{fine2}
\end{figure}

In Figure \ref{fine2}, we show the distribution of the CORNISH-PNe and known PNe. Compared to the CORNISH-PNe, the detection of known PNe (optical detections) drops towards $b=0^\circ$, whereas the CORNISH-PNe clearly peak about $b=0^\circ$. This is likely due to the extinction bias associated with optically detected PNe and increased reddening in this region (see the extinction map of  \citealt{gonza2012}).

\begin{figure}
	\includegraphics[width=\columnwidth, height=6.5cm]{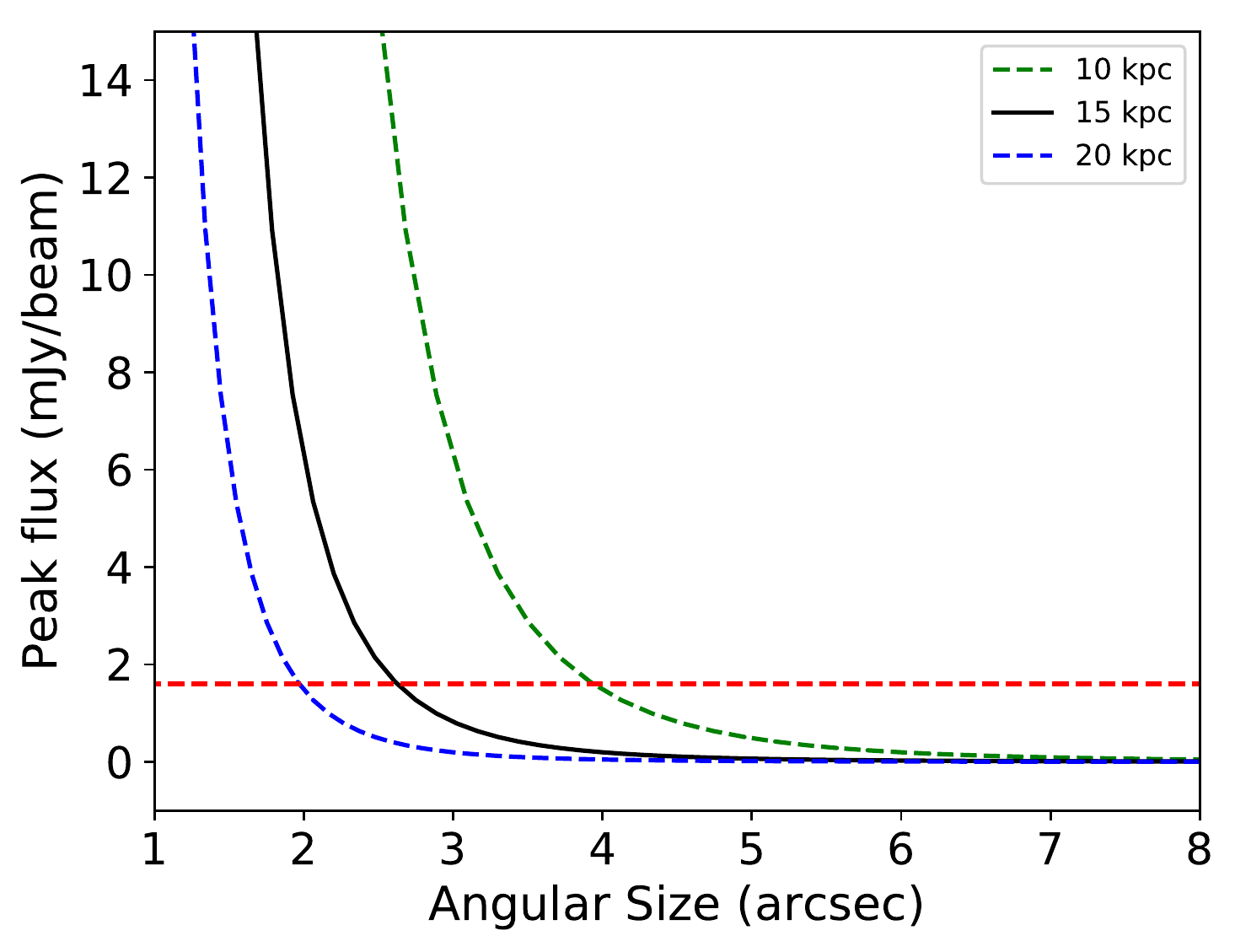}
    \caption{Flux densities at 5 GHz against angular sizes (arcsec) for PNe with physical radii > 0.12 pc (optically thin PNe), at a distance of 10, 15 and 20 kpc. The red horizontal line represents the 4$\sigma$ detection limit of the CORNISH survey.}
    \label{depth}
\end{figure}

Furthermore, if we assume a distance of 20 kpc and that PNe become optically thin at physical diameters > 0.12 pc \citep{zij1990}, we would expect corresponding 5 GHz peak fluxes (Equation 1 in \citealt{stang2008}) as shown in Figure \ref{depth}. At this distance of 20 kpc (see 4$\sigma$ detection limit line in Figure \ref{depth}) we could have detected all sources with angular sizes < 2.5$^{\arcsec}$ and above the CORNISH survey detection limit. This is reflected in Figure \ref{dist_size}, where CORNISH-PNe with small angular sizes are detected at larger distances. At closer distances (10 and 15 kpc are shown in Figure \ref{depth}), we would detect sources with larger angular sizes if their peak fluxes is above the detection limit.

\subsection{Combined colour-colour plots}\label{Combined}
The difficulty in spectroscopic confirmation of PNe due to extinction has informed the need for other methods of confirmation. The works of \cite{parker_2_2012} and \cite{cohenparker}, using the MASH PNe have shown that the use of multi-wavelength analysis can be robust in identifying PNe not detected at optical wavelengths, especially when they have reliable MIR and radio emission measurements. We expect that multi-wavelength colour-colour plots, together with the MIR/radio ratio used in this analysis, should flag different contaminants.

In the NIR colour-colour plot (see Figure \ref{fig:nir}), the CORNISH-PNe show a wide range of colours, which agrees with the wide range of unreddened PNe colours analysed by  \cite{phil2005}. In the MIR (Figure \ref{fig:mid}), the CORNISH-PNe seem well-separated from the symbiotic stars, except for the D-types (dusty). In the FIR, a few CORNISH-PNe extend into the H II regions area on both colour-colour plots. The three outliers in the FIR colour-colour plots (Figure \ref{higal2}) are G019.2356$+$00.4951, G052.1498$-$00.3758 and G058.1591$-$00.5499. G019.2356$+$00.4951 also has a negative spectral index and is discussed in Section \ref{individual_sources}.

\begin{figure*}
	\includegraphics[width=\textwidth, height=3.8cm]{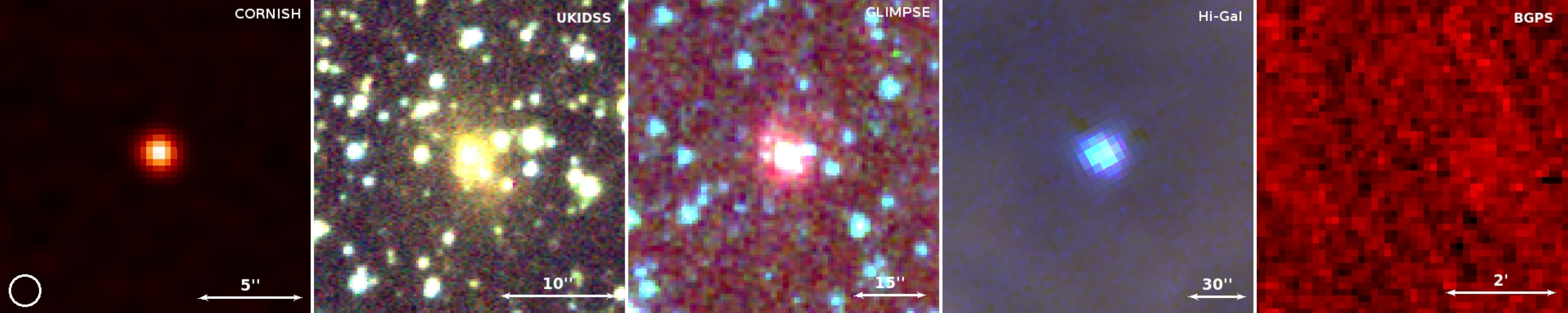}
    \caption{ Multi-wavelength images of G052.1498$-$00.3758. We show the CORNISH 5 GHz radio image (left), followed by NIR 3-colour images : J band is blue, H is green and K is red; MIR 3-colour image : 3.6 $\umu m$ is blue, 4.5 $\umu m$ is green and 8.0 $\umu m$ is red; FIR 3-colour image (right): 70 $\umu m$ is blue, 160 $\umu m$ is green and 250 $\umu m$ is red and the BGPS 1.1 mm. G052.1498$-$00.3758 has FIR colours that are similar to H II regions.}
    \label{G052.14}
\end{figure*}

\begin{figure}
	\includegraphics[width=\columnwidth, height=6.5cm]{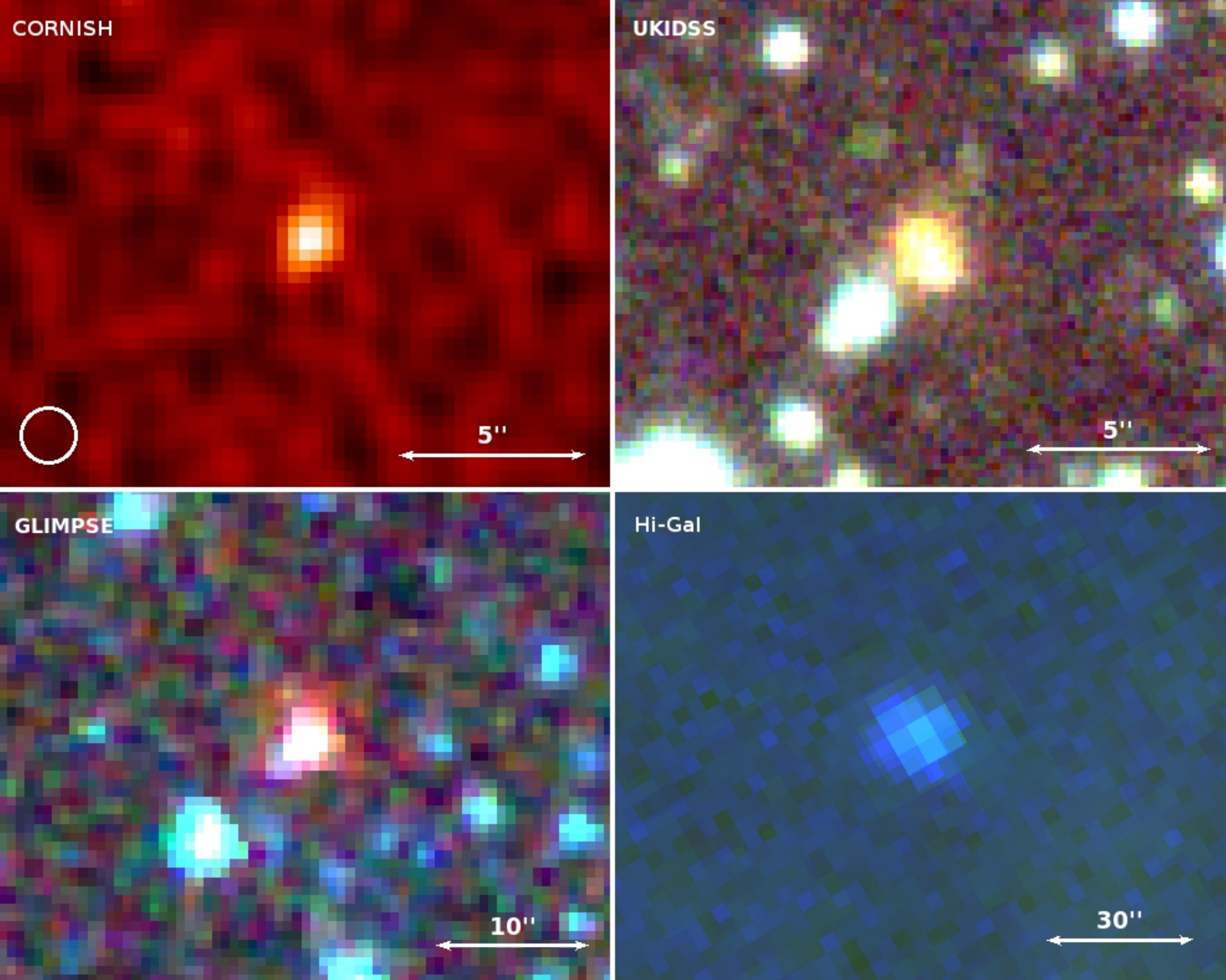}
    \caption{Multi-wavelength images of G058.1591$-$00.5499. We show the CORNISH 5 GHz radio image (top left), NIR 3-colour images (top right): J band is blue, H is green and K is red; MIR 3-colour image (bottom left): 3.6 $\umu m$ is blue, 4.5 $\umu m$ is green and 8.0 $\umu m$ is red and FIR 3-colour image (bottom right): 70 $\umu m$ is blue, 160 $\umu m$ is green and 250 $\umu m$ is red. G058.1591$-$00.5499 has FIR colours that are similar to H II regions.}
    \label{G058_18}
\end{figure}

Despite its FIR colours ($Log_{10}[160/24] \sim 1.54$; $Log_{10}[70/22] \sim 1.83$), G052.1498$-$00.3758 has no millimeter emission and so it is not likely an H II region (see Figure \ref{G052.14}), with a spectral index of $-0.36 \pm 0.09$. G058.1591$-$00.5499 has no millimetre data but, its multi-wavelength images (Figure \ref{G058_18}), NIR and MIR colours are consistent with PNe colours. 

\subsection{Sources With Negative Spectral Indices}\label{individual_sources}

The dominant radio continuum emission in an ionized nebulae is expected to be thermal, although some conditions and processes such as jets and magnetic fields in very young PNe could provide an environment for non-thermal emission. Variability at radio wavelengths could also result in non-thermal emission (see \citealt{cer2017,cerr2011}).

The presence of high velocity collimated jets, believed to be the primary shaping mechanism in PNe, could induce shock fronts and accelerate electrons to a relativistic velocity in a magnetic field, which could result in non-thermal emission. Observational evidence of this is seen in a post-AGB source ($IRAS\ 15445-5449$) with strong magnetic fields, where ionization has not started, but jets are present \citep{perez2013}. Here, we discuss the two CORNISH-PNe with significant negative spectral indices (see Figure \ref{fig:spec}; middle panel). 

\subsubsection{G019.2356$+$00.4951}

\begin{figure*}
\centering
	\includegraphics[width=\textwidth, height=3.8cm]{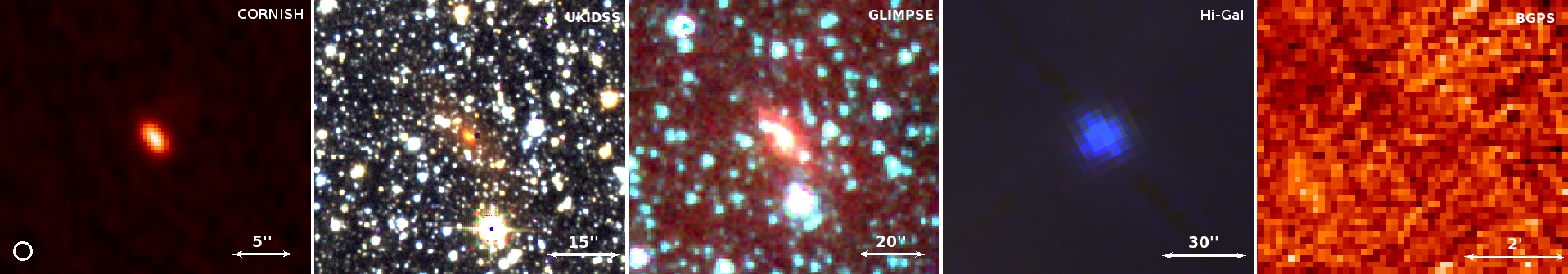}
    \caption{Multi-wavelength images of G019.2356$+$00.4951. We show the CORNISH 5 GHz radio image (left), followed by NIR 3-colour images : J band is blue, H is green and K is red; MIR 3-colour image : 3.6 $\umu m$ is blue, 4.5 $\umu m$ is green and 8.0 $\umu m$ is red; FIR 3-colour image (right): 70 $\umu m$ is blue, 160 $\umu m$ is green and 250 $\umu m$ is red and, the BGPS 1.1 mm. G019.2356$+$00.4951 has a negative spectral index of $-0.66\pm 0.09$ that is compatible with synchrotron emission.}
    \label{G019}
\end{figure*}

G019.2356$+$00.4951 is classified in the SIMBAD database as a radio source (MAGPIS and NVSS surveys). It is associated with X-ray emission (XGPS-I J182416-115554; 3XMM J182416.7-115558\footnote{http://xmm-catalog.irap.omp.eu/source/200519401010016} \citealt{rosen2016}) in the XMM-Newton Galactic Plane Survey (within 4$^{\arcsec}$ search radius), having a hardness ratio of 0.72 \citep{hands2004}. Although it is isolated in the far-infrared image data (Figure \ref{G019}), it has FIR colours ($Log_{10}[160/24] \sim 1.28$; $Log_{10}[70/22] \sim 1.32$) expected of H II regions but with no millimetre emission.  It can be seen in Figure \ref{G019} that this is an extended source in the MIR and NIR but the CORNISH survey has detected only the bright core.

One of the mechanisms through which X-ray emission is produced in PNe is thought to be related to their shaping mechanisms (wind-wind interaction), resulting in hot bubble formation. Such X-ray emissions is observed within the bright innermost shell. The presence of collimated jets and outflows can also result in shocks that emit X-rays. X-ray emission from such processes are thought to be characteristics of young and compact PNe \citep{kast2007,kast2008,chu2000,freeman2014}. 

According to \cite{freeman2014}, PNe that emit X-rays from such interactions are mainly elliptical. G019.2356$+$00.4951 has an elliptical morphology in the radio, NIR and MIR (see multi-wavelength images in Figure \ref{G019}) with a spectral index of $-0.66\pm 0.09$ that is compatible with synchrotron emission. We estimated a distance of $\sim 9\pm 1$ kpc and a diametre of $\sim 0.18\pm 0.02$ pc. We classified this source as a PN and it is likely a young PN, based on the X-ray emission and elliptical morphology. The SED of this source is shown in Figure \ref{sed} (Left panel). 
 
\subsubsection{G030.2335$-$00.1385}

\begin{figure*}
\centering
	\includegraphics[width=\textwidth, height=3.8cm]{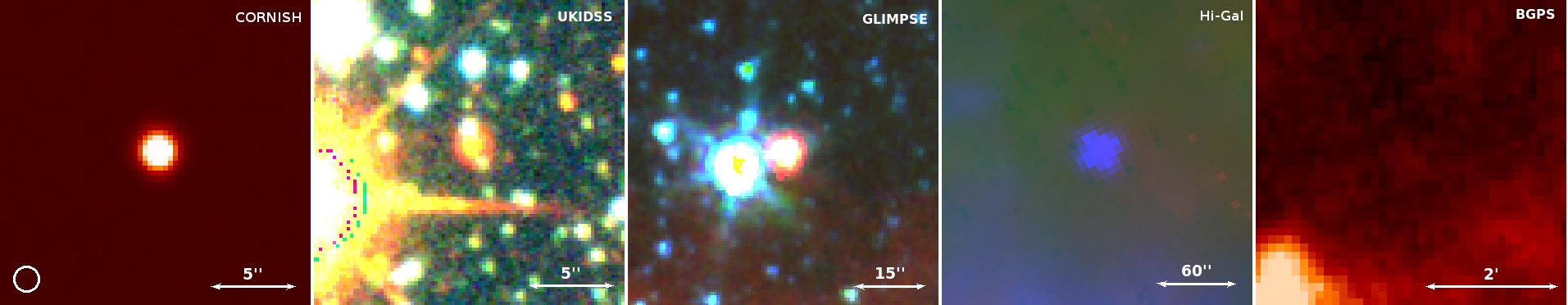}
    \caption{Multi-wavelength images of G030.2335$-$00.1385. We show the CORNISH 5 GHz radio image (left), followed by NIR 3-colour images : J band is blue, H is green and K is red, MIR 3-colour image : 3.6 $\umu m$ is blue, 4.5 $\umu m$ is green and 8.0 $\umu m$ is red; FIR 3-colour image (right): 70 $\umu m$ is blue, 160 ${\umu m}$ is green and 250 $\umu m$ is red and, the BGPS 1.1 mm. G030.2335$-$00.1385 has a negative spectral index of $-0.55\pm 0.08$.}
    \label{G030_23}
\end{figure*}

\begin{figure*}
	\includegraphics[height=6.8cm, width=\columnwidth]{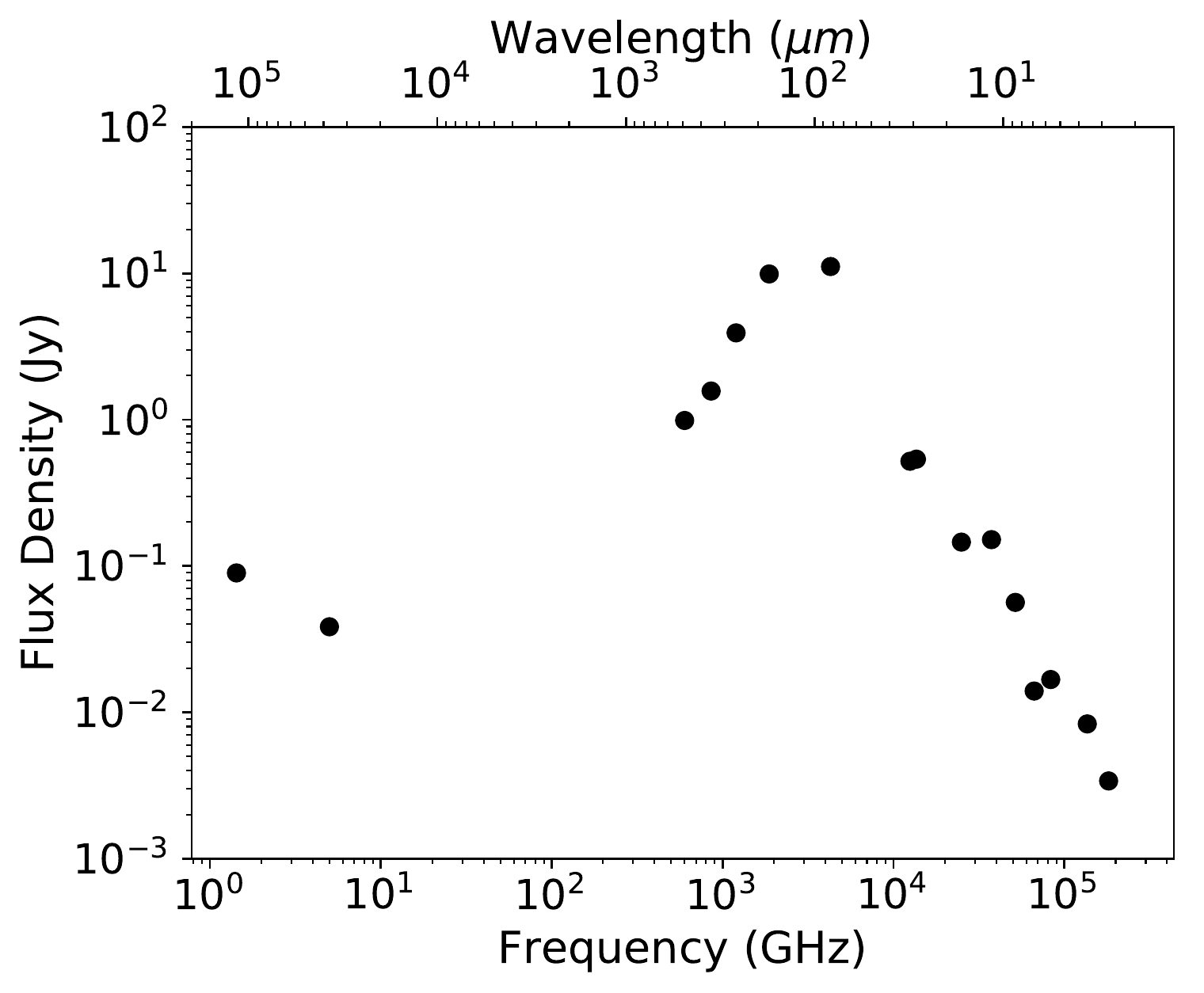}
	\includegraphics[height=6.8cm,width=\columnwidth]{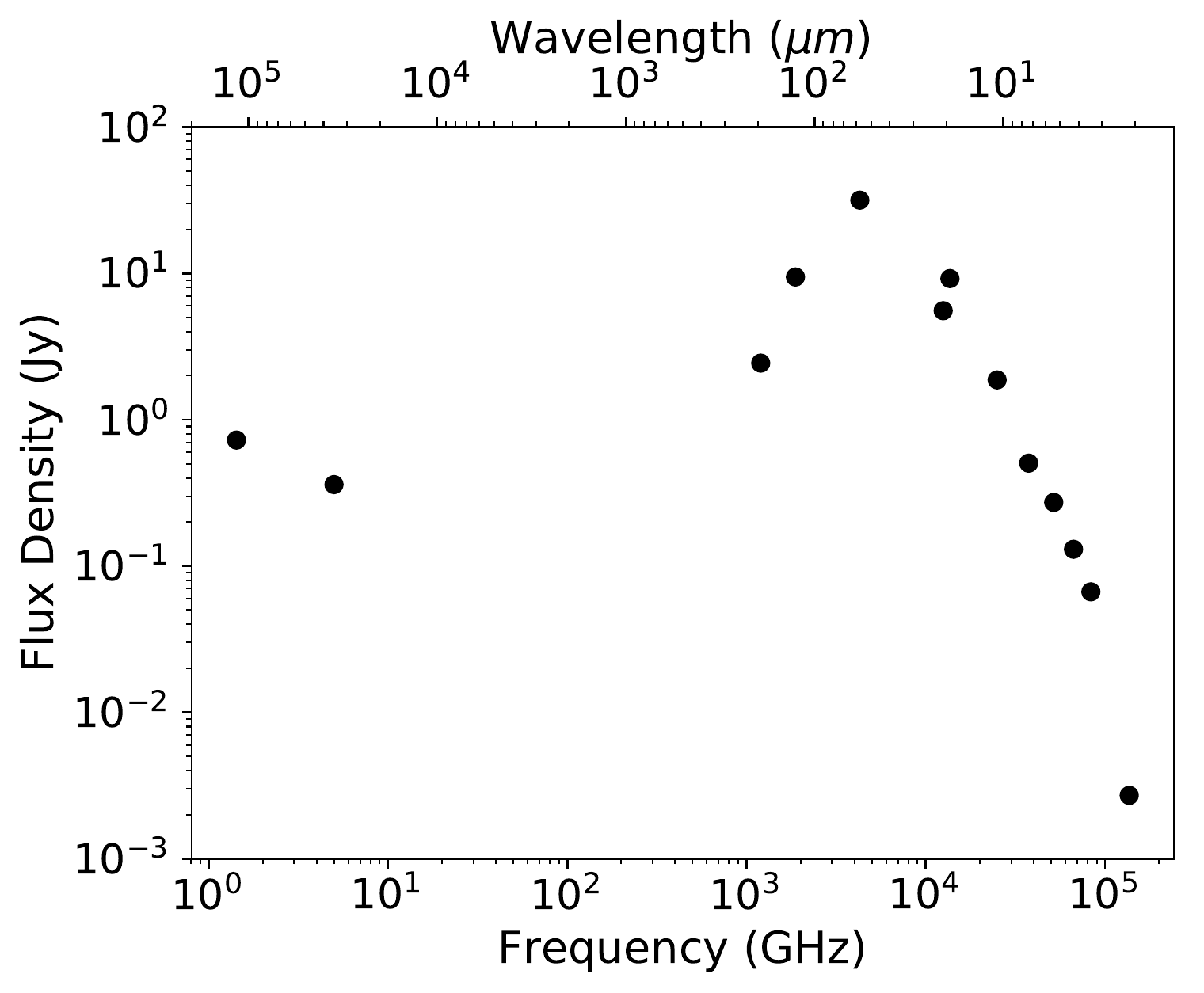}
    \caption{SED (spectral energy distribution) of G019.2356$+$00.4951 (top) and G030.2335$-$00.1385 (bottom).}
    \label{sed}
\end{figure*}

This is an unresolved source in the CORNISH survey with  $T_b$ close to $10^4$ K and a spectral index of $-0.55\pm 0.08$. This negative spectral index is not due to variability as it has a MAGPIS 6 cm integrated flux density of 348.57 mJy (360 $\pm$ 32.91 mJy in the CORNISH). It is not detected in the IPHAS survey. Massive stars are known to evolve fast enough to ionize their circumstellar envelope while the envelopes are still dense \citep{kwok1993,phil2003}.
This source was previously classified as an H \rom{2} region in \cite{leto2009}, but it is isolated in the FIR image, with no millimetre dust emission. It is classified as a PN by \cite{cop2013} from its NIR spectrum and its multi-wavelength images (Figure \ref{G030_23}) supports this classification.  It is also associated with water ($H_{2}O$) maser emission, having a spread of $\sim 70$ km/s, which supports a young nature. We estimate a distance of $\sim 7.0\pm 0.6$ kpc and a diameter of $\sim 0.06$ pc. The SED of this source is shown in Figure \ref{sed} (right panel). 

Collimated jets have been observed to be traced by $H_{2}O$ maser components in post-AGB sources. The velocity spread of these masers can be as wide as $\sim 500\ km/s$, as observed towards $IRAS\ 18113-2503$ \citep{gomez2011}. The presence of $H_{2}O$ masers in PNe is indicative of a young nature. A few PNe with negative spectral indices have also been reported in \cite{boj2011}. The first observational evidence of non-thermal emission in a PN with $H_{2}O$ maser emission was seen in $IRAS\ 15103-5754$, with a spectral index of $\simeq-0.54 \pm 0.08$ \citep{suarez2015} that is compatible with synchrotron emission. $IRAS\ 15103-5754$ is presently considered to be the youngest PN. G030.2335-00.1385 shares similar characteristics with $IRAS\ 15103-5754$ and will be a good source to investigate further. 

Sources with such negative spectral indices could be confused with supernovae remnants (SNRs), having typical spectral indices $\leq -0.5$ and X-ray emission \citep{and2017,bozz2017}. However, there is no known detection of the 22 GHz $H_2 O$ maser emission towards SNRs \citep{woodall2007,cla1999}, which rules out the possibility of G030.2335$-$00.1385 being a SNR.

Young SNRs peak at shorter wavelengths (20-50$\mu m$) \citep{williams2016} than PNe, while older or more evolved SNRs have FIR and sub-millimeter emission \citep{mats2015,laki2015}. The colours (NIR to FIR) of G019.2356$+$00.4951 are rather consistent with PNe (see \citealt{reachwil2006,pin2011}). If it were a core-collapse SNR or PWNe (pulsar wind nebula), we would expect some sub-millimetre emission or a flatter radio spectrum (see \citealt{GS2006,bie1997}) compared to its steep spectral index of $-0.66\pm 0.09$.

\newpage
\section{CORNISH-PNe Catalogue}\label{5}

PNe in the HASH database \citep{parker2017} have mean and median angular sizes of 44$\arcsec$ and 15$\arcsec$, respectively. The median angular size of PNe for both the MASH and IPHAS catalogues is $\sim 22^{\arcsec}$ \citep{parker2006,sabin2014}. It can be seen that known PNe, which are optically selected, are biased towards extended and/or more evolved PNe that may be absent or resolved out in the CORNISH survey. For the CORNISH-PNe, the mean and median angular size are $\sim 3^{\arcsec}$ and $\sim 2.5^{\arcsec}$, indicating a compact PNe sample that is probably more complete within the $|b|< 1^\circ$ region.

The classification of the CORNISH-PNe was done based on the radio properties, MIR/radio ratio, optical colours, infrared colours, extinction, physical sizes and distances. In the absence of IPHAS images, we have used the H$\alpha$ images from the SuperCosmos H$\alpha$ survey (SHS; \citealt{parker2005})\footnote{http://www-wfau.roe.ac.uk/sss/}, where available, to aid classification. In cases where we are unable to draw conclusions from this analysis, we have indicated possible PNe with a question mark `?', and where truly in doubt, we have tagged `unknown' or other source type. Sources that meet all requirements for PNe in all considered multi-wavelength analysis and images are classified as PNe. We indicate if PNe are newly confirmed, newly classified or newly discovered under the status column. The newly confirmed PNe are sources previously classified as possible PNe or PNe in the SIMBAD database, newly classified are sources previously identified as radio sources or YSO (young stellar objects) in the SIMBAD database, with no previous classification as PN, while newly discovered are sources with no astronomical record in the SIMBAD database. The final catalogue is presented in Table \ref{cataloguetable3} and the columns are defined as follows: CORNISH name (1), SIMBAD identification (2), $H\alpha$ detection (3), classification (4), status (5) and comments (6).

A cross-match of the 169 CORNISH-PNe with known PNe (True PNe) from HASH database \citep{parker2017} returns 24 matches (Table \ref{known_PNe}), excluding the CORNISH candidates and likely PNe. A further 47 suspected PNe, as classified in the SIMBAD database, are confirmed as such from the analysis here (see Table \ref{cataloguetable3} for references). 90 out of the remaining 98 CORNISH-PNe are new PNe (12 newly discovered and 78 are newly classified) and the remaining 8 are classified as possible PNe or other source types.

Independently, \cite{frag2016} have used visual inspection of the multi-wavelength images on the HASH database \citep{parker2017}, the MIR/Radio ratio and MIR colours to identify 70 of the sources in CORNISH catalogue as candidate PNe.

Some of the intrinsically red sources based on MIR colours from the GLIMPSE I and II surveys that were classified as candidate YSO by \cite{robi2008} turn out to be PNe (Also see \citealt{parker_2_2012}). YSO are embedded objects with weak radio emission from ionized stellar winds that is usually less than a few mJy \citep{hoare1994,hoare2002}. The 4.8 to 15 GHz radio integrated flux density of the YSO sample from the Red MSX Source Survey \citep{lum2013} are mostly upperlimits (97$\%$). We estimated a median MIR/Radio ratio in order of $10^3$, which is high compared to the MASH PNe \citep{cohenparker} and CORNISH-PNe value of $4.7\pm 1.1$ and $3.9\pm 0.90$, respectively. We expect them to also have FIR colours similar to H II regions as they emit more in the MIR to FIR.

\begin{figure*}

	\includegraphics[height=4.1cm, width=8cm]{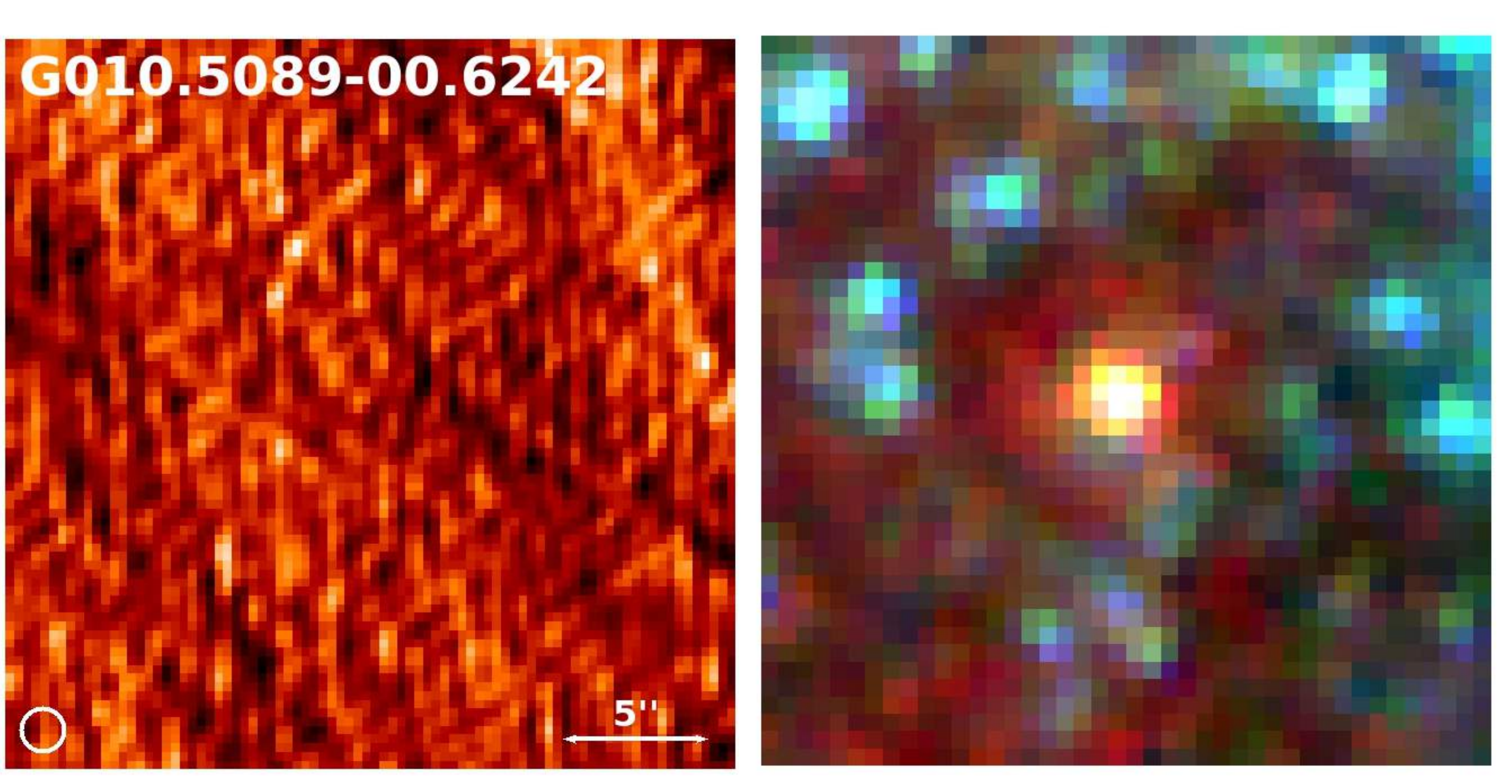}	
\hspace{0.5mm}
	\includegraphics[height=4.1cm, width=8cm]{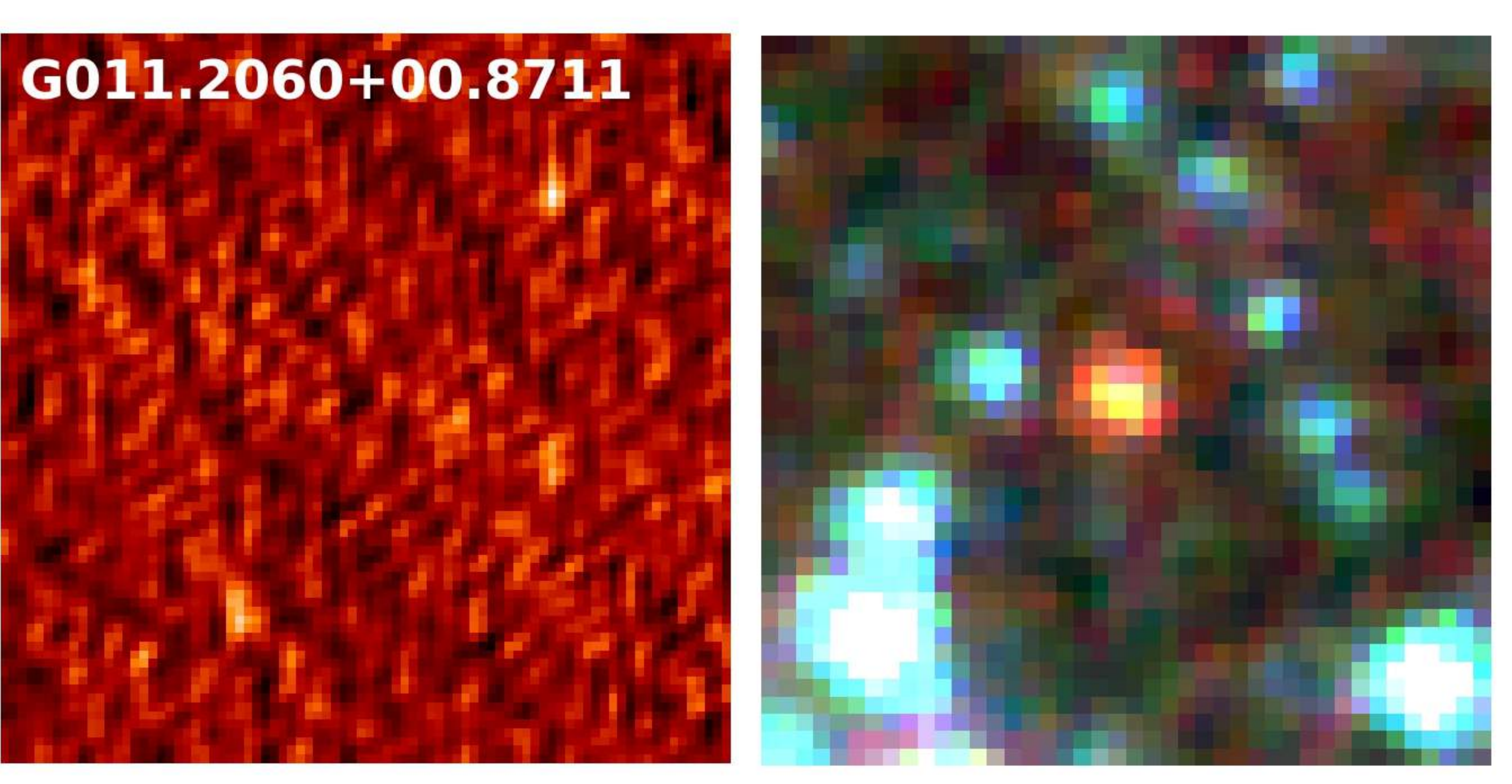}

	\includegraphics[height=4.1cm, width=8cm]{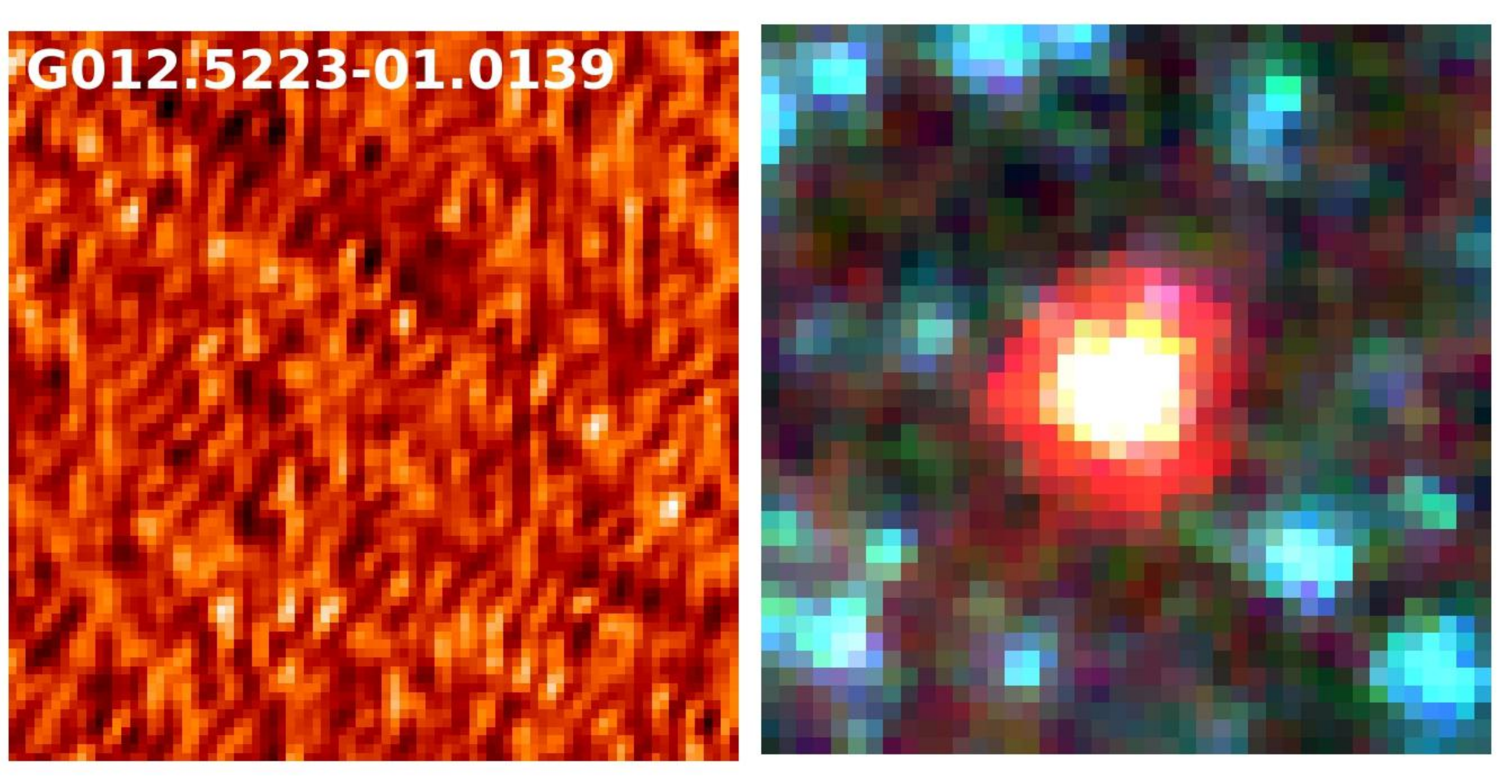}
\hspace{0.5mm}	
	\includegraphics[height=4.1cm, width=8cm]{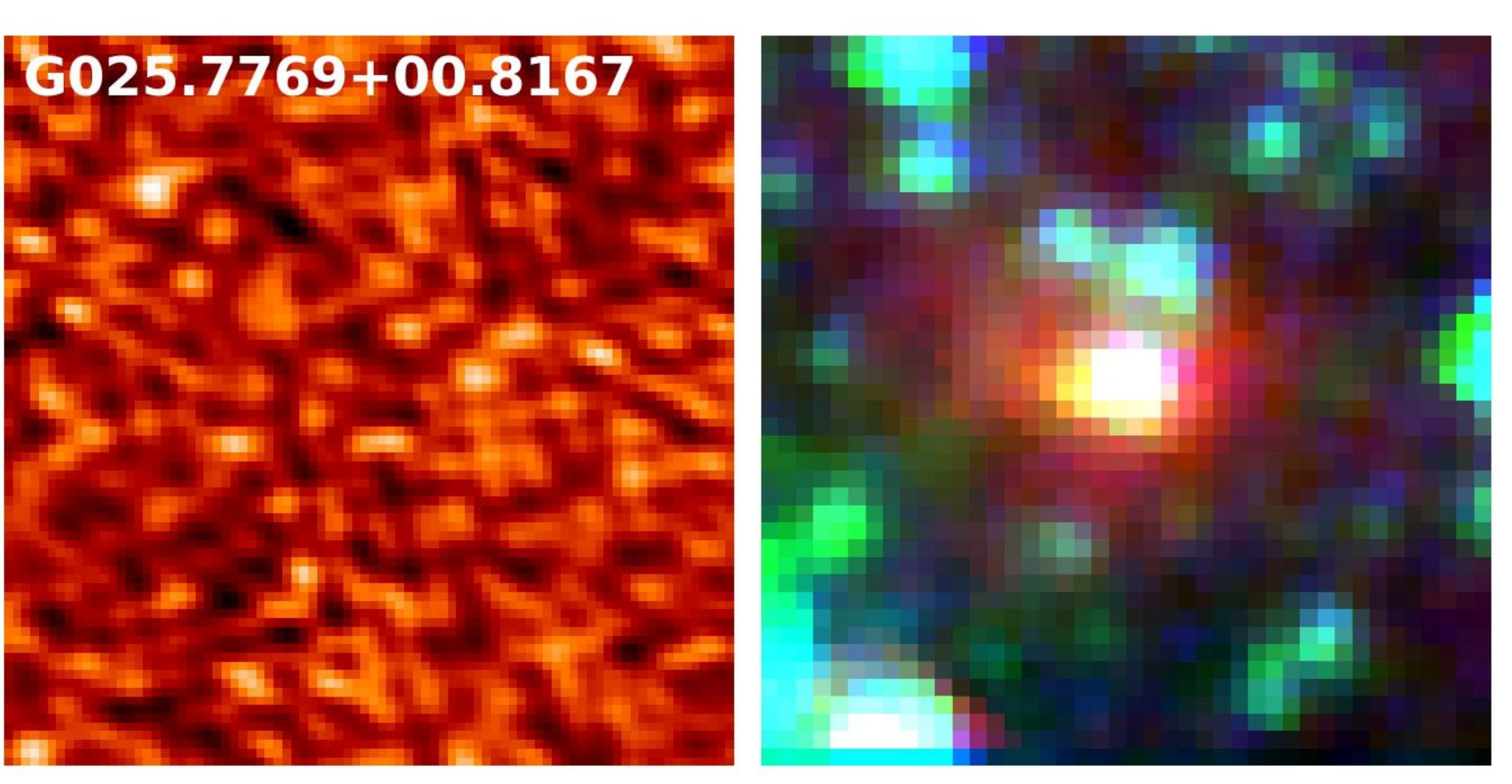}
	
\vspace{0.01cm}

	\includegraphics[height=4.1cm, width=8cm]{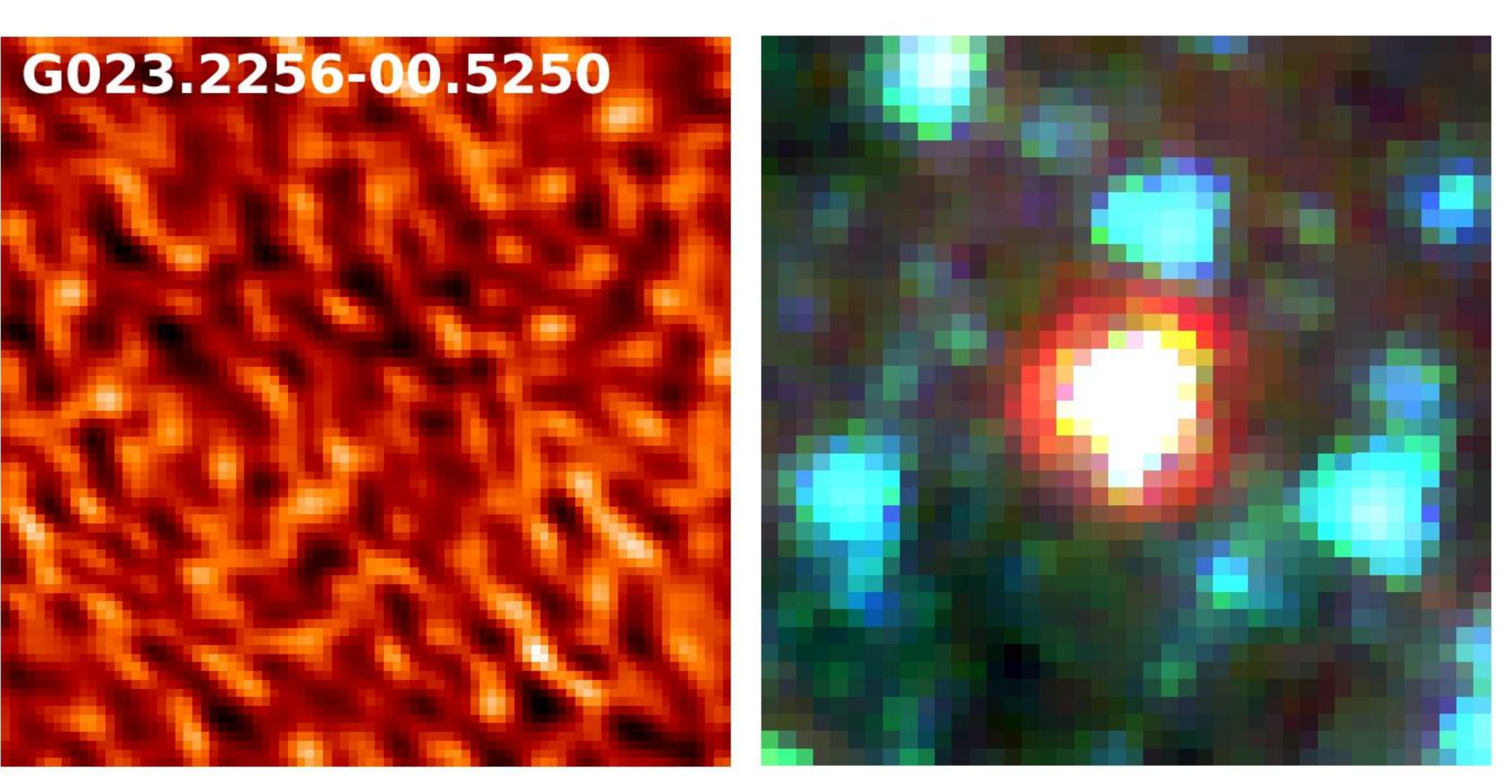}
\hspace{0.5mm}
	\includegraphics[height=4.1cm, width=8cm]{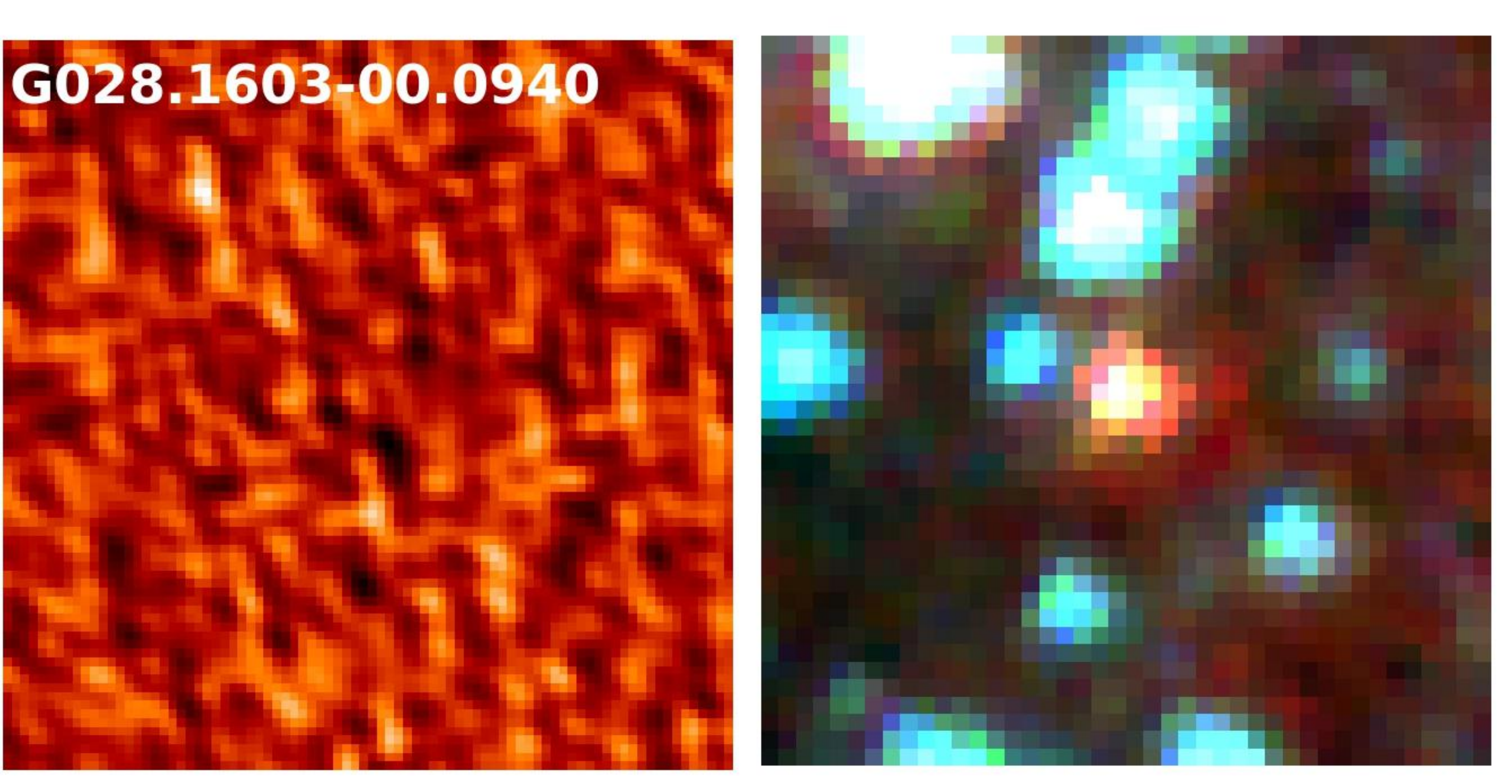}

	\includegraphics[height=4.1cm, width=8cm]{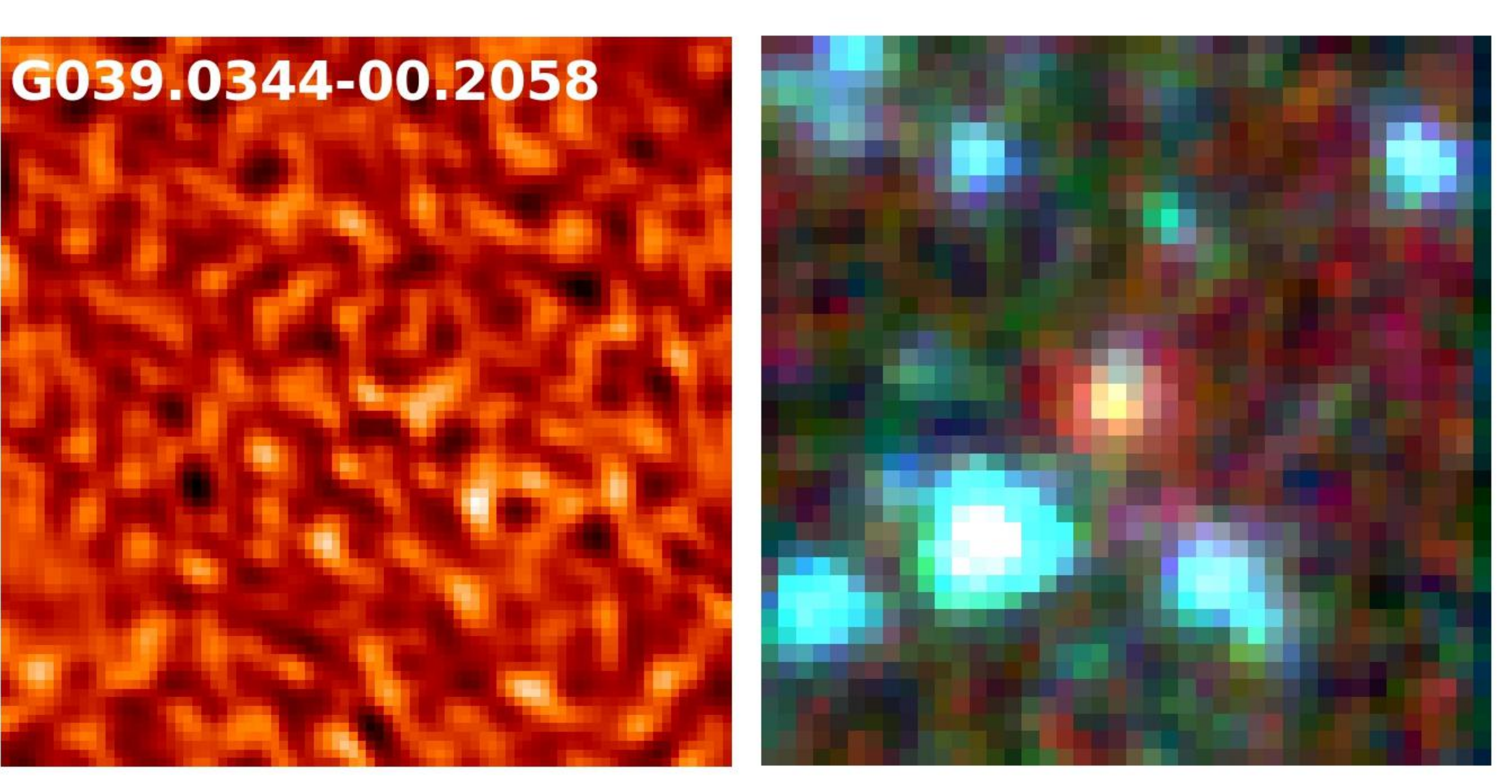}
\hspace{0.5mm}
	\includegraphics[height=4.1cm, width=8cm]{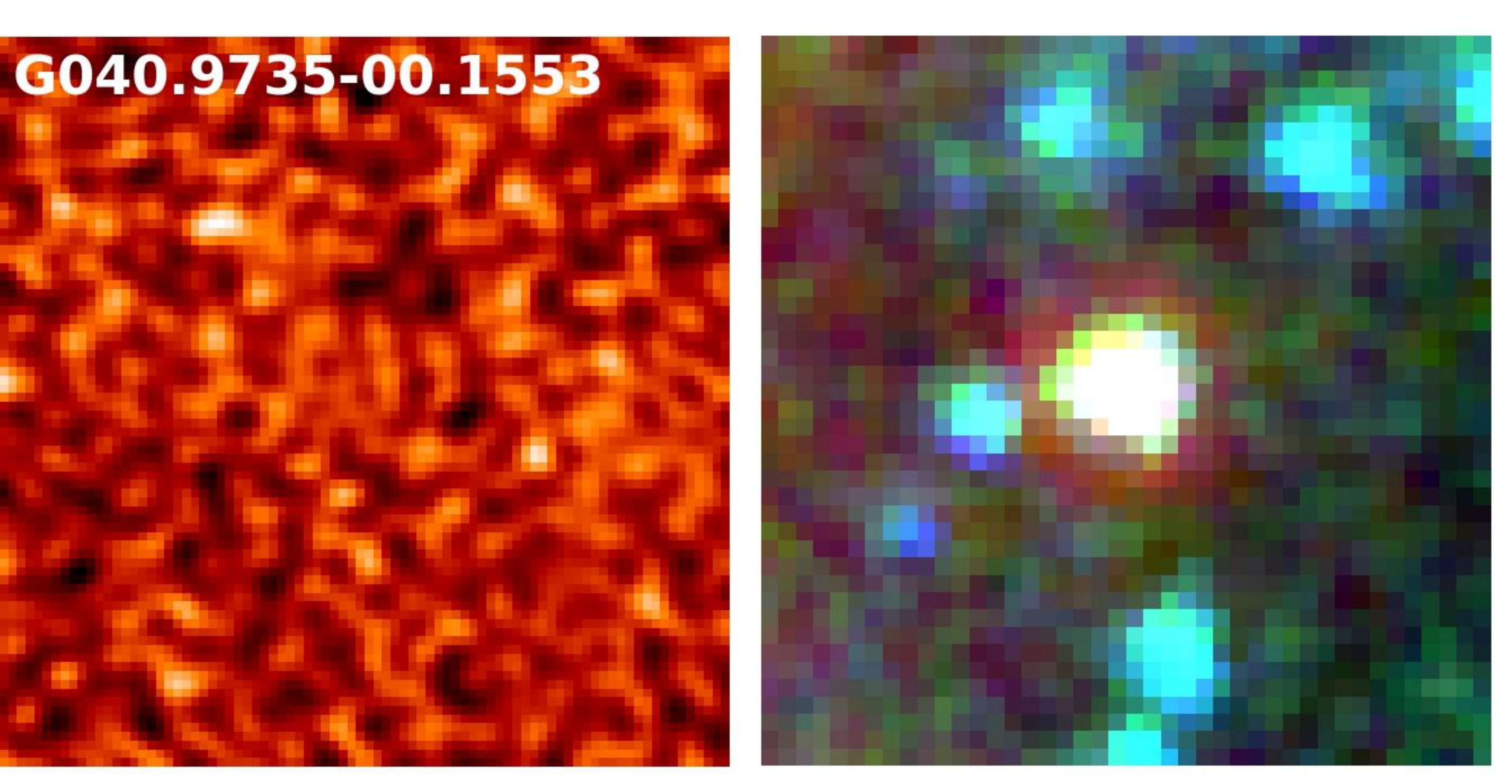}
	
\vspace{0.1cm}

	\includegraphics[height=4.1cm, width=8cm]{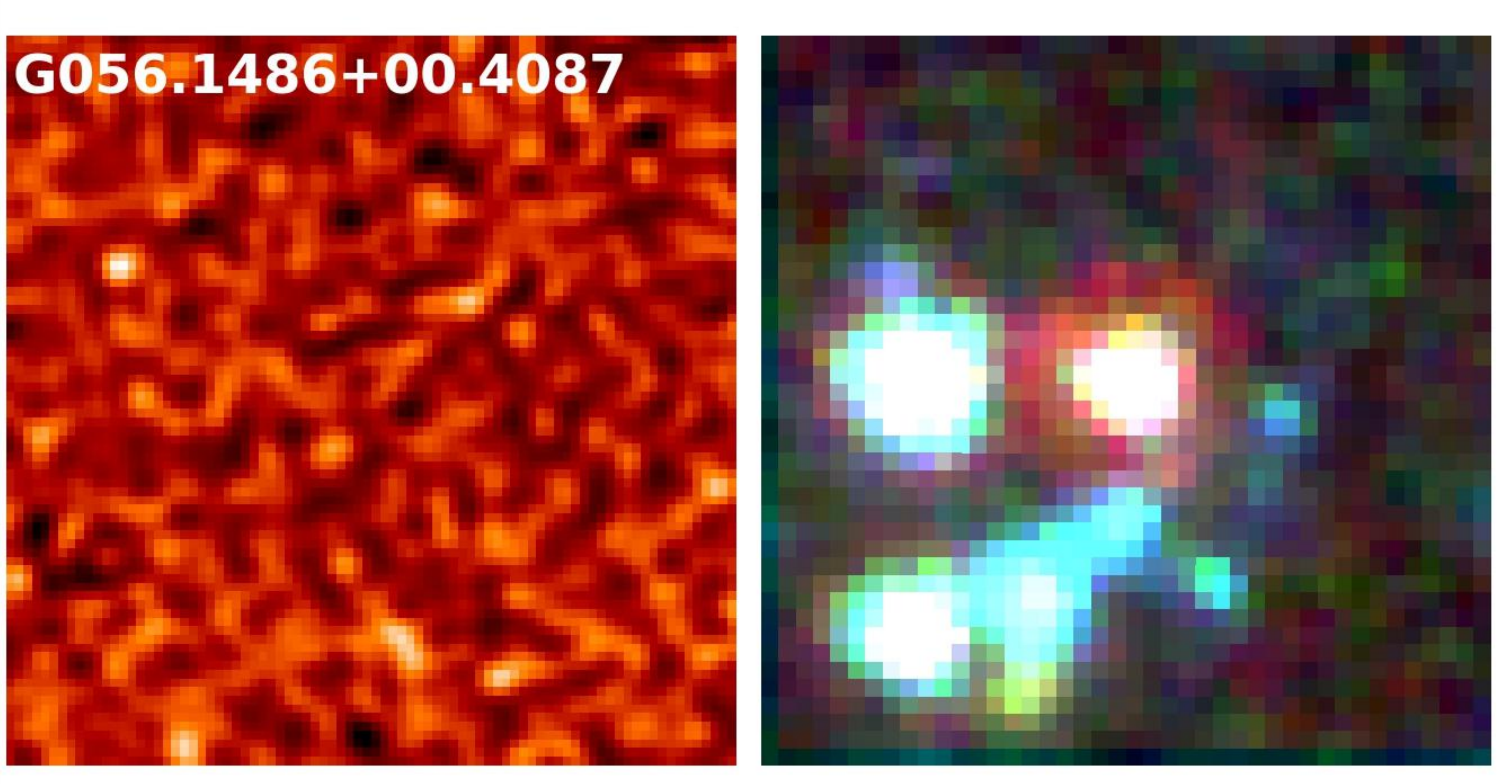}
\hspace{0.5mm}
	\includegraphics[height=4.1cm, width=8cm]{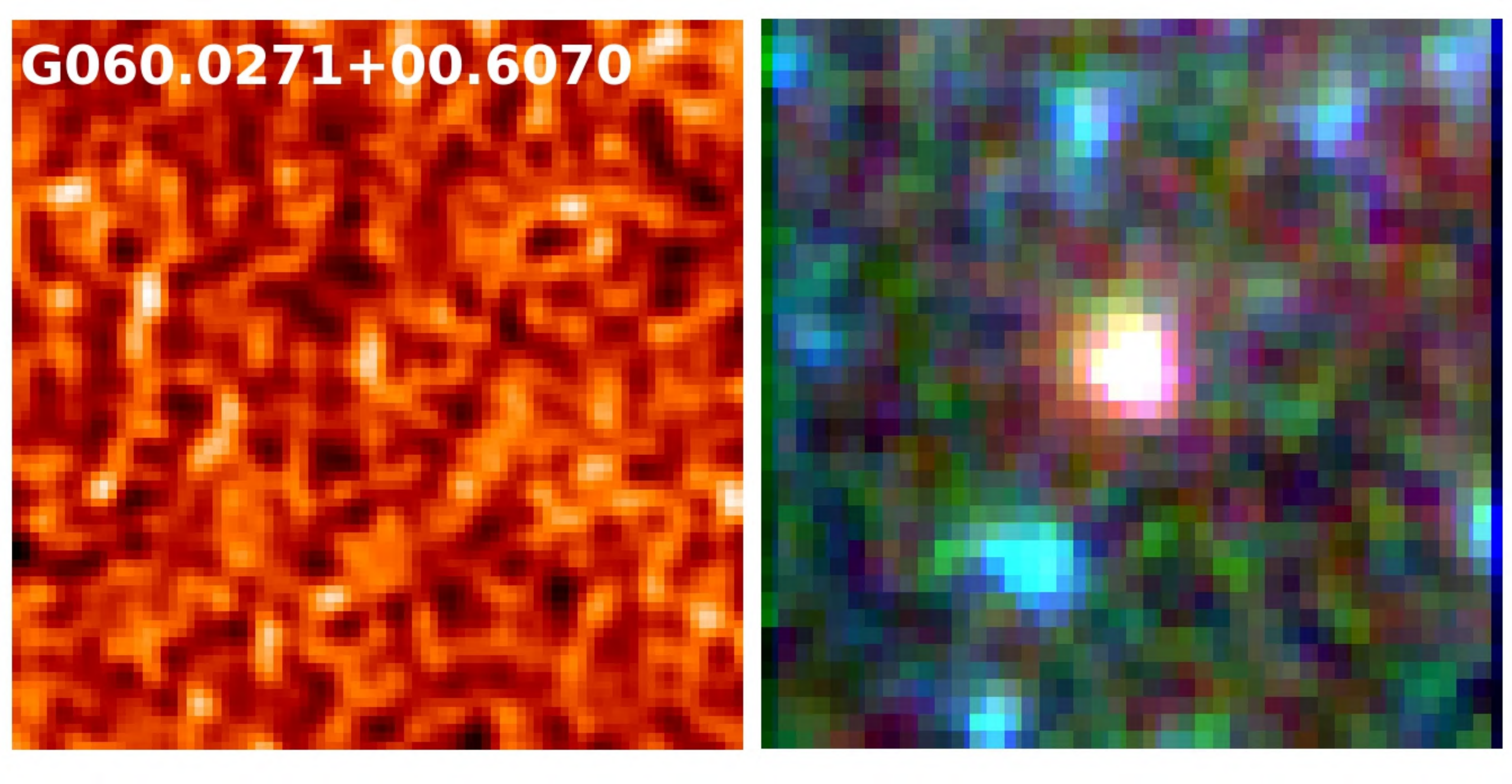}
    \caption{CORNISH 6 cm and 3-colour (3.6 $\umu m$ is blue, 5.8 $\umu m$ is green and 8.0 $\umu m$ is red) GLIMPSE images of the 10 high quality PNe candidates from \protect\cite{parker_2_2012}, within the CORNISH survey region. All images are $25^{\arcsec} \times 25^{\arcsec}$ in size and the CORNISH beam size is shown on the first image.}
    \label{G10}
\end{figure*}

Similarly, 10 out of the 27 high-quality candidate PNe identified by \cite{parker_2_2012}, within the CORNISH survey region, are not in the 7$\sigma$ catalogue. \cite{parker_2_2012} classified these as high quality PNe based on the MIR colours of previously known PNe, MIR environment and MIR false-colour images. Inspection of the CORNISH images shows that they are not reliable PNe candidates (see Figure \ref{G10}). Because these sources are compact in the GLIMPSE survey, we can rule out the possibility that a null detection by the CORNISH survey is due to completely resolved out emission.  These could possibly be evolved stellar objects or YSO \citep{oliver2013}.

\begin{table}
\caption{24 Known PNe present in the CORNISH$-$PNe sample.}\label{known_PNe} 
\begin{tabular}{p{2.4cm}p{1.3cm}p{3cm}}
\hline
\hline
 CORNISH Name& PNG&Other Names\\
\hline
G010.0989$+$00.7393&010.1$+$00.7&NGC 6537\\
G011.7434$-$00.6502& 011.7$-$00.6 &NGC 6567\\
G011.7900$-$00.1022&  011.7$+$00.0&M 1$-$43\\
G012.3315$-$00.1806& 012.3$-$00.1 &F2BP 1\\
G018.0661$+$00.8535& 018.0$+$00.8 &IRAS18179$-$1249\\
G019.5326$+$00.7308&019.5$+$00.7  &GLIPN1823$-$1133\\
G019.9448$+$00.9126& 019.9$+$00.9 &M 3$-$53\\
G020.4681$+$00.6793& 020.4$+$00.6 &PM 1$-$231\\
G021.8201$-$00.4779& 021.8$-$00.4 &M 3$-$28\\
G022.2211$+$00.9009& 022.2$+$00.9 &IRAS18257$-$0908\\
G026.8327$-$00.1516&026.8$-$00.1  &MPAJ1840$-$0529\\
G027.6635$-$00.8267& 027.6$-$00.8 &PHRJ1844$-$0503\\
G027.7016$+$00.7048& 027.7$+$00.7 &M 2$-$45\\
G029.2113$-$00.0689& 029.2$+$00.0 &TDC1\\
G032.5485$-$00.4739&032.5$-$00.4  &MPAJ1852$-$0033\\
G033.4543$-$00.6149& 033.4$-$00.6 &GLMP 844\\
G050.0405$+$01.0961& 050.0$+$01.0 &IRAS19171$+$1536\\
G051.5095$+$00.1686& 051.5$+$00.2 &KLW 1\\
G051.8341$+$00.2838& 051.8$+$00.2 &IPHASJ192553.5$+$165331\\
G055.5070$-$00.5579& 055.5$-$00.5 &M 1$-$71\\
G056.4016$-$00.9033& 056.4$-$00.9 &K 3$-$42\\
G059.3987$-$00.7880&059.4$-$00.7  &PM 1$-$313\\
G062.4936$-$00.2699&  062.4$-$00.2&M 2$-$48\\
G063.8893$+$00.1229&  063.8$+$00.1&K 3$-$48\\

\hline		  			     	    	       		 	 	
\end{tabular}	  			     	    	       				 	\end{table}

\clearpage
\newpage

\section{Conclusion}
Following the visual selection of the CORNISH-PNe (169) from the CORNISH survey by the CORNISH team, we have focused on their multi-wavelength properties, including their radio properties, optical to far-infrared colours, extinction and distances. The sensitivity of the CORNISH survey combined with multi-wavelength data, at infrared and millimetre wavelengths, provides an uncontaminated sample of PNe that is free from extinction biases associated with optical based surveys. At 6 cm, the CORNISH survey is more complete within the $|b|< 1^\circ$ region and has uncovered 90 new and compact Galactic PNe within 110 $deg^2$ area. Radio selected samples from such surveys, unaffected by extinction, will be excellent to compare with models of population synthesis. This would contribute to better understanding of the formation and evolution of Galactic PNe. In future work, we will compare this sample, although small, with model predictions from population synthesis.  In summary:
\vspace{-3cm}
\begin{itemize}
\item[--] We have used multi-wavelength properties to certify the CORNISH-PNe classification and confirm a clean sample, although one will need an optical or NIR spectra to fully confirm the status of a PN. 
\item[--] Radio properties show a CORNISH-PNe sample that is compact, with some PNe possessing $T_b$ > 1000 K. We were able to estimate spectral indices for 127 of the CORNISH-PNe (75$\%$). 98$\%$ of these have spectral indices that are consistent with thermal free-free emission. The 2$\%$ exhibiting non-thermal emission at a 3$\sigma$ significance level are likely very young PNe.
\item[--] 22 out of the 76 CORNISH-PNe within the IPHAS survey region were found to have H$\alpha$ emitting counterparts. The PNe not detected by IPHAS are likely due to line-of-sight extinction.
\item[--] The broad distribution of the CORNISH-PNe in the MIR shows a sample dominated by different emission mechanisms across the different IRAC bands. This broad distribution is also observed in NIR colours.
\item[--] Some of the CORNISH-PNe, as expected, show higher reddening compared to previous PNe samples studied in the NIR. 
\item[--] In the FIR, the CORNISH-PNe show a wider range of colours compared to the optically detected PNe studied by \cite{anderson2012}.
\item[--] We estimated extinction using the methods in Section \ref{ext_sec}. Extinction variations towards the CORNISH-PNe using the different methods reflect a sample dominated by different emission mechanisms across the near-infrared bands, hence, a possible range of intrinsic colours.
\item[--] Heliocentric distances were estimated using the \cite{frew2016} calibration. The average distance for the CORNISH-PNe is 15 kpc with a corresponding average physical diameter of 0.17 pc.
\item[--] A catalogue of 169 CORNISH-PNe is presented. We find 90 new PNe, out of which 12 are newly discovered and 78 are newly classified as PN. A further 47 suspected PNe are confirmed as such from the analysis here and 24 known PNe were detected. Eight sources are classified as possible PNe or other source types.
\item[--] The clean sample of PNe, together with the extensive photometry and analysis presented here should form good training sets in using machine learning to build classification/identification models for PNe in future large radio surveys.
\end{itemize}

\section*{Acknowledgements}
The authors thank the referee for the excellent suggestions and comments that have improved the quality of this paper. TI acknowledges OSAPND, for funding through the award of a scholarship. This research made use of the NASA/ IPAC Infrared Science Archive, which is operated by the Jet Propulsion Laboratory, California Institute of Technology, under contract with the National Aeronautics and Space Administration; data products from the SuperCOSMOS H-alpha Survey (SHS); the HASH PN database at hashpn.space. This research also made use of the SIMBAD database and the VizieR services, operated at CDS, Strasbourg, France and, SAOImage dS9, developed by the Smithsonian Astrophysical Observatory.
\clearpage
\newpage
\onecolumn

\clearpage

\clearpage
\begin{landscape}
\begin{longtable}{lllllp{6cm}}

\caption{ Final catalogue of the CORNISH$-$PNe with classifications.}\label{cataloguetable3} \\	
\hline
\hline
  \multicolumn{1}{l}{Name} &
  \multicolumn{1}{l}{SIMBAD} &
  \multicolumn{1}{l}{$H\alpha$} &
  \multicolumn{1}{l}{class} &
  \multicolumn{1}{l}{Status}&
  \multicolumn{1}{p{6cm}}{Comments}  \\
 &ID&detection&&& \\
\hline
\endfirsthead

\multicolumn{6}{c}%
{{\bfseries \tablename\ \thetable{} -- continued from previous page}} \\
\hline
\hline
  \multicolumn{1}{l}{Name} &
  \multicolumn{1}{l}{SIMBAD} & 
  \multicolumn{1}{l}{$H\alpha$} &
  \multicolumn{1}{l}{class} &
  \multicolumn{1}{l}{Status}&
  \multicolumn{1}{p{6cm}}{Comments}  \\
 &ID&detection&&& \\
\hline 
\endhead

\hline \multicolumn{6}{r}{{Continued on next page}} \\ \hline
\endfoot

\hline 
\endlastfoot

G009.9702$-$00.5292& pAG$^1$ (IRAS 18066$-$2034)&$-$&PN&Newly classified&\\
G010.4168$+$00.9356&pAG$^2$ (IRAS 18021$-$1928)& SHS&PN	&Newly classified&\\
G010.5960$-$00.8733& star (IRAS 18092$-$2012) &$-$&PN& Newly classified &\\
G011.3266$-$00.371&YSO?$^3$&$-$&PN& Newly classified&\\
G011.4290$-$01.0091&$-$&$-$&PN& Newly discovered\\
G011.4581$+$01.0736&pAG$^1$ (IRAS 18038$-$1830)&$-$&PN& Newly classified &\\
G011.7210$-$00.4916&-&$-$&PN?&Newly discovered&FIR colours are upper limits\\
G012.0438$-$00.5077&$-$&$-$&PN?&Newly discovered&FIR colours are upper limits\\
G012.1157$+$00.0757&Radio source$^5$.&SHS&PN& Newly classified &\\
G012.1528$-$00.3304&YSO?(J18132367$-$1833495)$^3$&$-$&PN& Newly classified &\\

G012.3830$+$00.7990&PN?$^7$ (IRAS 18067$-$1749)&$-$&PN& Newly confirmed\\
G012.6012$+$00.5592&YSO?$^3$ (IRAS 18081$-$1745)&$-$&PN& Newly classified &\\
G013.3565$-$00.7559&YSO?$^3$&$-$&PN& Newly classified &\\
G013.6313$-$00.6023&Radio source$^5$ &$-$&PN& Newly classified &\\
G013.9166$+$00.6500&PN? $^8$&$-$&PN& Newly confirmed &\\
G014.2365$+$00.2117&PN?$^9$&SHS&PN& Newly confirmed&\\
G014.4573$-$00.1847&Radio source$^{10}$&$-$&PN&Newly classified&\\

G014.5851$+$00.4613&PN?$^{26}$&IPHAS&PN & Newly confirmed\\&\\

G014.7503$-$00.2496&Radio source$^{10}$&-&PN& Newly classified &\\

G014.8960$+$00.4837&YSO?$^3$&SHS&PN& Newly classified &\\
G015.1999$-$00.0863&Radio source$^{10}$&$-$&PN& Newly classified &\\
G015.5410$+$00.3359&Radio source$^{10}$&$-$&PN& Newly classified &\\
G015.5847$+$00.4002&Radio source$^{10}$&$-$&PN& Newly classified &\\
G015.7993$-$00.0063&PN$^9$&$-$&PN&Newly confirmed&\\
G016.0550$+$00.8280&PN$^9$&SHS&PN&Newly confirmed&\\
G016.4034$-$00.5740&Radio source$^5$&$-$&PN& Newly classified &\\
G016.4276$+$01.0072&$-$&$-$&PN& Newly discovered\\
G016.4999$+$00.1152&Radio source$^5$&$-$&PN&Newly classified& NIR colours have stellar contamination\\
G016.6002$-$00.2754&Radio source$^{10}$&$-$&PN& Newly classified &\\
G017.0152$-$00.1906&Radio source$^{10}$&$-$&PN&Newly classified\\
G017.3669$+$00.5224&PN?$^{11}$&$-$&PN& Newly confirmed &\\

G017.4147$+$00.3791&Radio source$^5$&$-$&PN& Newly classified &\\
G017.4487$+$00.1146&Radio source$^{10}$&$-$&PN& Newly classified &\\
G017.7250$-$00.2427&Radio source$^{10}$&SHS&PN& Newly classified &\\
G017.8222$+$00.9866&$-$&SHS&PN& Newly discovered\\
G017.8645$+$00.2120&YSO?$^3$ (IRAS 18198$-$1318)&$-$&PN& Newly classified &\\

G018.1286$-$00.2189&$-$&$-$&PN?&Newly discovered&FIR are upper limits\\

G018.2402$-$00.9152&PN $^9$&IPHAS&PN&Newly confirmed&\\
G018.2413$-$00.5552&YSO?$^3$&$-$&PN&Newly classified&Has no FIR colours\\

G018.5242$+$00.1519&YSO?$^6$&$-$&PN& Newly classified &\\
G018.5776$-$00.7484&$-$&$-$&PN&Newly confirmed&\\
G019.2356$+$00.4951&Radio source$^5$ &$-$&PN& Newly classified &\\
G019.4676$-$00.0154&Radio source$^{10}$&$-$&PN& Newly classified &\\

G019.9298$-$00.6639&IR (IRAS 18270$-$1153)&$-$&PN& Newly classified &\\
G020.5176$+$00.4778&Radio source$^{5}$&SHS&PN& Newly classified &\\

G020.6015$+$00.0206&-&-&PN&Newly discovered&Very resolved in the CORNISH survey compared to the  MAGPIS but a clear nebula in the 24 $\mu m$. Heavy stellar contamination in the NIR and MIR \\
G020.9782$+$00.9253&PN?&SHS&PN& Newly confirmed\\
G021.1653$+$00.4755&PN$^9$ (IRAS 18252$-$1016) &SHS&PN&Newly confirmed&\\
G021.3425$-$00.8423&PN$^9$ (IRAS 18303$-$1043)&$-$&PN&Newly confirmed&\\

G021.6657$+$00.8110&PN$^{25}$&IPHAS&PN&Newly confirmed&\\

G021.6849$-$00.7381&Radio source $^5$&SHS&PN& Newly confirmed\\
G021.9972$-$00.8838&-&$-$&PN& Newly discovered &\\

G022.5477$-$00.1061&Radio source$^{10}$&$-$&PN& Newly classified &\\
G022.6429$-$00.4422&YSO?$^3$&$-$&PN& Newly classified &\\
G022.6580$+$00.2959&Radio source$^{10}$ &$-$&PN& Newly classified &\\
G023.2321$+$00.0809&Radio source$^{10}$&$-$&PN& Newly classified &\\
G023.4181$-$00.3940&PN$^9$ (IRAS 18326$-$0840) &$-$&PN&Newly confirmed&\\
G023.5044$-$00.5245&Star (IRAS 18332$-$0839)$^5$&$-$&PN& Newly classified &\\
G023.8214$-$00.5788&PN$^9$&$-$&PN&Newly confirmed&\\
G023.8897$-$00.7379&PN$^{13}$ (IRAS 18347$-$0825)&$-$&PN&Newly confirmed&\\
G024.0943$-$01.0992&$-$&$-$&Radio star?&Newly discovered&Upper limits on FIR colours\\

G024.1659$+$00.2502&H II region$^{14}$&$-$&PN& Newly classified &\\
G024.3852$+$00.2869&PN$^9$&$-$&PN&Newly confirmed&\\
G024.7921$-$01.0043&IR(IRAS 18373$-$0744)&SHS&PN& Newly classified &\\
G024.8959$+$00.4586&YSO?$^3$&$-$&PN& Newly classified &\\
G025.0485$-$00.6621&PN$^8$&$-$&PN& Newly confirmed &\\
G025.5190$+$00.2165&H II region $^{14}$&$-$&PN& Newly classified &\\
G025.5769$+$00.1389&Radio source$^{10}$&$-$&PN& Newly classified &\\
G025.8466$+$01.1718&PN?$^{13}$&$-$&PN& Newly confirmed&No MIR data but PN colours in NIR and FIR\\
G026.0823$-$00.0347&Radio source$^{10}$&$-$&PN?&Newly classified&Near dark filaments in the MIR but PN colours in the MIR and FIR \\
G026.1620$+$00.5926&$-$&$-$&PN& Newly discovered\\
G026.2268$+$00.7685&$-$&$-$&PN& Newly discovered\\
G026.6529$+$00.2874&Radio source$^{10}$&$-$&PN& Newly classified &\\
G026.7145$+$00.1319&Radio source$^{10}$&$-$&PN& Newly classified &\\
G027.4265$-$00.2499&YSO?$^3$&SHS&PN& Newly classified &\\
G027.6595$-$00.3835&PN$^9$ (IRAS 18404$-$0455)&SHS&PN&Newly confirmed&\\

G027.6640$-$00.2485&$-$&$-$&PN?Radio star?&Newly classified&\\

G029.0538$+$00.9915&YSO?$^3$&SHS&PN& Newly classified &\\
G029.1652$-$00.0168&Radio source$^{10}$&$-$&PN& Newly classified &\\

G029.5780$-$00.2686&PN$^{24}$&IPHAS&PN&Newly confirmed&    \\

G029.8742$-$00.8190&PN$^9$ (IRAS 18461$-$0309)&$-$&PN&Newly confirmed&\\
G030.0226$+$00.1570&Radio source$^{10}$&$-$&PN&Newly classified&\\
G030.0294$-$00.3318&Radio source$^{10}$&$-$&PN& Newly classified &\\
G030.2335$-$00.1385&PN?$^{16}$ (IRAS 18443$-$0231)&$-$&PN& Newly confirmed\\
G030.5302$+$00.1315&Radio source$^{10}$&$-$&PN?&Newly classified&\\
G030.6670$-$00.3319&PN$^9$  (IRAS 18458$-$0213)&$-$&PN&Newly confirmed&\\
G030.8560$+$00.3826&Radio source$^{10}$&$-$&PN& Newly classified &\\

G031.2131$-$00.1803&OH $\&$ H$_2$O maser$^{17}$&$-$&PN& Newly classified &\\

G031.3724$-$00.7514&PN? (J18511825$-$0143487)&$-$&PN& Newly confirmed\\
G032.3076$+$00.1536&PN$^{16}$ (IRAS 18471$-$0032)&$-$&PN&Newly confirmed&\\
G032.6136$+$00.7971&PN?$^{9}$&IPHAS&PN& Newly confirmed &\\
G032.8082$-$00.3159&Radio source$^{10}$&$-$&PN& Newly classified &\\
G032.8177$-$00.1165&$-$&$-$&PN& Newly discovered &\\
G033.1198$-$00.8949&$-$&$-$&PN& Newly discovered\\
G033.3526$+$00.4043&YSO?$^{3}$(IRAS 18482$+$0029)&$-$&PN& Newly classified &\\

G033.7952$+$00.4307&YSO?$^3$&$-$&PN& Newly classified &\\
G033.9059$-$00.0436&PN$^9$&$-$&PN&Newly confirmed&\\

G034.1792$-$00.1777&Radio source$^{10}$&$-$&PN& Newly classified &\\
G034.4200$-$00.3183&PN$^9$&$-$&PN&Newly confirmed&\\
G034.8624$-$00.0630&PN$^9$&$-$&PN&Newly confirmed&\\

G035.2162$+$00.4280&Radio source$^{10}$&$-$&PN& Newly classified &\\
G035.4719$-$00.4365&Radio source$^{10}$&IPHAS&PN& Newly classified &\\

G035.5654$-$00.4922&PN?$^{27}$&IPHAS&PN&Newly confirmed&\\

G036.0116$-$00.2562&Radio source$^{10}$&$-$&PN& Newly classified &\\
G036.5393$+$00.2003&Radio source$^{10}$&$-$&PN& Newly classified &\\
G037.9031$-$00.2754&YSO?$^3$&$-$&PN& Newly classified &\\
G037.9601$+$00.4534&PN$^9$&$-$&PN&Newly confirmed&\\
G038.9237$-$00.0807&Radio source$^{10}$&$-$&PN& Newly classified &\\
G039.1617$+$00.7826&PN?$^8$&$-$&PN& Newly confirmed &\\
G039.5911$-$00.3785&$-$&$-$&PN& Newly discovered\\
G040.2606$-$00.2755&YSO$^9$ (IRAS 19034$+$0618)&$-$&PN& Newly classified &\\
G040.3359$-$01.0102&$-$&$-$&PN& Newly discovered\\
G041.1982$+$00.0348&YSO?$^3$ (IRAS 19040$+$0717)&$-$&PN& Newly classified &\\
G041.3540$+$00.5390&Radio source$^5$&IPHAS&PN& Newly classified &\\
G041.7871$+$00.4884&YSO?$^3$&$-$&PN& Newly classified &\\
G042.6629$-$00.8648&PN?$^{20,8}$ (PM 1$-$288)&$-$&PN& Newly confirmed\\
G043.0281$+$00.1399&PN$^9$ (IRAS 19071$+$0857)&$-$&PN&Newly confirmed&\\
G043.2946$-$00.6455&PN$^9$&$-$&PN&Newly confirmed&\\
G043.5793$+$00.0261&PN$^{16}$&$-$&PN&Newly confirmed&\\
G043.6554$-$00.8279&IR (IRAS 19117$+$0903)&IPHAS&PN& Newly classified &\\
G044.6375$+$00.4827&YSO?$^3$ (IRAS 19089$+$1032)&IPHAS&PN& Newly classified &\\
G045.1801$+$00.9893&Radio source$^{21}$&$-$&PN& Newly classified &\\
G045.2830$-$00.6278&PN$^{16}$ (J19163050$+$1040563)&$-$&PN&Newly confirmed&\\
G046.9747$+$00.2702&YSO?$^3$ (IRAS 19141$+$1230)&$-$&PN& Newly classified &\\

G047.6884$-$00.3024&PN$^{9,1}$&$-$&PN& Newly confirmed &\\

G047.9875$+$00.2026&star (IRAS 19163$+$1322)&$-$&PN& Newly classified &\\
G048.5619$+$00.9029&YSO?$^3$&IPHAS&PN& Newly classified &\\
G048.6748$-$00.5350&YSO?$^3$&$-$&PN& Newly classified &\\

G048.7319$+$00.9305&PN$^{24}$&IPHAS&PN&Newly confirmed&    \\

G049.6948$+$00.8642&PN?$^{22}$ (J19193387$+$1516421)&$-$&PN& Newly confirmed&\\
G050.0003$+$00.5072&YSO? $^3$&$-$&PN& Newly classified &\\

G050.0457$+$00.7683&PN?$^9$&$-$&H II region?& Newly classified &\\
G050.4802$+$00.7056&PN?$^{24}$ &IPHAS&PN&Newly confirmed &\\
G050.5556$+$00.0448&PN$^9$&IPHAS&PN&Newly confirmed&\\
G050.8950$+$00.0572&YSO?$^3$ (J19245196$+$1557287)&$-$&PN& Newly classified &\\
G051.6061$+$00.9140&YSO?$^3$ (IRAS 19209$+$1653)&$-$&PN& Newly classified &\\
G052.1498$-$00.3758&Radio source$^5$&$-$&PN& Newly classified &\\
G057.5352$+$00.2266&Radio source$^{21}$&$-$&PN& Newly classified &\\
G058.1591$-$00.5499&Radio source$^{21}$&$-$&PN& Newly classified &\\
G058.6410$+$00.9196&PN$^{23}$ $-$ IRAS 19353$+$2302&IPHAS&PN& Newly confirmed\\
G059.8236$-$00.5361&PN?$^{16}$ (IRAS 19434$+$2320)&IPHAS&PN& Newly confirmed\\
G060.8480$-$00.8954&YSO?$^3$&$-$&PN& Newly classified &\\
G060.9866$-$00.5698&Radio source$^5$ &IPHAS&PN& Newly classified&\\
G062.7551$-$00.7262&PN$^{24}$&IPHAS&PN&Newly confirmed&\\
G063.0455$+$00.5977&$-$&$-$&PN& Newly discovered&\\
\hline		  			     	    	       		 	 		  			     	    	       				 	 	    	    	       		 		
\end{longtable}
\begin{minipage}{\textwidth}
   	\small
    pAG: post$-$AGB star, SgrS: Super$-$giant star, IR source: Infrared source. 
    The questionmark (?) in front of some of the source classification 
    indicates 'possible'. e.g pAG? is a possible post-AGB star

	$^1$\citep{koh2001}, $^2$ \citep{ratag1991},$^3$ \citep{robi2008}, $^4$ \citep{rob22010},$^5$ \citep{zoo1990}, $^6$ \citep{fell2002},$^7$ \citep{cornish2012}, $^8$ \citep{condon1999}, $^9$ \citep{Urquhart2009},$^{10}$ \citep{becker1994}, $^{11}$ \citep{andber2011},$^{12}$ \citep{parker_2_2012},$^{13}$ \citep{kist1995},$^{14}$ \citep{andy2015}, $^{15}$ \citep{Sewilo2004},$^{16}$ \citep{Kanarek2015}, $^{17}$ \citep{Caswell2001}, $^{18}$ \citep{Szczerba2007}, $^{19}$ \citep{su2006}, $^{20}$ \citep{pm1988}, $^{21}$ \citep{Taylor1996}, $^{22}$  \citep{phillips2008}	, $^{23}$  \citep{vande1995},$^{24}$  \citep{sabin2014}, $^{25}$  \citep{van1996},$^{26}$  \citep{mis2008},$^{27}$  \citep{parker2006}
\end{minipage}
	
\end{landscape}

\twocolumn
\clearpage
\newpage

\clearpage

\bibliography{me3}

\newcommand{\noop}[1]{}
\begin{thebibliography}{}
\makeatletter
\relax
\def\mn@urlcharsother{\let\do\@makeother \do\$\do\&\do\#\do\^\do\_\do\%\do\~}
\def\mn@doi{\begingroup\mn@urlcharsother \@ifnextchar [ {\mn@doi@}
  {\mn@doi@[]}}
\def\mn@doi@[#1]#2{\def\@tempa{#1}\ifx\@tempa\@empty \href
  {http://dx.doi.org/#2} {doi:#2}\else \href {http://dx.doi.org/#2} {#1}\fi
  \endgroup}
\def\mn@eprint#1#2{\mn@eprint@#1:#2::\@nil}
\def\mn@eprint@arXiv#1{\href {http://arxiv.org/abs/#1} {{\tt arXiv:#1}}}
\def\mn@eprint@dblp#1{\href {http://dblp.uni-trier.de/rec/bibtex/#1.xml}
  {dblp:#1}}
\def\mn@eprint@#1:#2:#3:#4\@nil{\def\@tempa {#1}\def\@tempb {#2}\def\@tempc
  {#3}\ifx \@tempc \@empty \let \@tempc \@tempb \let \@tempb \@tempa \fi \ifx
  \@tempb \@empty \def\@tempb {arXiv}\fi \@ifundefined
  {mn@eprint@\@tempb}{\@tempb:\@tempc}{\expandafter \expandafter \csname
  mn@eprint@\@tempb\endcsname \expandafter{\@tempc}}}

\bibitem[\protect\citeauthoryear{{Acker}, {Marcout}, {Ochsenbein}, {Stenholm},
  {Tylenda}  \& {Schohn}}{{Acker} et~al.}{1992}]{acker1992}
{Acker} A.,  {Marcout} J.,  {Ochsenbein} F.,  {Stenholm} B.,  {Tylenda} R.,
  {Schohn} C.,  1992, {The Strasbourg-ESO Catalogue of Galactic Planetary
  Nebulae. Parts I, II.}

\bibitem[\protect\citeauthoryear{{Anderson}, {Bania}, {Balser}  \&
  {Rood}}{{Anderson} et~al.}{2011}]{andber2011}
{Anderson} L.~D.,  {Bania} T.~M.,  {Balser} D.~S.,   {Rood} R.~T.,  2011,
  \mn@doi [\apjs] {10.1088/0067-0049/194/2/32}, \href
  {http://adsabs.harvard.edu/abs/2011ApJS..194...32A} {194, 32}

\bibitem[\protect\citeauthoryear{{Anderson} et~al.}{{Anderson}
  et~al.}{2012}]{anderson2012}
{Anderson} L.~D.,  et~al., 2012, \mn@doi [\aap] {10.1051/0004-6361/201117640},
  \href {http://adsabs.harvard.edu/abs/2012A%26A...537A...1A} {537, A1}

\bibitem[\protect\citeauthoryear{{Anderson}, {Armentrout}, {Johnstone},
  {Bania}, {Balser}, {Wenger}  \& {Cunningham}}{{Anderson}
  et~al.}{2015}]{andy2015}
{Anderson} L.~D.,  {Armentrout} W.~P.,  {Johnstone} B.~M.,  {Bania} T.~M.,
  {Balser} D.~S.,  {Wenger} T.~V.,   {Cunningham} V.,  2015, \mn@doi [\apjs]
  {10.1088/0067-0049/221/2/26}, \href
  {http://adsabs.harvard.edu/abs/2015ApJS..221...26A} {221, 26}

\bibitem[\protect\citeauthoryear{{Anderson} et~al.,}{{Anderson}
  et~al.}{2017}]{and2017}
{Anderson} L.~D.,  et~al., 2017, \mn@doi [\aap] {10.1051/0004-6361/201731019},
  \href {http://adsabs.harvard.edu/abs/2017A%26A...605A..58A} {605, A58}

\bibitem[\protect\citeauthoryear{{Balick}}{{Balick}}{1987}]{bal1987}
{Balick} B.,  1987, \mn@doi [\aj] {10.1086/114504}, \href
  {http://adsabs.harvard.edu/abs/1987AJ.....94..671B} {94, 671}

\bibitem[\protect\citeauthoryear{{Barentsen} et~al.}{{Barentsen}
  et~al.}{2014}]{barensten2014}
{Barentsen} G.,  et~al., 2014, \mn@doi [\mnras] {10.1093/mnras/stu1651}, \href
  {http://adsabs.harvard.edu/abs/2014MNRAS.444.3230B} {444, 3230}

\bibitem[\protect\citeauthoryear{{Bear} \& {Soker}}{{Bear} \&
  {Soker}}{2016}]{soker2016}
{Bear} E.,  {Soker} N.,  2016, preprint, \href
  {http://adsabs.harvard.edu/abs/2016arXiv160608149B} {} (\mn@eprint {arXiv}
  {1606.08149})

\bibitem[\protect\citeauthoryear{{Becker}, {White}, {Helfand}  \&
  {Zoonematkermani}}{{Becker} et~al.}{1994}]{becker1994}
{Becker} R.~H.,  {White} R.~L.,  {Helfand} D.~J.,   {Zoonematkermani} S.,
  1994, \mn@doi [\apjs] {10.1086/191941}, \href
  {http://adsabs.harvard.edu/abs/1994ApJS...91..347B} {91, 347}

\bibitem[\protect\citeauthoryear{{Benjamin} et~al.}{{Benjamin}
  et~al.}{2003}]{benjamin2003}
{Benjamin} R.~A.,  et~al., 2003, \mn@doi [\pasp] {10.1086/376696}, \href
  {http://adsabs.harvard.edu/abs/2003PASP..115..953B} {115, 953}

\bibitem[\protect\citeauthoryear{{Bietenholz}, {Kassim}, {Frail}, {Perley},
  {Erickson}  \& {Hajian}}{{Bietenholz} et~al.}{1997}]{bie1997}
{Bietenholz} M.~F.,  {Kassim} N.,  {Frail} D.~A.,  {Perley} R.~A.,  {Erickson}
  W.~C.,   {Hajian} A.~R.,  1997, \mn@doi [\apj] {10.1086/304853}, \href
  {http://adsabs.harvard.edu/abs/1997ApJ...490..291B} {490, 291}

\bibitem[\protect\citeauthoryear{{Boji{\v c}i{\'c}}, {Parker}, {Filipovi{\'c}}
  \& {Frew}}{{Boji{\v c}i{\'c}} et~al.}{2011}]{boj2011}
{Boji{\v c}i{\'c}} I.~S.,  {Parker} Q.~A.,  {Filipovi{\'c}} M.~D.,   {Frew}
  D.~J.,  2011, \mn@doi [\mnras] {10.1111/j.1365-2966.2010.17900.x}, \href
  {http://adsabs.harvard.edu/abs/2011MNRAS.412..223B} {412, 223}

\bibitem[\protect\citeauthoryear{{Bozzetto} et~al.,}{{Bozzetto}
  et~al.}{2017}]{bozz2017}
{Bozzetto} L.~M.,  et~al., 2017, \mn@doi [\apjs] {10.3847/1538-4365/aa653c},
  \href {http://adsabs.harvard.edu/abs/2017ApJS..230....2B} {230, 2}

\bibitem[\protect\citeauthoryear{{Cardelli}, {Clayton}  \& {Mathis}}{{Cardelli}
  et~al.}{1989}]{card1989}
{Cardelli} J.~A.,  {Clayton} G.~C.,   {Mathis} J.~S.,  1989, \mn@doi [\apj]
  {10.1086/167900}, \href {http://adsabs.harvard.edu/abs/1989ApJ...345..245C}
  {345, 245}

\bibitem[\protect\citeauthoryear{{Carey} et~al.}{{Carey}
  et~al.}{2009}]{carey2009}
{Carey} S.~J.,  et~al., 2009, \mn@doi [\pasp] {10.1086/596581}, \href
  {http://adsabs.harvard.edu/abs/2009PASP..121...76C} {121, 76}

\bibitem[\protect\citeauthoryear{{Caswell}}{{Caswell}}{2001}]{Caswell2001}
{Caswell} J.~L.,  2001, \mn@doi [\mnras] {10.1046/j.1365-8711.2001.04745.x},
  \href {http://adsabs.harvard.edu/abs/2001MNRAS.326..805C} {326, 805}

\bibitem[\protect\citeauthoryear{{Cerrigone} et~al.}{{Cerrigone}
  et~al.}{2009}]{cerr2009}
{Cerrigone} L.,  et~al., 2009, \mn@doi [\apj] {10.1088/0004-637X/703/1/585},
  \href {http://adsabs.harvard.edu/abs/2009ApJ...703..585C} {703, 585}

\bibitem[\protect\citeauthoryear{{Cerrigone} et~al.}{{Cerrigone}
  et~al.}{2011}]{cerr2011}
{Cerrigone} L.,  et~al., 2011, \mn@doi [\mnras]
  {10.1111/j.1365-2966.2010.17968.x}, \href
  {http://adsabs.harvard.edu/abs/2011MNRAS.412.1137C} {412, 1137}

\bibitem[\protect\citeauthoryear{{Cerrigone}, {Umana}, {Trigilio}, {Leto},
  {Buemi}  \& {Ingallinera}}{{Cerrigone} et~al.}{2017}]{cer2017}
{Cerrigone} L.,  {Umana} G.,  {Trigilio} C.,  {Leto} P.,  {Buemi} C.~S.,
  {Ingallinera} A.,  2017, \mn@doi [\mnras] {10.1093/mnras/stx690}, \href
  {http://adsabs.harvard.edu/abs/2017MNRAS.468.3450C} {468, 3450}

\bibitem[\protect\citeauthoryear{{Chu} et~al.}{{Chu} et~al.}{2000}]{chu2000}
{Chu} Y.-H.,  et~al., 2000, in {Kastner} J.~H.,  {Soker} N.,   {Rappaport} S.,
  eds,  Astronomical Society of the Pacific Conference Series Vol. 199,
  Asymmetrical Planetary Nebulae II: From Origins to Microstructures. p.~419
  (\mn@eprint {} {astro-ph/9909106})

\bibitem[\protect\citeauthoryear{{Churchwell} et~al.}{{Churchwell}
  et~al.}{2009}]{churchwell2009}
{Churchwell} E.,  et~al., 2009, \mn@doi [\pasp] {10.1086/597811}, \href
  {http://adsabs.harvard.edu/abs/2009PASP..121..213C} {121, 213}

\bibitem[\protect\citeauthoryear{{Claussen}, {Goss}, {Frail}  \&
  {Seta}}{{Claussen} et~al.}{1999}]{cla1999}
{Claussen} M.~J.,  {Goss} W.~M.,  {Frail} D.~A.,   {Seta} M.,  1999, \mn@doi
  [\aj] {10.1086/300779}, \href
  {http://adsabs.harvard.edu/abs/1999AJ....117.1387C} {117, 1387}

\bibitem[\protect\citeauthoryear{{Cohen} et~al.}{{Cohen}
  et~al.}{2007a}]{cohen2007a}
{Cohen} M.,  et~al., 2007a, \mn@doi [\apj] {10.1086/521427}, \href
  {http://adsabs.harvard.edu/abs/2007ApJ...669..343C} {669, 343}

\bibitem[\protect\citeauthoryear{{Cohen} et~al.}{{Cohen}
  et~al.}{2007b}]{cohen2007}
{Cohen} M.,  et~al., 2007b, \mn@doi [\apj] {10.1086/521427}, \href
  {http://adsabs.harvard.edu/abs/2007ApJ...669..343C} {669, 343}

\bibitem[\protect\citeauthoryear{{Cohen} et~al.}{{Cohen}
  et~al.}{2011}]{cohenparker}
{Cohen} M.,  et~al., 2011, \mn@doi [\mnras] {10.1111/j.1365-2966.2010.18157.x},
  \href {http://adsabs.harvard.edu/abs/2011MNRAS.413..514C} {413, 514}

\bibitem[\protect\citeauthoryear{{Condon} \& {Kaplan}}{{Condon} \&
  {Kaplan}}{1998}]{condon1998_2}
{Condon} J.~J.,  {Kaplan} D.~L.,  1998, \mn@doi [\apjs] {10.1086/313128}, \href
  {http://adsabs.harvard.edu/abs/1998ApJS..117..361C} {117, 361}

\bibitem[\protect\citeauthoryear{{Condon}, {Cotton}, {Greisen}, {Yin},
  {Perley}, {Taylor}  \& {Broderick}}{{Condon} et~al.}{1998}]{condon1998}
{Condon} J.~J.,  {Cotton} W.~D.,  {Greisen} E.~W.,  {Yin} Q.~F.,  {Perley}
  R.~A.,  {Taylor} G.~B.,   {Broderick} J.~J.,  1998, \mn@doi [\aj]
  {10.1086/300337}, \href {http://adsabs.harvard.edu/abs/1998AJ....115.1693C}
  {115, 1693}

\bibitem[\protect\citeauthoryear{{Condon}, {Kaplan}  \& {Terzian}}{{Condon}
  et~al.}{1999}]{condon1999}
{Condon} J.~J.,  {Kaplan} D.~L.,   {Terzian} Y.,  1999, \mn@doi [\apjs]
  {10.1086/313236}, \href {http://adsabs.harvard.edu/abs/1999ApJS..123..219C}
  {123, 219}

\bibitem[\protect\citeauthoryear{{Cooper} et~al.}{{Cooper}
  et~al.}{2013}]{cop2013}
{Cooper} H.~D.~B.,  et~al., 2013, \mn@doi [\mnras] {10.1093/mnras/sts681},
  \href {http://adsabs.harvard.edu/abs/2013MNRAS.430.1125C} {430, 1125}

\bibitem[\protect\citeauthoryear{{Corradi} et~al.}{{Corradi}
  et~al.}{2008}]{corradi2008}
{Corradi} R.~L.~M.,  et~al., 2008, \mn@doi [\aap] {10.1051/0004-6361:20078989},
  \href {http://adsabs.harvard.edu/abs/2008A%26A...480..409C} {480, 409}

\bibitem[\protect\citeauthoryear{{Corradi} et~al.}{{Corradi}
  et~al.}{2010}]{corradi2010}
{Corradi} R.~L.~M.,  et~al., 2010, \mn@doi [\aap]
  {10.1051/0004-6361/200913231}, \href
  {http://adsabs.harvard.edu/abs/2010A%26A...509A..41C} {509, A41}

\bibitem[\protect\citeauthoryear{{Cox}, {Garc{\'{\i}}a-Hern{\'a}ndez},
  {Garc{\'{\i}}a-Lario}  \& {Manchado}}{{Cox} et~al.}{2011}]{cox2011}
{Cox} N.~L.~J.,  {Garc{\'{\i}}a-Hern{\'a}ndez} D.~A.,  {Garc{\'{\i}}a-Lario}
  P.,   {Manchado} A.,  2011, \mn@doi [\aj] {10.1088/0004-6256/141/4/111},
  \href {http://adsabs.harvard.edu/abs/2011AJ....141..111C} {141, 111}

\bibitem[\protect\citeauthoryear{{Cox}, {Pilleri}, {Bern{\'e}}, {Cernicharo}
  \& {Joblin}}{{Cox} et~al.}{2016}]{cox2016}
{Cox} N.~L.~J.,  {Pilleri} P.,  {Bern{\'e}} O.,  {Cernicharo} J.,   {Joblin}
  C.,  2016, \mn@doi [\mnras] {10.1093/mnrasl/slv184}, \href
  {http://adsabs.harvard.edu/abs/2016MNRAS.456L..89C} {456, L89}

\bibitem[\protect\citeauthoryear{{De Marco}}{{De Marco}}{2009}]{orsola2009}
{De Marco} O.,  2009, \mn@doi [\pasp] {10.1086/597765}, \href
  {http://adsabs.harvard.edu/abs/2009PASP..121..316D} {121, 316}

\bibitem[\protect\citeauthoryear{{De Marco} et~al.}{{De Marco}
  et~al.}{2015}]{orsola2015}
{De Marco} O.,  et~al., 2015, \mn@doi [\mnras] {10.1093/mnras/stv249}, \href
  {http://adsabs.harvard.edu/abs/2015MNRAS.448.3587D} {448, 3587}

\bibitem[\protect\citeauthoryear{{Drew} et~al.}{{Drew} et~al.}{2005}]{drew2005}
{Drew} J.~E.,  et~al., 2005, \mn@doi [\mnras]
  {10.1111/j.1365-2966.2005.09330.x}, \href
  {http://adsabs.harvard.edu/abs/2005MNRAS.362..753D} {362, 753}

\bibitem[\protect\citeauthoryear{{Ducati}, {Bevilacqua}, {Rembold}  \&
  {Ribeiro}}{{Ducati} et~al.}{2001}]{ducati2001}
{Ducati} J.~R.,  {Bevilacqua} C.~M.,  {Rembold} S.~B.,   {Ribeiro} D.,  2001,
  \mn@doi [\apj] {10.1086/322439}, \href
  {http://adsabs.harvard.edu/abs/2001ApJ...558..309D} {558, 309}

\bibitem[\protect\citeauthoryear{{Felli}, {Testi}, {Schuller}  \&
  {Omont}}{{Felli} et~al.}{2002}]{fell2002}
{Felli} M.,  {Testi} L.,  {Schuller} F.,   {Omont} A.,  2002, \mn@doi [\aap]
  {10.1051/0004-6361:20020973}, \href
  {http://adsabs.harvard.edu/abs/2002A%26A...392..971F} {392, 971}

\bibitem[\protect\citeauthoryear{{Filipovi{\'c}} et~al.,}{{Filipovi{\'c}}
  et~al.}{2009}]{filli2009}
{Filipovi{\'c}} M.~D.,  et~al., 2009, \mn@doi [\mnras]
  {10.1111/j.1365-2966.2009.15307.x}, \href
  {http://adsabs.harvard.edu/abs/2009MNRAS.399..769F} {399, 769}

\bibitem[\protect\citeauthoryear{{Fragkou}, {Boji{\v c}i{\'c}}, {Frew}  \&
  {Parker}}{{Fragkou} et~al.}{2017}]{frag2016}
{Fragkou} V.,  {Boji{\v c}i{\'c}} I.,  {Frew} D.,   {Parker} Q.,  2017, in
  {Liu} X.,  {Stanghellini} L.,   {Karakas} A.,  eds,  IAU Symposium Vol. 323,
  Planetary Nebulae: Multi-Wavelength Probes of Stellar and Galactic Evolution.
  pp 329--330 (\mn@eprint {arXiv} {1611.08590}),
  \mn@doi{10.1017/S1743921317000473}

\bibitem[\protect\citeauthoryear{{Freeman} et~al.}{{Freeman}
  et~al.}{2014}]{freeman2014}
{Freeman} M.,  et~al., 2014, \mn@doi [\apj] {10.1088/0004-637X/794/2/99}, \href
  {http://adsabs.harvard.edu/abs/2014ApJ...794...99F} {794, 99}

\bibitem[\protect\citeauthoryear{{Frew} \& {Parker}}{{Frew} \&
  {Parker}}{2010}]{parker2010}
{Frew} D.~J.,  {Parker} Q.~A.,  2010, \mn@doi [\pasa] {10.1071/AS09040}, \href
  {http://adsabs.harvard.edu/abs/2010PASA...27..129F} {27, 129}

\bibitem[\protect\citeauthoryear{{Frew}, {Parker}  \& {Boji{\v
  c}i{\'c}}}{{Frew} et~al.}{2016}]{frew2016}
{Frew} D.~J.,  {Parker} Q.~A.,   {Boji{\v c}i{\'c}} I.~S.,  2016, \mn@doi
  [\mnras] {10.1093/mnras/stv1516}, \href
  {http://adsabs.harvard.edu/abs/2016MNRAS.455.1459F} {455, 1459}

\bibitem[\protect\citeauthoryear{{Gaensler} \& {Slane}}{{Gaensler} \&
  {Slane}}{2006}]{GS2006}
{Gaensler} B.~M.,  {Slane} P.~O.,  2006, \mn@doi [\araa]
  {10.1146/annurev.astro.44.051905.092528}, \href
  {http://adsabs.harvard.edu/abs/2006ARA%26A..44...17G} {44, 17}

\bibitem[\protect\citeauthoryear{{Garcia-Lario} et~al.}{{Garcia-Lario}
  et~al.}{1997}]{garcia1997}
{Garcia-Lario} P.,  et~al., 1997, \mn@doi [\aaps] {10.1051/aas:1997277}, \href
  {http://adsabs.harvard.edu/abs/1997A%26AS..126..479G} {126}

\bibitem[\protect\citeauthoryear{{Garc{\'{\i}}a-Segura}
  et~al.}{{Garc{\'{\i}}a-Segura} et~al.}{1999}]{garcia1999}
{Garc{\'{\i}}a-Segura} G.,  et~al., 1999, \mn@doi [\apj] {10.1086/307205},
  \href {http://adsabs.harvard.edu/abs/1999ApJ...517..767G} {517, 767}

\bibitem[\protect\citeauthoryear{{Garc{\'{\i}}a-Segura}
  et~al.}{{Garc{\'{\i}}a-Segura} et~al.}{2005}]{garcia2005}
{Garc{\'{\i}}a-Segura} G.,  et~al., 2005, \mn@doi [\apj] {10.1086/426110},
  \href {http://adsabs.harvard.edu/abs/2005ApJ...618..919G} {618, 919}

\bibitem[\protect\citeauthoryear{{G{\'o}mez} et~al.}{{G{\'o}mez}
  et~al.}{2011}]{gomez2011}
{G{\'o}mez} J.~F.,  et~al., 2011, \mn@doi [\apjl]
  {10.1088/2041-8205/739/1/L14}, \href
  {http://adsabs.harvard.edu/abs/2011ApJ...739L..14G} {739, L14}

\bibitem[\protect\citeauthoryear{{Gonz{\'a}lez-Solares}
  et~al.}{{Gonz{\'a}lez-Solares} et~al.}{2008}]{gon2008}
{Gonz{\'a}lez-Solares} E.~A.,  et~al., 2008, \mn@doi [\mnras]
  {10.1111/j.1365-2966.2008.13399.x}, \href
  {http://adsabs.harvard.edu/abs/2008MNRAS.388...89G} {388, 89}

\bibitem[\protect\citeauthoryear{{Gonzalez}, {Rejkuba}, {Zoccali}, {Valenti},
  {Minniti}, {Schultheis}, {Tobar}  \& {Chen}}{{Gonzalez}
  et~al.}{2012}]{gonza2012}
{Gonzalez} O.~A.,  {Rejkuba} M.,  {Zoccali} M.,  {Valenti} E.,  {Minniti} D.,
  {Schultheis} M.,  {Tobar} R.,   {Chen} B.,  2012, \mn@doi [\aap]
  {10.1051/0004-6361/201219222}, \href
  {http://adsabs.harvard.edu/abs/2012A%26A...543A..13G} {543, A13}

\bibitem[\protect\citeauthoryear{{Griffin} et~al.}{{Griffin}
  et~al.}{2009}]{griffin2009}
{Griffin} M.,  et~al., 2009, in {Pagani} L.,  {Gerin} M.,  eds,  EAS
  Publications Series Vol. 34, EAS Publications Series. pp 33--42,
  \mn@doi{10.1051/eas:0934003}

\bibitem[\protect\citeauthoryear{{Gutermuth} \& {Heyer}}{{Gutermuth} \&
  {Heyer}}{2015}]{rob2015}
{Gutermuth} R.~A.,  {Heyer} M.,  2015, \mn@doi [\aj]
  {10.1088/0004-6256/149/2/64}, \href
  {http://adsabs.harvard.edu/abs/2015AJ....149...64G} {149, 64}

\bibitem[\protect\citeauthoryear{{Guzman-Ramirez} et~al.}{{Guzman-Ramirez}
  et~al.}{2014}]{guz2014}
{Guzman-Ramirez} L.,  et~al., 2014, \mn@doi [\mnras] {10.1093/mnras/stu454},
  \href {http://adsabs.harvard.edu/abs/2014MNRAS.441..364G} {441, 364}

\bibitem[\protect\citeauthoryear{{Hands} et~al.}{{Hands}
  et~al.}{2004}]{hands2004}
{Hands} A.~D.~P.,  et~al., 2004, \mn@doi [\mnras]
  {10.1111/j.1365-2966.2004.07777.x}, \href
  {http://adsabs.harvard.edu/abs/2004MNRAS.351...31H} {351, 31}

\bibitem[\protect\citeauthoryear{{Helfand}, {Becker}, {White}, {Fallon}  \&
  {Tuttle}}{{Helfand} et~al.}{2006}]{hel2006}
{Helfand} D.~J.,  {Becker} R.~H.,  {White} R.~L.,  {Fallon} A.,   {Tuttle} S.,
  2006, \mn@doi [\aj] {10.1086/503253}, \href
  {http://adsabs.harvard.edu/abs/2006AJ....131.2525H} {131, 2525}

\bibitem[\protect\citeauthoryear{{Hoare}}{{Hoare}}{1990}]{hoare1990}
{Hoare} M.~G.,  1990, \mnras, \href
  {http://adsabs.harvard.edu/abs/1990MNRAS.244..193H} {244, 193}

\bibitem[\protect\citeauthoryear{{Hoare}}{{Hoare}}{2002}]{hoare2002}
{Hoare} M.~G.,  2002, in {Crowther} P.,  ed.,  Astronomical Society of the
  Pacific Conference Series Vol. 267, Hot Star Workshop III: The Earliest
  Phases of Massive Star Birth. p.~137

\bibitem[\protect\citeauthoryear{{Hoare}, {Drew}, {Muxlow}  \& {Davis}}{{Hoare}
  et~al.}{1994}]{hoare1994}
{Hoare} M.~G.,  {Drew} J.~E.,  {Muxlow} T.~B.,   {Davis} R.~J.,  1994, \mn@doi
  [\apjl] {10.1086/187185}, \href
  {http://adsabs.harvard.edu/abs/1994ApJ...421L..51H} {421, L51}

\bibitem[\protect\citeauthoryear{{Hoare} et~al.}{{Hoare}
  et~al.}{2012}]{cornish2012}
{Hoare} M.~G.,  et~al., 2012, \mn@doi [\pasp] {10.1086/668058}, \href
  {http://cdsads.u-strasbg.fr/abs/2012PASP..124..939H} {124, 939}

\bibitem[\protect\citeauthoryear{{Hora} et~al.}{{Hora} et~al.}{2004}]{hora2004}
{Hora} J.~L.,  et~al., 2004, \mn@doi [\apjs] {10.1086/422820}, \href
  {http://adsabs.harvard.edu/abs/2004ApJS..154..296H} {154, 296}

\bibitem[\protect\citeauthoryear{{Indebetouw} et~al.,}{{Indebetouw}
  et~al.}{2005}]{inde2005}
{Indebetouw} R.,  et~al., 2005, \mn@doi [\apj] {10.1086/426679}, \href
  {http://adsabs.harvard.edu/abs/2005ApJ...619..931I} {619, 931}

\bibitem[\protect\citeauthoryear{{Kalcheva}, {Hoare}, {Urquhart}, {Kurtz},
  {Lumsden}, {Purcell}  \& {Zijlstra}}{{Kalcheva} et~al.}{2018}]{kalprep}
{Kalcheva} I.~E.,  {Hoare} M.~G.,  {Urquhart} J.~S.,  {Kurtz} S.,  {Lumsden}
  S.~L.,  {Purcell} C.~R.,   {Zijlstra} A.~A.,  2018, preprint, \href
  {http://adsabs.harvard.edu/abs/2018arXiv180309334K} {} (\mn@eprint {arXiv}
  {1803.09334})

\bibitem[\protect\citeauthoryear{{Kaler}}{{Kaler}}{1983}]{kaler1983}
{Kaler} J.~B.,  1983, \mn@doi [\apj] {10.1086/160629}, \href
  {http://adsabs.harvard.edu/abs/1983ApJ...264..594K} {264, 594}

\bibitem[\protect\citeauthoryear{{Kanarek}, {Shara}, {Faherty}, {Zurek}  \&
  {Moffat}}{{Kanarek} et~al.}{2015}]{Kanarek2015}
{Kanarek} G.,  {Shara} M.,  {Faherty} J.,  {Zurek} D.,   {Moffat} A.,  2015,
  \mn@doi [\mnras] {10.1093/mnras/stv1342}, \href
  {http://adsabs.harvard.edu/abs/2015MNRAS.452.2858K} {452, 2858}

\bibitem[\protect\citeauthoryear{{Kastner}}{{Kastner}}{2007}]{kast2007}
{Kastner} J.~H.,  2007, in Asymmetrical Planetary Nebulae IV. p.~5 (\mn@eprint
  {arXiv} {0709.4136})

\bibitem[\protect\citeauthoryear{{Kastner}, {Montez}, {Balick}  \& {De
  Marco}}{{Kastner} et~al.}{2008}]{kast2008}
{Kastner} J.~H.,  {Montez} Jr. R.,  {Balick} B.,   {De Marco} O.,  2008,
  \mn@doi [\apj] {10.1086/523890}, \href
  {http://adsabs.harvard.edu/abs/2008ApJ...672..957K} {672, 957}

\bibitem[\protect\citeauthoryear{{Kistiakowsky} \& {Helfand}}{{Kistiakowsky} \&
  {Helfand}}{1995}]{kist1995}
{Kistiakowsky} V.,  {Helfand} D.~J.,  1995, \mn@doi [\aj] {10.1086/117683},
  \href {http://adsabs.harvard.edu/abs/1995AJ....110.2225K} {110, 2225}

\bibitem[\protect\citeauthoryear{{Kohoutek}}{{Kohoutek}}{2001}]{koh2001}
{Kohoutek} L.,  2001, VizieR Online Data Catalog, \href
  {http://adsabs.harvard.edu/abs/2001yCat.4024....0K} {4024}

\bibitem[\protect\citeauthoryear{{Kronberger}, {Jacoby}, {Harmer}  \&
  {Patchick}}{{Kronberger} et~al.}{2014}]{kron2014}
{Kronberger} M.,  {Jacoby} G.~H.,  {Harmer} D.,   {Patchick} D.,  2014, in
  Asymmetrical Planetary Nebulae VI Conference. p.~47

\bibitem[\protect\citeauthoryear{{Kwok}}{{Kwok}}{1982}]{kwok1982}
{Kwok} S.,  1982, \mn@doi [\apj] {10.1086/160078}, \href
  {http://adsabs.harvard.edu/abs/1982ApJ...258..280K} {258, 280}

\bibitem[\protect\citeauthoryear{{Kwok}}{{Kwok}}{1985}]{kwok1985}
{Kwok} S.,  1985, \mn@doi [\aj] {10.1086/113707}, \href
  {http://adsabs.harvard.edu/abs/1985AJ.....90...49K} {90, 49}

\bibitem[\protect\citeauthoryear{{Kwok}}{{Kwok}}{1993}]{kwok1993}
{Kwok} S.,  1993, \actaa, \href
  {http://adsabs.harvard.edu/abs/1993AcA....43..359K} {43, 359}

\bibitem[\protect\citeauthoryear{{Kwok}}{{Kwok}}{2003}]{kwok2003}
{Kwok} S.,  2003, in {Corradi} R.~L.~M.,  {Mikolajewska} J.,   {Mahoney} T.~J.,
   eds,  Astronomical Society of the Pacific Conference Series Vol. 303,
  Symbiotic Stars Probing Stellar Evolution. p.~428

\bibitem[\protect\citeauthoryear{{Kwok}}{{Kwok}}{2007}]{kwok2007}
{Kwok} S.,  2007, in Asymmetrical Planetary Nebulae IV.

\bibitem[\protect\citeauthoryear{{Kwok}}{{Kwok}}{2010}]{kwok2010}
{Kwok} S.,  2010, \mn@doi [\pasa] {10.1071/AS09027}, \href
  {http://adsabs.harvard.edu/abs/2010PASA...27..174K} {27, 174}

\bibitem[\protect\citeauthoryear{{Kwok}, {Hrivnak}  \& {Milone}}{{Kwok}
  et~al.}{1986}]{kwok1986}
{Kwok} S.,  {Hrivnak} B.~J.,   {Milone} E.~F.,  1986, \mn@doi [\apj]
  {10.1086/164090}, \href {http://adsabs.harvard.edu/abs/1986ApJ...303..451K}
  {303, 451}

\bibitem[\protect\citeauthoryear{{Laki{\'c}evi{\'c}}
  et~al.,}{{Laki{\'c}evi{\'c}} et~al.}{2015}]{laki2015}
{Laki{\'c}evi{\'c}} M.,  et~al., 2015, \mn@doi [\apj]
  {10.1088/0004-637X/799/1/50}, \href
  {http://adsabs.harvard.edu/abs/2015ApJ...799...50L} {799, 50}

\bibitem[\protect\citeauthoryear{{Leto} et~al.}{{Leto} et~al.}{2009}]{leto2009}
{Leto} P.,  et~al., 2009, \mn@doi [\aap] {10.1051/0004-6361/200911894}, \href
  {http://adsabs.harvard.edu/abs/2009A%26A...507.1467L} {507, 1467}

\bibitem[\protect\citeauthoryear{{Lucas} et~al.}{{Lucas}
  et~al.}{2008}]{lucas2008}
{Lucas} P.~W.,  et~al., 2008, \mn@doi [\mnras]
  {10.1111/j.1365-2966.2008.13924.x}, \href
  {http://adsabs.harvard.edu/abs/2008MNRAS.391..136L} {391, 136}

\bibitem[\protect\citeauthoryear{{Lumsden}, {Hoare}, {Urquhart}, {Oudmaijer},
  {Davies}, {Mottram}, {Cooper}  \& {Moore}}{{Lumsden} et~al.}{2013}]{lum2013}
{Lumsden} S.~L.,  {Hoare} M.~G.,  {Urquhart} J.~S.,  {Oudmaijer} R.~D.,
  {Davies} B.,  {Mottram} J.~C.,  {Cooper} H.~D.~B.,   {Moore} T.~J.~T.,  2013,
  \mn@doi [\apjs] {10.1088/0067-0049/208/1/11}, \href
  {http://adsabs.harvard.edu/abs/2013ApJS..208...11L} {208, 11}

\bibitem[\protect\citeauthoryear{{Mampaso} et~al.}{{Mampaso}
  et~al.}{2006}]{mampaso2006}
{Mampaso} A.,  et~al., 2006, \mn@doi [\aap] {10.1051/0004-6361:20054778}, \href
  {http://adsabs.harvard.edu/abs/2006A%26A...458..203M} {458, 203}

\bibitem[\protect\citeauthoryear{{Marquez-Lugo} et~al.}{{Marquez-Lugo}
  et~al.}{2015}]{mar2015}
{Marquez-Lugo} R.~A.,  et~al., 2015, \mn@doi [\mnras] {10.1093/mnras/stv1783},
  \href {http://adsabs.harvard.edu/abs/2015MNRAS.453.1888M} {453, 1888}

\bibitem[\protect\citeauthoryear{{Matsuura}, {Zijlstra}, {Gray}, {Molster}  \&
  {Waters}}{{Matsuura} et~al.}{2005}]{mat2005}
{Matsuura} M.,  {Zijlstra} A.~A.,  {Gray} M.~D.,  {Molster} F.~J.,   {Waters}
  L.~B.~F.~M.,  2005, \mn@doi [\mnras] {10.1111/j.1365-2966.2005.09464.x},
  \href {http://adsabs.harvard.edu/abs/2005MNRAS.363..628M} {363, 628}

\bibitem[\protect\citeauthoryear{{Matsuura}, {Indebetouw}, {Kamenetzky},
  {McCray}, {Zanardo}, {Barlow}  \& {Dwek}}{{Matsuura} et~al.}{2015}]{mats2015}
{Matsuura} M.,  {Indebetouw} R.,  {Kamenetzky} J.,  {McCray} R.,  {Zanardo} G.,
   {Barlow} M.~J.,   {Dwek} E.,  2015, in {Iono} D.,  {Tatematsu} K.,
  {Wootten} A.,   {Testi} L.,  eds,  Astronomical Society of the Pacific
  Conference Series Vol. 499, Revolution in Astronomy with ALMA: The Third
  Year. p.~323

\bibitem[\protect\citeauthoryear{{Miszalski}, {Parker}, {Acker}, {Birkby},
  {Frew}  \& {Kovacevic}}{{Miszalski} et~al.}{2008}]{mis2008}
{Miszalski} B.,  {Parker} Q.~A.,  {Acker} A.,  {Birkby} J.~L.,  {Frew} D.~J.,
  {Kovacevic} A.,  2008, \mn@doi [\mnras] {10.1111/j.1365-2966.2007.12727.x},
  \href {http://adsabs.harvard.edu/abs/2008MNRAS.384..525M} {384, 525}

\bibitem[\protect\citeauthoryear{{Miszalski} et~al.}{{Miszalski}
  et~al.}{2013}]{misz2013}
{Miszalski} B.,  et~al., 2013, \mn@doi [\mnras] {10.1093/mnras/stt673}, \href
  {http://adsabs.harvard.edu/abs/2013MNRAS.432.3186M} {432, 3186}

\bibitem[\protect\citeauthoryear{{Moe} \& {De Marco}}{{Moe} \& {De
  Marco}}{2006}]{moe2006}
{Moe} M.,  {De Marco} O.,  2006, \mn@doi [\apj] {10.1086/506900}, \href
  {http://adsabs.harvard.edu/abs/2006ApJ...650..916M} {650, 916}

\bibitem[\protect\citeauthoryear{{Molinari} et~al.}{{Molinari}
  et~al.}{2010}]{molinari2010}
{Molinari} S.,  et~al., 2010, \mn@doi [\pasp] {10.1086/651314}, \href
  {http://adsabs.harvard.edu/abs/2010PASP..122..314M} {122, 314}

\bibitem[\protect\citeauthoryear{{Molinari} et~al.}{{Molinari}
  et~al.}{2016}]{molinari2016}
{Molinari} S.,  et~al., 2016, \mn@doi [\aap] {10.1051/0004-6361/201526380},
  \href {http://adsabs.harvard.edu/abs/2016A%26A...591A.149M} {591, A149}

\bibitem[\protect\citeauthoryear{{Oliveira} et~al.,}{{Oliveira}
  et~al.}{2013}]{oliver2013}
{Oliveira} J.~M.,  et~al., 2013, \mn@doi [\mnras] {10.1093/mnras/sts250}, \href
  {http://adsabs.harvard.edu/abs/2013MNRAS.428.3001O} {428, 3001}

\bibitem[\protect\citeauthoryear{{Parker} et~al.,}{{Parker}
  et~al.}{2005}]{parker2005}
{Parker} Q.~A.,  et~al., 2005, \mn@doi [\mnras]
  {10.1111/j.1365-2966.2005.09350.x}, \href
  {http://adsabs.harvard.edu/abs/2005MNRAS.362..689P} {362, 689}

\bibitem[\protect\citeauthoryear{{Parker} et~al.}{{Parker}
  et~al.}{2006}]{parker2006}
{Parker} Q.~A.,  et~al., 2006, \mn@doi [\mnras]
  {10.1111/j.1365-2966.2006.10950.x}, \href
  {http://adsabs.harvard.edu/abs/2006MNRAS.373...79P} {373, 79}

\bibitem[\protect\citeauthoryear{{Parker} et~al.,}{{Parker}
  et~al.}{2012}]{parker_2_2012}
{Parker} Q.~A.,  et~al., 2012, \mn@doi [\mnras]
  {10.1111/j.1365-2966.2012.21927.x}, \href
  {http://adsabs.harvard.edu/abs/2012MNRAS.427.3016P} {427, 3016}

\bibitem[\protect\citeauthoryear{{Parker}, {Boji{\v c}i{\'c}}  \&
  {Frew}}{{Parker} et~al.}{2017}]{parker2017}
{Parker} Q.~A.,  {Boji{\v c}i{\'c}} I.,   {Frew} D.~J.,  2017, in {Liu} X.,
  {Stanghellini} L.,   {Karakas} A.,  eds,  IAU Symposium Vol. 323, Planetary
  Nebulae: Multi-Wavelength Probes of Stellar and Galactic Evolution. pp 36--39
  (\mn@eprint {arXiv} {1612.00167}), \mn@doi{10.1017/S1743921317000904}

\bibitem[\protect\citeauthoryear{{Pena} \& {Torres-Peimbert}}{{Pena} \&
  {Torres-Peimbert}}{1987}]{pena1987}
{Pena} M.,  {Torres-Peimbert} S.,  1987, \rmxaa, \href
  {http://adsabs.harvard.edu/abs/1987RMxAA..14..534P} {14, 534}

\bibitem[\protect\citeauthoryear{{P{\'e}rez-S{\'a}nchez}
  et~al.}{{P{\'e}rez-S{\'a}nchez} et~al.}{2013}]{perez2013}
{P{\'e}rez-S{\'a}nchez} A.~F.,  et~al., 2013, \mn@doi [\mnras]
  {10.1093/mnrasl/slt117}, \href
  {http://adsabs.harvard.edu/abs/2013MNRAS.436L..79P} {436, L79}

\bibitem[\protect\citeauthoryear{{Phillips}}{{Phillips}}{2003}]{phil2003}
{Phillips} J.~P.,  2003, \mn@doi [\aap] {10.1051/0004-6361:20031246}, \href
  {http://adsabs.harvard.edu/abs/2003A%26A...412..791P} {412, 791}

\bibitem[\protect\citeauthoryear{{Phillips} \& {M{\'a}rquez-Lugo}}{{Phillips}
  \& {M{\'a}rquez-Lugo}}{2011}]{phillips2011}
{Phillips} J.~P.,  {M{\'a}rquez-Lugo} R.~A.,  2011, \rmxaa, \href
  {http://adsabs.harvard.edu/abs/2011RMxAA..47...83P} {47, 83}

\bibitem[\protect\citeauthoryear{{Phillips} \& {Ramos-Larios}}{{Phillips} \&
  {Ramos-Larios}}{2008}]{phillips2008}
{Phillips} J.~P.,  {Ramos-Larios} G.,  2008, \mn@doi [\mnras]
  {10.1111/j.1365-2966.2008.13083.x}, \href
  {http://adsabs.harvard.edu/abs/2008MNRAS.386..995P} {386, 995}

\bibitem[\protect\citeauthoryear{{Phillips} \& {Ramos-Larios}}{{Phillips} \&
  {Ramos-Larios}}{2009}]{phil_b2009}
{Phillips} J.~P.,  {Ramos-Larios} G.,  2009, \mn@doi [\mnras]
  {10.1111/j.1365-2966.2009.14846.x}, \href
  {http://adsabs.harvard.edu/abs/2009MNRAS.396.1915P} {396, 1915}

\bibitem[\protect\citeauthoryear{{Phillips} \& {Zepeda-Garcia}}{{Phillips} \&
  {Zepeda-Garcia}}{2009}]{phil2009}
{Phillips} J.~P.,  {Zepeda-Garcia} D.,  2009, \mn@doi [\mnras]
  {10.1111/j.1365-2966.2009.14495.x}, \href
  {http://adsabs.harvard.edu/abs/2009MNRAS.394.1875P} {394, 1875}

\bibitem[\protect\citeauthoryear{{Pinheiro Gon{\c c}alves}, {Noriega-Crespo},
  {Paladini}, {Martin}  \& {Carey}}{{Pinheiro Gon{\c c}alves}
  et~al.}{2011}]{pin2011}
{Pinheiro Gon{\c c}alves} D.,  {Noriega-Crespo} A.,  {Paladini} R.,  {Martin}
  P.~G.,   {Carey} S.~J.,  2011, \mn@doi [\aj] {10.1088/0004-6256/142/2/47},
  \href {http://adsabs.harvard.edu/abs/2011AJ....142...47P} {142, 47}

\bibitem[\protect\citeauthoryear{{Poglitsch} et~al.}{{Poglitsch}
  et~al.}{2008}]{pog2008}
{Poglitsch} A.,  et~al., 2008, in Space Telescopes and Instrumentation 2008:
  Optical, Infrared, and Millimeter. p. 701005, \mn@doi{10.1117/12.790016}

\bibitem[\protect\citeauthoryear{{Pottasch}}{{Pottasch}}{1986}]{pottasch1986}
{Pottasch} S.~R.,  1986, Comptes Rendus sur les Journees de Strasbourg, \href
  {http://adsabs.harvard.edu/abs/1986CRJS....8...97P} {8, 97}

\bibitem[\protect\citeauthoryear{{Pottasch} et~al.,}{{Pottasch}
  et~al.}{1984}]{pottasch1984}
{Pottasch} S.~R.,  et~al., 1984, \aap, \href
  {http://adsabs.harvard.edu/abs/1984A%26A...138...10P} {138, 10}

\bibitem[\protect\citeauthoryear{{Pottasch}, {Beintema}  \&
  {Feibelman}}{{Pottasch} et~al.}{2000}]{pottasch2000}
{Pottasch} S.~R.,  {Beintema} D.~A.,   {Feibelman} W.~A.,  2000, \aap, \href
  {http://adsabs.harvard.edu/abs/2000A%26A...363..767P} {363, 767}

\bibitem[\protect\citeauthoryear{{Preite-Martinez}}{{Preite-Martinez}}{1988}]{pm1988}
{Preite-Martinez} A.,  1988, \aaps, \href
  {http://adsabs.harvard.edu/abs/1988A%26AS...76..317P} {76, 317}

\bibitem[\protect\citeauthoryear{{Purcell} et~al.}{{Purcell}
  et~al.}{2013}]{cornissh2013}
{Purcell} C.~R.,  et~al., 2013, \mn@doi [\apjs] {10.1088/0067-0049/205/1/1},
  \href {http://cdsads.u-strasbg.fr/abs/2013ApJS..205....1P} {205, 1}

\bibitem[\protect\citeauthoryear{{Ramos-Larios} \& {Phillips}}{{Ramos-Larios}
  \& {Phillips}}{2005}]{phil2005}
{Ramos-Larios} G.,  {Phillips} J.~P.,  2005, \mn@doi [\mnras]
  {10.1111/j.1365-2966.2005.08713.x}, \href
  {http://adsabs.harvard.edu/abs/2005MNRAS.357..732R} {357, 732}

\bibitem[\protect\citeauthoryear{{Ramos-Larios}, {Guerrero}, {Sabin}  \&
  {Santamar{\'{\i}}a}}{{Ramos-Larios} et~al.}{2017}]{ramos2017}
{Ramos-Larios} G.,  {Guerrero} M.~A.,  {Sabin} L.,   {Santamar{\'{\i}}a} E.,
  2017, \mn@doi [\mnras] {10.1093/mnras/stx1519}, \href
  {http://adsabs.harvard.edu/abs/2017MNRAS.470.3707R} {470, 3707}

\bibitem[\protect\citeauthoryear{{Ratag} \& {Pottasch}}{{Ratag} \&
  {Pottasch}}{1991}]{ratag1991}
{Ratag} M.~A.,  {Pottasch} S.~R.,  1991, \aaps, \href
  {http://adsabs.harvard.edu/abs/1991A%26AS...91..481R} {91, 481}

\bibitem[\protect\citeauthoryear{{Reach} et~al.}{{Reach}
  et~al.}{2005}]{reach2005}
{Reach} W.~T.,  et~al., 2005, \mn@doi [\pasp] {10.1086/432670}, \href
  {http://adsabs.harvard.edu/abs/2005PASP..117..978R} {117, 978}

\bibitem[\protect\citeauthoryear{{Reach} et~al.,}{{Reach}
  et~al.}{2006}]{reachwil2006}
{Reach} W.~T.,  et~al., 2006, \mn@doi [\aj] {10.1086/499306}, \href
  {http://adsabs.harvard.edu/abs/2006AJ....131.1479R} {131, 1479}

\bibitem[\protect\citeauthoryear{{Ressler}, {Cohen}, {Wachter}, {Hoard},
  {Mainzer}  \& {Wright}}{{Ressler} et~al.}{2010}]{ress2010}
{Ressler} M.~E.,  {Cohen} M.,  {Wachter} S.,  {Hoard} D.~W.,  {Mainzer} A.~K.,
   {Wright} E.~L.,  2010, \mn@doi [\aj] {10.1088/0004-6256/140/6/1882}, \href
  {http://adsabs.harvard.edu/abs/2010AJ....140.1882R} {140, 1882}

\bibitem[\protect\citeauthoryear{{Robitaille} et~al.}{{Robitaille}
  et~al.}{2008}]{robi2008}
{Robitaille} T.~P.,  et~al., 2008, \mn@doi [\aj]
  {10.1088/0004-6256/136/6/2413}, \href
  {http://cdsads.u-strasbg.fr/abs/2008AJ....136.2413R} {136, 2413}

\bibitem[\protect\citeauthoryear{{Rodr{\'{\i}}guez-Flores}
  et~al.}{{Rodr{\'{\i}}guez-Flores} et~al.}{2014}]{rog2014}
{Rodr{\'{\i}}guez-Flores} E.~R.,  et~al., 2014, \mn@doi [\aap]
  {10.1051/0004-6361/201323182}, \href
  {http://adsabs.harvard.edu/abs/2014A%26A...567A..49R} {567, A49}

\bibitem[\protect\citeauthoryear{{Rosen} et~al.,}{{Rosen}
  et~al.}{2016}]{rosen2016}
{Rosen} S.~R.,  et~al., 2016, \mn@doi [\aap] {10.1051/0004-6361/201526416},
  \href {http://adsabs.harvard.edu/abs/2016A%26A...590A...1R} {590, A1}

\bibitem[\protect\citeauthoryear{{Rosolowsky} et~al.}{{Rosolowsky}
  et~al.}{2010a}]{rob2010}
{Rosolowsky} E.,  et~al., 2010a, \mn@doi [\apjs] {10.1088/0067-0049/188/1/123},
  \href {http://adsabs.harvard.edu/abs/2010ApJS..188..123R} {188, 123}

\bibitem[\protect\citeauthoryear{{Rosolowsky} et~al.,}{{Rosolowsky}
  et~al.}{2010b}]{rob22010}
{Rosolowsky} E.,  et~al., 2010b, \mn@doi [\apjs] {10.1088/0067-0049/188/1/123},
  \href {http://adsabs.harvard.edu/abs/2010ApJS..188..123R} {188, 123}

\bibitem[\protect\citeauthoryear{{Ruffle} et~al.}{{Ruffle}
  et~al.}{2004}]{ruf2004}
{Ruffle} P.~M.~E.,  et~al., 2004, \mn@doi [\mnras]
  {10.1111/j.1365-2966.2004.08113.x}, \href
  {http://adsabs.harvard.edu/abs/2004MNRAS.353..796R} {353, 796}

\bibitem[\protect\citeauthoryear{{Sabin} et~al.}{{Sabin}
  et~al.}{2014}]{sabin2014}
{Sabin} L.,  et~al., 2014, \mn@doi [\mnras] {10.1093/mnras/stu1404}, \href
  {http://adsabs.harvard.edu/abs/2014MNRAS.443.3388S} {443, 3388}

\bibitem[\protect\citeauthoryear{{Sahai} \& {Trauger}}{{Sahai} \&
  {Trauger}}{1998}]{SahaiTrauger1998}
{Sahai} R.,  {Trauger} J.~T.,  1998, \mn@doi [\aj] {10.1086/300504}, \href
  {http://adsabs.harvard.edu/abs/1998AJ....116.1357S} {116, 1357}

\bibitem[\protect\citeauthoryear{{Sahai} et~al.}{{Sahai}
  et~al.}{2007}]{sahai2007}
{Sahai} R.,  et~al., 2007, \mn@doi [\aj] {10.1086/522944}, \href
  {http://adsabs.harvard.edu/abs/2007AJ....134.2200S} {134, 2200}

\bibitem[\protect\citeauthoryear{{Sahai} et~al.}{{Sahai}
  et~al.}{2011}]{sahai2011}
{Sahai} R.,  et~al., 2011, \mn@doi [\aj] {10.1088/0004-6256/141/4/134}, \href
  {http://adsabs.harvard.edu/abs/2011AJ....141..134S} {141, 134}

\bibitem[\protect\citeauthoryear{{Sewilo}, {Churchwell}, {Kurtz}, {Goss}  \&
  {Hofner}}{{Sewilo} et~al.}{2004}]{Sewilo2004}
{Sewilo} M.,  {Churchwell} E.,  {Kurtz} S.,  {Goss} W.~M.,   {Hofner} P.,
  2004, in American Astronomical Society Meeting Abstracts. p.~1574

\bibitem[\protect\citeauthoryear{{Si{\'o}dmiak} \& {Tylenda}}{{Si{\'o}dmiak} \&
  {Tylenda}}{2001}]{siodmiak2001}
{Si{\'o}dmiak} N.,  {Tylenda} R.,  2001, \mn@doi [\aap]
  {10.1051/0004-6361:20010664}, \href
  {http://adsabs.harvard.edu/abs/2001A%26A...373.1032S} {373, 1032}

\bibitem[\protect\citeauthoryear{{Stanghellini}, {Shaw}  \&
  {Villaver}}{{Stanghellini} et~al.}{2008}]{stang2008}
{Stanghellini} L.,  {Shaw} R.~A.,   {Villaver} E.,  2008, \mn@doi [\apj]
  {10.1086/592395}, \href {http://adsabs.harvard.edu/abs/2008ApJ...689..194S}
  {689, 194}

\bibitem[\protect\citeauthoryear{{Su{\'a}rez}, {Garc{\'{\i}}a-Lario},
  {Manchado}, {Manteiga}, {Ulla}  \& {Pottasch}}{{Su{\'a}rez}
  et~al.}{2006}]{su2006}
{Su{\'a}rez} O.,  {Garc{\'{\i}}a-Lario} P.,  {Manchado} A.,  {Manteiga} M.,
  {Ulla} A.,   {Pottasch} S.~R.,  2006, \mn@doi [\aap]
  {10.1051/0004-6361:20054108}, \href
  {http://adsabs.harvard.edu/abs/2006A%26A...458..173S} {458, 173}

\bibitem[\protect\citeauthoryear{{Suarez} et~al.}{{Suarez}
  et~al.}{2015}]{suarez2015}
{Suarez} O.,  et~al., 2015, preprint, \href
  {http://adsabs.harvard.edu/abs/2015arXiv150404277S} {} (\mn@eprint {arXiv}
  {1504.04277})

\bibitem[\protect\citeauthoryear{{Szczerba}, {Si{\'o}dmiak}, {Stasi{\'n}ska}
  \& {Borkowski}}{{Szczerba} et~al.}{2007}]{Szczerba2007}
{Szczerba} R.,  {Si{\'o}dmiak} N.,  {Stasi{\'n}ska} G.,   {Borkowski} J.,
  2007, \mn@doi [\aap] {10.1051/0004-6361:20067035}, \href
  {http://adsabs.harvard.edu/abs/2007A%26A...469..799S} {469, 799}

\bibitem[\protect\citeauthoryear{{Taylor}, {Goss}, {Coleman}, {van Leeuwen}  \&
  {Wallace}}{{Taylor} et~al.}{1996}]{Taylor1996}
{Taylor} A.~R.,  {Goss} W.~M.,  {Coleman} P.~H.,  {van Leeuwen} J.,   {Wallace}
  B.~J.,  1996, \mn@doi [\apjs] {10.1086/192363}, \href
  {http://adsabs.harvard.edu/abs/1996ApJS..107..239T} {107, 239}

\bibitem[\protect\citeauthoryear{{Tylenda}, {Acker}, {Stenholm}  \&
  {Koeppen}}{{Tylenda} et~al.}{1992}]{tylenda1992}
{Tylenda} R.,  {Acker} A.,  {Stenholm} B.,   {Koeppen} J.,  1992, \aaps, \href
  {http://adsabs.harvard.edu/abs/1992A%26AS...95..337T} {95, 337}

\bibitem[\protect\citeauthoryear{{Urquhart} et~al.}{{Urquhart}
  et~al.}{2009a}]{urq2009}
{Urquhart} J.~S.,  et~al., 2009a, \mn@doi [\aap] {10.1051/0004-6361/200912108},
  \href {http://cdsads.u-strasbg.fr/abs/2009A%26A...501..539U} {501, 539}

\bibitem[\protect\citeauthoryear{{Urquhart} et~al.,}{{Urquhart}
  et~al.}{2009b}]{Urquhart2009}
{Urquhart} J.~S.,  et~al., 2009b, \mn@doi [\aap] {10.1051/0004-6361/200912108},
  \href {http://adsabs.harvard.edu/abs/2009A%26A...501..539U} {501, 539}

\bibitem[\protect\citeauthoryear{{Urquhart} et~al.,}{{Urquhart}
  et~al.}{2013}]{urq2013}
{Urquhart} J.~S.,  et~al., 2013, \mn@doi [\mnras] {10.1093/mnras/stt1310},
  \href {http://adsabs.harvard.edu/abs/2013MNRAS.435..400U} {435, 400}

\bibitem[\protect\citeauthoryear{{Van de Steene}}{{Van de
  Steene}}{2017}]{van2017}
{Van de Steene} G.~C.,  2017, preprint, \href
  {http://adsabs.harvard.edu/abs/2017arXiv170106797V} {} (\mn@eprint {arXiv}
  {1701.06797})

\bibitem[\protect\citeauthoryear{{Van de Steene} et~al.,}{{Van de Steene}
  et~al.}{2015}]{van2015}
{Van de Steene} G.~C.,  et~al., 2015, \mn@doi [\aap]
  {10.1051/0004-6361/201424189}, \href
  {http://adsabs.harvard.edu/abs/2015A%26A...574A.134V} {574, A134}

\bibitem[\protect\citeauthoryear{{Viironen} et~al.}{{Viironen}
  et~al.}{2009a}]{v2009}
{Viironen} K.,  et~al., 2009a, VizieR Online Data Catalog, \href
  {http://adsabs.harvard.edu/abs/2009yCat..35040291V} {350}

\bibitem[\protect\citeauthoryear{{Viironen} et~al.}{{Viironen}
  et~al.}{2009b}]{vii2009}
{Viironen} K.,  et~al., 2009b, \mn@doi [\aap] {10.1051/0004-6361/200811575},
  \href {http://ukads.nottingham.ac.uk/abs/2009A%26A...502..113V} {502, 113}

\bibitem[\protect\citeauthoryear{{Viironen} et~al.}{{Viironen}
  et~al.}{2009c}]{viironen2009}
{Viironen} K.,  et~al., 2009c, \mn@doi [\aap] {10.1051/0004-6361/200912002},
  \href {http://adsabs.harvard.edu/abs/2009A%26A...504..291V} {504, 291}

\bibitem[\protect\citeauthoryear{{Villaver}, {Garc{\'{\i}}a-Segura}  \&
  {Manchado}}{{Villaver} et~al.}{2002}]{vill2002}
{Villaver} E.,  {Garc{\'{\i}}a-Segura} G.,   {Manchado} A.,  2002, \mn@doi
  [\apj] {10.1086/340022}, \href
  {http://adsabs.harvard.edu/abs/2002ApJ...571..880V} {571, 880}

\bibitem[\protect\citeauthoryear{{Weidmann}, {Gamen}, {van Hoof}, {Zijlstra},
  {Minniti}  \& {Volpe}}{{Weidmann} et~al.}{2013}]{Weid2013}
{Weidmann} W.~A.,  {Gamen} R.,  {van Hoof} P.~A.~M.,  {Zijlstra} A.,  {Minniti}
  D.,   {Volpe} M.~G.,  2013, \mn@doi [\aap] {10.1051/0004-6361/201220492},
  \href {http://adsabs.harvard.edu/abs/2013A%26A...552A..74W} {552, A74}

\bibitem[\protect\citeauthoryear{{White} et~al.}{{White}
  et~al.}{2005}]{white2005}
{White} R.~L.,  et~al., 2005, \mn@doi [\aj] {10.1086/431249}, \href
  {http://adsabs.harvard.edu/abs/2005AJ....130..586W} {130, 586}

\bibitem[\protect\citeauthoryear{{Whitelock}}{{Whitelock}}{1985}]{white1985}
{Whitelock} P.~A.,  1985, \mn@doi [\mnras] {10.1093/mnras/213.1.59}, \href
  {http://adsabs.harvard.edu/abs/1985MNRAS.213...59W} {213, 59}

\bibitem[\protect\citeauthoryear{{Williams} \& {Temim}}{{Williams} \&
  {Temim}}{2016}]{williams2016}
{Williams} B.~J.,  {Temim} T.,  2016, {Infrared Emission from Supernova
  Remnants: Formation and Destruction of Dust},
  \mn@doi{10.1007/978-3-319-20794-0_94-1.
}

\bibitem[\protect\citeauthoryear{{Willner}, {Becklin}  \&
  {Visvanathan}}{{Willner} et~al.}{1972}]{willner1972}
{Willner} S.~P.,  {Becklin} E.~E.,   {Visvanathan} N.,  1972, \mn@doi [\apj]
  {10.1086/151591}, \href {http://adsabs.harvard.edu/abs/1972ApJ...175..699W}
  {175, 699}

\bibitem[\protect\citeauthoryear{{Wood} \& {Churchwell}}{{Wood} \&
  {Churchwell}}{1989a}]{wood21989}
{Wood} D.~O.~S.,  {Churchwell} E.,  1989a, \mn@doi [\apjs] {10.1086/191329},
  \href {http://adsabs.harvard.edu/abs/1989ApJS...69..831W} {69, 831}

\bibitem[\protect\citeauthoryear{{Wood} \& {Churchwell}}{{Wood} \&
  {Churchwell}}{1989b}]{wood11989}
{Wood} D.~O.~S.,  {Churchwell} E.,  1989b, \mn@doi [\apj] {10.1086/167390},
  \href {http://adsabs.harvard.edu/abs/1989ApJ...340..265W} {340, 265}

\bibitem[\protect\citeauthoryear{{Woodall} \& {Gray}}{{Woodall} \&
  {Gray}}{2007}]{woodall2007}
{Woodall} J.~M.,  {Gray} M.~D.,  2007, \mn@doi [\mnras]
  {10.1111/j.1745-3933.2007.00311.x}, \href
  {http://adsabs.harvard.edu/abs/2007MNRAS.378L..20W} {378, L20}

\bibitem[\protect\citeauthoryear{{Wright} et~al.,}{{Wright}
  et~al.}{2010}]{wright2010}
{Wright} E.~L.,  et~al., 2010, \mn@doi [\aj] {10.1088/0004-6256/140/6/1868},
  \href {http://adsabs.harvard.edu/abs/2010AJ....140.1868W} {140, 1868}

\bibitem[\protect\citeauthoryear{{Zijlstra}}{{Zijlstra}}{1990}]{zij1990}
{Zijlstra} A.~A.,  1990, \aap, \href
  {http://adsabs.harvard.edu/abs/1990A%26A...234..387Z} {234, 387}

\bibitem[\protect\citeauthoryear{{Zijlstra} \& {Pottasch}}{{Zijlstra} \&
  {Pottasch}}{1991}]{zi1991}
{Zijlstra} A.~A.,  {Pottasch} S.~R.,  1991, \aap, \href
  {http://adsabs.harvard.edu/abs/1991A%26A...243..478Z} {243, 478}

\bibitem[\protect\citeauthoryear{{Zoonematkermani} et~al.}{{Zoonematkermani}
  et~al.}{1990}]{zoo1990}
{Zoonematkermani} S.,  et~al., 1990, \mn@doi [\apjs] {10.1086/191496}, \href
  {http://adsabs.harvard.edu/abs/1990ApJS...74..181Z} {74, 181}

\bibitem[\protect\citeauthoryear{{van Hoof}}{{van Hoof}}{2000}]{vanhoof2000}
{van Hoof} P.~A.~M.,  2000, \mn@doi [\mnras]
  {10.1046/j.1365-8711.2000.03281.x}, \href
  {http://adsabs.harvard.edu/abs/2000MNRAS.314...99V} {314, 99}

\bibitem[\protect\citeauthoryear{{van Hoof}, {Oudmaijer}  \& {Waters}}{{van
  Hoof} et~al.}{1997}]{van1997}
{van Hoof} P.~A.~M.,  {Oudmaijer} R.~D.,   {Waters} L.~B.~F.~M.,  1997, \mn@doi
  [\mnras] {10.1093/mnras/289.2.371}, \href
  {http://adsabs.harvard.edu/abs/1997MNRAS.289..371V} {289, 371}

\bibitem[\protect\citeauthoryear{{van de Steene} \& {Pottasch}}{{van de Steene}
  \& {Pottasch}}{1995}]{vande1995}
{van de Steene} G.~C.,  {Pottasch} S.~R.,  1995, \aap, \href
  {http://adsabs.harvard.edu/abs/1995A%26A...299..238V} {299, 238}

\bibitem[\protect\citeauthoryear{{van de Steene}, {Jacoby}  \& {Pottasch}}{{van
  de Steene} et~al.}{1996}]{van1996}
{van de Steene} G.~C.,  {Jacoby} G.~H.,   {Pottasch} S.~R.,  1996, \aaps, \href
  {http://adsabs.harvard.edu/abs/1996A%26AS..118..243V} {118, 243}

\makeatother
\end{thebibliography}
\bibliographystyle{mnras}



\bsp	
\label{lastpage}
\end{document}